\documentclass[notitlepage,prd,showkeys,preprintnumbers,floatfix,superscriptaddress,nofootinbib]{revtex4-1}
\usepackage{float}
\usepackage{amsmath,amssymb}
\usepackage{graphicx}
\usepackage{mathtools,slashed}
\usepackage{bm}
\usepackage{comment}
\usepackage{xcolor}
\usepackage{subfigure}
\usepackage{array}
\usepackage{multirow}
\usepackage{url}
\usepackage{color}
\usepackage[T1]{fontenc}
\usepackage[latin9]{inputenc}
\usepackage{graphicx}
\usepackage{esint}
\usepackage{hyperref}
\usepackage{atbegshi}
\usepackage{lipsum}

\newcommand{\nn}[0]{\nonumber}

\newcommand{\W}[0]{{\mathcal{W}}}
\newcommand{\cW}[0]{\mathcal{W}}
\newcommand{\df}[0]{\mathrm{df}}
\newcommand{\M}[0]{{\mathcal{M}}}
\newcommand{\Wtildf}[0]{{\mathcal{W}}_{L,{\rm df}}}
\newcommand{\Wdf}[0]{{\mathcal{W}}_{\rm df}}

\newcommand{\w}[0]{w}

\newcommand{\KSS}[0]{Kim:2005gf}
\newcommand{\origTwoTwo}[0]{Briceno:2015tza}

\begin{document}

\preprint{\vbox{\hbox{JLAB-THY-18-2878} }}
\preprint{\vbox{\hbox{CERN-TH-2018-263} }}

\title{
Form factors of two-hadron states from a covariant finite-volume formalism
}

\author{Alessandro Baroni}
\email[]{abaro008@odu.edu}
\affiliation{
Department of Physics and Astronomy University of South Carolina, \\ 
712 Main Street, Columbia, South Carolina  29208, USA
}
\author{Ra\'ul A.~Brice\~no}
\email[]{rbriceno@jlab.org}
\affiliation{Thomas Jefferson National Accelerator Facility, 12000 Jefferson Avenue, Newport News, Virginia 23606, USA}
\affiliation{ Department of Physics, Old Dominion University, Norfolk, Virginia 23529, USA}
\author{Maxwell T. Hansen}
\email[]{maxwell.hansen@cern.ch}
\affiliation{Theoretical Physics Department, CERN, 1211 Geneva 23, Switzerland}
\author{Felipe G. Ortega-Gama}
\email[]{felortga@jlab.org}
\affiliation{Thomas Jefferson National Accelerator Facility, 12000 Jefferson Avenue, Newport News, Virginia 23606, USA}
\affiliation{Department of Physics, College of William and Mary, Williamsburg, Virginia 23187, USA}

\date{\today}

\begin{abstract}
In this work we develop a Lorentz-covariant version of the previously derived formalism for relating finite-volume matrix elements to 
$\textbf 2 + \mathcal J \to \textbf 2$ transition amplitudes. We also give various details relevant for the implementation of this formalism in a realistic numerical lattice QCD calculation. Particular focus is given to the role of single-particle form factors
in disentangling finite-volume effects from the triangle diagram that arise when $\mathcal J$ couples to one of the two hadrons. This also leads to a new finite-volume function, denoted $G$, the numerical evaluation of which is described in detail. As an example we discuss the determination of the $\pi \pi + \mathcal J \to \pi \pi$ amplitude in the $\rho$ channel, for which the single-pion form factor, $F_\pi(Q^2)$, as well as the scattering phase, $\delta_{\pi\pi}$, are required to remove all power-law finite-volume effects. The formalism presented here holds for local currents with arbitrary Lorentz structure, and we give specific examples of insertions with up to two Lorentz indices.

 \end{abstract}

\keywords{finite volume, lattice QCD}

\nopagebreak
\maketitle

\section{Introduction\label{sec:intro}}

In recent years, interest in hadron spectroscopy has increased significantly, primarily due to various experimental discoveries of unconventional excitations.\footnote{\label{footnote1}For recent reviews of the experimental and theoretical understanding of these states we point the reader to Refs.~\cite{Guo:2017jvc, Lebed:2016hpi, Liu:2013waa, Briceno:2015rlt, Chen:2016qju}.} This has led to an abundance of theoretical proposals as to the underlying nature of the unexpected states. Possible explanations range from multi-hadron molecules to compact multi-quark configurations, to kinematic singularities arising from specific Feynman-diagram topologies~\cite{Guo:2017jvc, Lebed:2016hpi, Liu:2013waa, Briceno:2015rlt, Chen:2016qju}. In many cases, experimental data alone is not sufficient to distinguish between available explanations, and thus many questions remain unresolved. 

In some cases, theoretical calculations can provide access to experimentally-unavailable quantities that may shed light onto the structure of the QCD spectrum. With this goal in mind, in this work we present a detailed framework that will allow for the rigorous lattice-QCD calculation of transition amplitudes, mediated by electroweak or other external currents, involving two hadrons each in the initial and final states. We abbreviate our process of interest by $\textbf 2 + \mathcal J \to \textbf 2$, where each bold-faced 2 counts the QCD-stable hadrons in the state and $\mathcal J$ represents a generic, local external current.

The approach discussed here is based on prior formalism developed by two of us in Ref.~\cite{Briceno:2015tza}.\footnote{This, in turn, was inspired and guided by the work of Refs.~\cite{Briceno:2012yi,Bernard:2012bi, Detmold:2004qn}.} In Sec.~\ref{sec:formalism} we present a slightly modified version of this formalism in which all infinite-volume quantities are Lorentz covariant and the single-particle matrix elements that enter, abbreviated $\textbf 1 + \mathcal J \to \textbf 1$, are expressed in terms of standard Lorentz-invariant form factors. After extracting the $\textbf 2 + \mathcal J \to \textbf 2$ transition amplitudes, one can proceed to determine form factors as well as distribution functions\footnote{Distribution functions are accessed in lattice calculations via spatially non-local operator insertions~\cite{Ji:2013dva}. These may suffer from further finite-volume effects associated with the size of the operators as discussed in Ref.~\cite{Briceno:2018lfj}. This class of effects is not addressed by the present formalism and must be treated separately} of bound states or resonances that couple to the asymptotic states. From the form factors and distribution functions, in turn, one can obtain various structural parameters, e.g.~the charge or even the gluonic~\cite{Winter:2017bfs, Detmold:2016gpy} radius of a given state.

\bigskip
 
The primary reason why a non-trivial formalism is required to extract multi-hadron observables from lattice QCD calculations is that the latter are performed in a finite spatial volume, usually a cube defined with periodic boundary conditions on the quark and gluon fields. This complicates the determination of scattering and transition amplitudes, because there is no simple relation between the finite-volume QCD eigenstates and the asymptotic multi-particle states that arise in the infinite-volume limit of the theory. However, in certain cases, it is possible to derive relations between finite- and infinite-volume observables. These have been implemented with great success to access a wide variety of scattering quantities directly from numerical LQCD calculations. See Ref.~\cite{Briceno:2017max} for a recent review.

The most well-established such relation is that derived by L\"uscher in Refs.~\cite{Luscher:1986pf, Luscher:1990ux} over three decades ago. In these seminal papers he showed how elastic two-particle scattering amplitudes can be extracted from the finite-volume energy spectrum below the lowest lying three- or four-particle threshold. Since then, the idea has been generalized to all possible two-body systems, in particular to multiple two-particle channels built form any number of particle species,  including particles with any intrinsic spin~\cite{Rummukainen:1995vs, Feng:2004ua, He:2005ey, Bedaque:2004kc, Liu:2005kr, Kim:2005gf, Christ:2005gi, Lage:2009zv, Bernard:2010fp, Fu:2011xz, Leskovec:2012gb,  Briceno:2012yi, Hansen:2012tf, Guo:2012hv, Li:2012bi, Briceno:2013hya, Briceno:2014oea}. These formal ideas, together with significant algorithmic developments, have resulted in a proliferation of scattering amplitudes determined directly from lattice QCD~\cite{Dudek:2012xn, Wilson:2015dqa, Bali:2015gji, Bulava:2016mks, Andersen:2017una, Alexandrou:2017mpi, Brett:2018jqw,Moir:2016srx, Woss:2018irj, Dudek:2014qha,Wilson:2014cna, Dudek:2016cru, Briceno:2017qmb, Briceno:2016mjc}. 
A key limitation to the methods currently being implemented is the restriction to two-particle states, but the formal extension to three-particle systems has received significant attention recently and is progressing \cite{Briceno:2012rv, Hansen:2014eka, Mai:2017bge, Hammer:2017kms, Briceno:2017tce,  Briceno:2018aml}.

Similar developments have resulted in both perturbative~\cite{Detmold:2014fpa} non-perturbative~\cite{Lellouch:2000pv, Kim:2005gf, Christ:2005gi, Hansen:2012tf, Meyer:2011um, Feng:2014gba, Meyer:2012wk, Agadjanov:2014kha, Briceno:2012yi, Bernard:2012bi, Briceno:2015csa, Briceno:2014uqa, Briceno:2015tza} relations between finite-volume matrix elements and electroweak amplitudes. These have already been implemented in a variety of LQCD calculations~\cite{Bai:2015nea, Blum:2012uk, Boyle:2012ys, Blum:2011pu, Blum:2011ng, Feng:2014gba, Briceno:2015dca, Briceno:2016kkp, Alexandrou:2018jbt}. In particular, Refs.~\cite{Briceno:2015dca, Briceno:2016kkp, Alexandrou:2018jbt} extracted the $\rho \gamma^\star \to \pi$ electromagnetic form factor by determining the energy dependence of the corresponding amplitude, $\pi \pi \gamma^\star \to \pi$.

\begin{figure}
\begin{center}
\includegraphics[width=.9\textwidth]{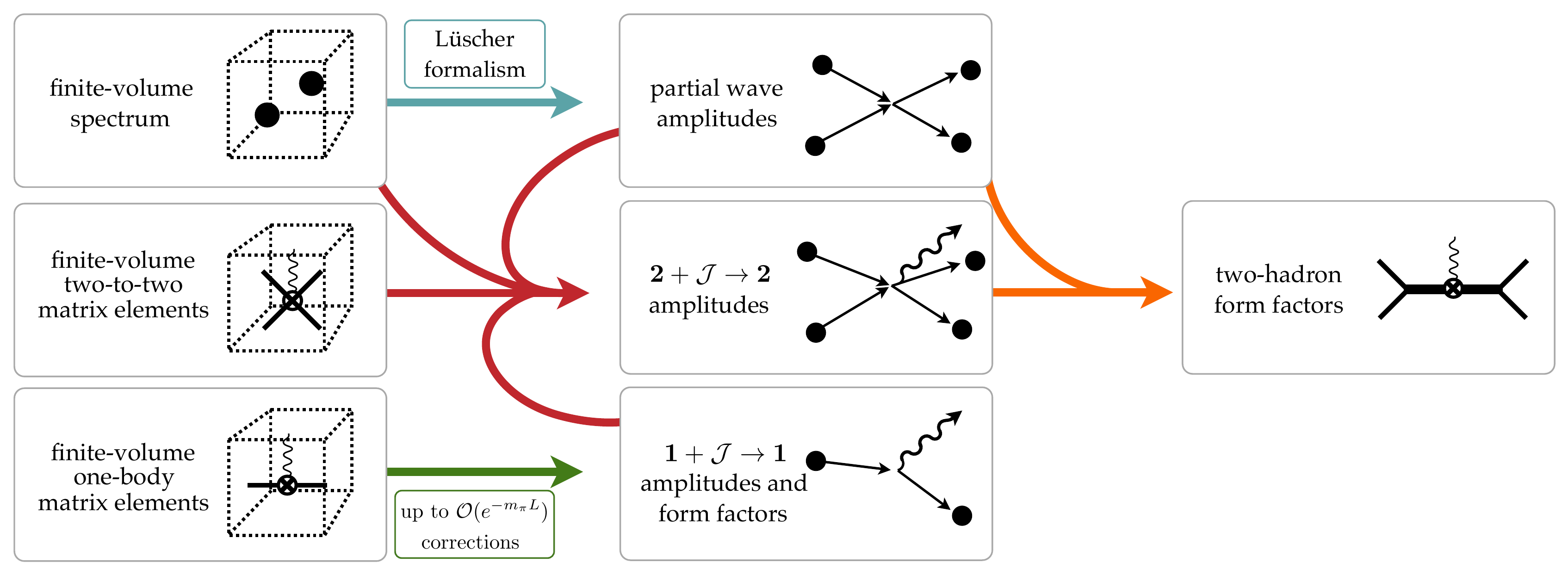}
\caption{Road map of the formal approach outlined in this work. See also Fig.~2 of Ref.~\cite{\origTwoTwo}. The four red arrows merging together represent how the present approach combines various finite- and infinite-volume information to extract the $\textbf 2 + \mathcal J \to \textbf 2$ amplitudes. Analytically continuing these to the resonance-pole location gives a robust, model-independent definition of the resonance form factor.\label{fig:road_map}}
\end{center}
\end{figure}

In this work we turn our attention to the prospect of determining $\textbf 2 + \mathcal J \to \textbf 2$ transition amplitudes from finite-volume matrix elements. This was previously considered in Ref.~\cite{Briceno:2015tza}. In contrast to that work, here we restrict attention to kinematics such that only one two-particle channel is open, and take the two particles in the channel to be scalars. In addition, we only consider flavor-conserving external currents, so that the initial and final two-particle states contain the same particles.  Just as in Ref.~\cite{Briceno:2015tza}, the two-particle states are composed of QCD-stable (pseudo-)scalars. Relaxing these restrictions to provide a Lorentz-covariant formalism for any number of two-particle channels, including those with intrinsic spin, is expected to be straightforward.

As in Ref.~\cite{Briceno:2015tza} in this work we derive a mapping between finite-volume matrix elements of two-particle states and the infinite-volume  $\textbf 2 + \mathcal J \to \textbf 2$ amplitude.
The result is summarized by the flow-chart shown in Fig.~\ref{fig:road_map}. We find that, given the following quantities:
\begin{itemize}
\item the two-particle finite-volume spectrum,
\item the $\textbf 1 + \mathcal J \to \textbf 1$ form factors, 
\item the finite-volume two-particle matrix elements of $\mathcal J$,
\end{itemize}
one can systematically constrain the $\textbf 2 + \mathcal J \to \textbf 2$ amplitude in the kinematic window in which only the accommodated channel contributes. Our relation requires the generalized Lellouch-L\"uscher factors \cite{Lellouch:2000pv, Briceno:2014uqa, Briceno:2015tza}, that enter multiplicatively in the conversion, as well as
a new finite-volume function, denoted $G$, that appears in an additive correction, together with the single-particle form factor as well as the two-to-two scattering amplitude.

A simple limiting case of our result is the one in which the single-particle form factors vanish. In this limit the finite-volume artifacts associated with the $G$ function also vanish and one recovers a Lellouch-L\"uscher-like relation in which the correction factor appears twice, once each for the initial and final two-particle states.
However, when the single-particle form factors are nonzero, the term containing $G$ is expected to give the dominant finite-volume effects. In particular, the analysis of Ref.~\cite{Detmold:2014fpa} showed that, in the case of weak interactions, the finite-volume effects on the ground-state $\textbf 2 + \mathcal J \to \textbf 2$ matrix element are given by an expansion in powers of $1/L$.\footnote{In fact, the authors of Ref.~\cite{Detmold:2014fpa} consider $\textbf n + \mathcal J \to \textbf n$ matrix elements.}  In these expressions, the diagrams that appear as our $G$ give $1/L^2$ corrections while all other terms contribute with additional powers of $1/L$.

\bigskip

 The purpose of this work is to improve certain technical details of the formalism and to provide more concrete information on the implementation procedure. We begin by providing a covariant version of the formalism in Sec.~\ref{sec:formalism}, where we also discuss three key examples involving $\pi\pi$ states. Then, in Sec.~\ref{sec:G}, we explain in detail our approach for evaluating $G$ and outline why this is more challenging than the more-standard finite-volume functions relevant for two-to-two scattering. 
In addition to the standard threshold singularities, in this section we discuss and illustrate the presence of triangle singularities in the $G$ function.  In Sec.~\ref{sec:conclusion} we conclude and provide an outlook for future studies. Finally we include two appendices: Appendix~\ref{sec:newder} gives various details relevant for the derivation of the improved formalism used here. Appendix \ref{sec:appFG} includes various technical aspects regarding the evaluation of the finite-volume functions discussed in the main text.

\section{Covariant representation of the formalism\label{sec:formalism}}

In this section we revisit the formalism derived in Ref.~\cite{Briceno:2015tza} and present a modified form in which all infinite-volume quantities are Lorentz covariant. %
We focus here only on the final result, and in Appendix \ref{sec:newder} we explain the (minimal) modifications to the original derivation that lead to this new form. 

This section is divided into three subsections: In Sec.~\ref{sec:KinQC} we review the required notation and give the quantization condition, as well as the generalized Lellouch-L\"uscher matrix, for two-particle states in a finite volume. Then, in Sec.~\ref{sec:cov2to2}, we provide a full description of our covariant $\textbf 2 + \mathcal J \to \textbf 2$ formalism. Finally, in Sec.~\ref{sec:2to2ex}, we consider a handful of specific examples to show how our general expressions reduce for a particular system with specified quantum numbers.

\subsection{Kinematics and the quantization condition}
\label{sec:KinQC}
 
We denote the 4-momentum of the incoming state in the finite-volume frame by $P^\mu_i \equiv (E_i , \textbf P_i)$ and that of the outgoing state by $P^\mu_f \equiv (E_f , \textbf P_f)$. The center-of-momentum (c.m.) energies corresponding to these are then given by
\begin{equation}
E_i^\star \equiv \sqrt{P_i^2} = \sqrt{E_i^2 - \textbf P_i^2} \,, \ \ \ \text{and} \ \ \ 
E_f^\star \equiv \sqrt{P_f^2} = \sqrt{E_f^2 - \textbf P_f^2} \,.
\end{equation}
This also defines the metric used for the Minkowski-signature 4-vector dot products throughout. Following the notation of Ref.~\cite{\KSS} we use $\star$ to indicate quantities defined in either the incoming or the outgoing c.m.~frame. As explained below, we often use an $i$ or an $f$ index in addition to the $\star$, in order to completely specify the frame.

\begin{figure}
\begin{center}
\includegraphics[width=0.75\textwidth]{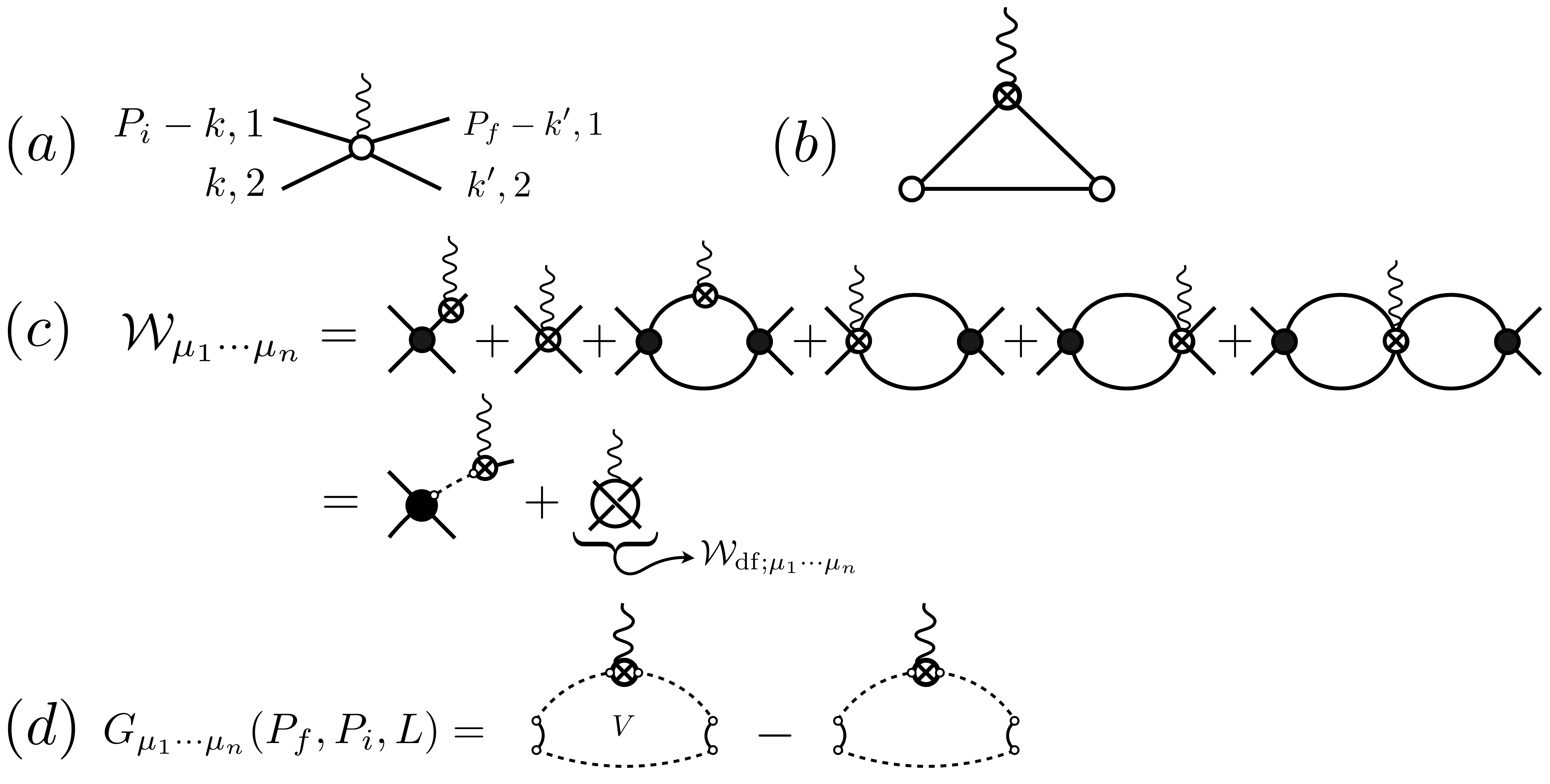}
\caption{
(a) The kinematics of the process considered, as described in the text.
(b) The triangle diagram that appears due to the single insertion of the external current.  
(c) The diagrammatic representation of the $\textbf 2 + \mathcal{J} \to  \textbf 2 $ transition amplitude. The black circles depict the $\textbf 2 \to \textbf 2$ scattering amplitude in the absence of the current. The crossed circles represent various couplings of the external current. Those with two hadronic external legs correspond to the standard $\textbf 1 + \mathcal J \to \textbf 1$ matrix element, while those with four external hadrons represent diagrams that are two-particle irreducible in the channel carrying the total momentum. The solid lines denote fully-dressed propagators of the low-energy degrees of freedom (the hadrons). In the second line we separate this into a contribution with on-shell singularities together with the divergence-free amplitude. 
(d) The diagrammatic representation of the new finite-volume function $G_{\mu_1\ldots\mu_n}$, defined in Eq.~(\ref{eq:Gmat_gen}). 
\label{fig:iW_Gfunc}
}
\end{center}
\end{figure}

In this work we accommodate all values of 3-momenta allowed by the periodic boundary conditions, i.e.~$\textbf P_i = 2 \pi \textbf d_i/L$ and $\textbf P_{\!f} = 2 \pi \textbf d_f/L$ where $\textbf d_i$ and $\textbf d_f$ are 3-vectors of integers. The energies and 3-momenta can differ between the initial and final states due to the momentum carried by the external current, $P_f^{\mu} - P_i^{\mu}$ [see Fig.~\ref{fig:iW_Gfunc}(a)]. The physical quantities discussed below depend on Lorentz scalars. For most systems we will primarily be sensitive to spacelike values of the momentum transfer, motivating us to introduce
\begin{align}
Q^2 \equiv  -(P_f - P_i)^2 \,,
\end{align}
which is positive for spacelike $P_f^\mu - P_i^\mu$.

As mentioned above, we restrict our attention here to values of $E_i^\star$ and $E_f^\star$ such that only a single two-particle channel can propagate. Within the single channel considered, we accommodate both identical and non-identical scalars and allow these to be non-degenerate in the latter case, with physical masses $m_1$ and $m_2$. We assume, however, that the current, $\mathcal J$, is flavor conserving so that the same two particles appear in the initial and final state.\footnote{Given the results presented below, implementing the covariant modification to the multi-channel expressions of Ref.~\cite{Briceno:2015tza} with flavor-changing currents should be straightforward, albeit tedious and likely leading to index-heavy notation.}

\bigskip

We now turn to the kinematic variables describing individual particles within the two-particle channel. For the remainder of this subsection, take $P^\mu \equiv (E, \textbf P)$ to simultaneously represent the initial and final state 4-momenta. Denoting the 3-momentum of particle 2 (with mass $m_2$) by $\textbf k$, the corresponding on-shell 4-vector is $k^\mu = (\omega_{k2}, \textbf k)$, where
\begin{equation}
\omega_{ k2} \equiv \sqrt{ k^2 + m_2^2} \,,
\end{equation}
with $k=\vert \textbf k \vert$. 

Next note that, in order to satisfy the specified total 4-momentum ($P^\mu$), particle 1 must carry $P^\mu - k^\mu \equiv (E - \omega_{k2}, \textbf P - \textbf k)$. Thus, for general $\textbf k$, one cannot simultaneously require that the particle momenta sum to $P^\mu$ and that particle 1 is on shell. The latter holds only when the temporal component, $E - \omega_{k2}$, is equal to
\begin{equation}
\omega_{P k1}  \equiv \sqrt{(\textbf P -\textbf k)^2 + m_1^2} \,.
\end{equation}

To better understand when the on-shell condition ($E - \omega_{k2} = \omega_{P k1}$) is satisfied, it is useful to introduce ${[\Lambda_{\pmb \beta}]^\mu}_\nu \equiv {\Lambda^\mu}_\nu(\pmb \beta)$ as a boost matrix with boost velocity $\pmb \beta \equiv \textbf P/E$. We then define
\begin{equation}
k^{\star \mu} \equiv (\omega_{k2}^\star, \textbf k^\star) \equiv {[\Lambda_{\pmb \beta}]^\mu}_\nu k^\nu \,, 
\end{equation}
and observe
\begin{equation}
\omega_{k2}^\star = \sqrt{k^{\star 2} + m_{2}^2} \,,
\end{equation}
where $k^\star \equiv \vert \textbf k^\star \vert$.%
\footnote{It is worth emphasizing that the definitions of $\textbf k^\star$ and $k^\star$ depend on $(E, \textbf P)$, $\textbf k$ and $m_2$ but not on $m_1$. This asymmetry in the definition is removed when both particles are on shell.} %
By contrast, the 4-momentum of particle 1 boosts to
\begin{equation}
P^{\star \mu} - k^{\star \mu} = (E^\star - \omega_{k2}^\star, - \textbf k^\star) \equiv   {[\Lambda_{\pmb \beta}]^\mu}_\nu (P^\nu - k^\nu) \,,
\end{equation}
where we have used ${[\Lambda_{\pmb \beta}]^\mu}_\nu P^\nu = (E^\star, \textbf 0)$. We deduce that the c.m.~frame on-shell condition is $E^\star = \omega_{k1}^\star +  \omega_{k2}^\star$ where
\begin{equation}
\omega_{k1}^\star \equiv \sqrt{k^{\star 2} + m_{1}^2} \,.
\end{equation}

The advantage of working in this frame is that the on-shell condition reduces to a constraint on the value of $k^\star$. In particular the particle pair is on shell if and only if $k^\star = q^\star$ with the latter defined by
\begin{equation}
E^\star \equiv \sqrt{q^{\star 2} + m_{1}^2} + \sqrt{q^{\star 2} + m_{2}^2} \,.
\end{equation}
Finally, the initial or final state-indices must be applied to all of these quantities once the total 4-momentum is associated with a particular state. In particular if we take $P^\mu \to P_i^\mu$, then the corresponding quantities above become $\omega_{P_i k2}$, ${[\Lambda_{\pmb \beta_i}]^\mu}_\nu$, $k^{\star \mu}_i$, $\textbf k^\star_i$, $k^\star_i$, $\omega_{k1i}^\star$, $\omega_{k2i}^\star$, $E^\star_i$, $q^\star_i$. The only quantities that do not inherit a frame index are the finite-volume frame momentum, $\textbf k$, and the corresponding energy, $\omega_{k2}$.

\bigskip

With this notation in hand we now give the quantization condition for two scalar particles in a finite volume. This is written as a determinant condition involving the on-shell two-particle scattering amplitude, $\mathcal M(P^2)$, represented as a diagonal matrix in angular-momentum space. For any fixed values of $\textbf P$ and $L$, the finite-volume energy spectrum is given by solutions to ~\cite{Kim:2005gf, Hansen:2012tf,  Briceno:2014oea}
\begin{align}
\det[\mathcal M(P^2)^{-1} + F(P,L) ]= 0 \,.
\label{eq:QC}
\end{align}
This holds in the region $0 < E^\star < E^\star_{\text{th}}$, where $E^\star_{\text{th}}$ is the energy of the lowest-lying multi-particle threshold that we ignore, which could be a two-, three-, or four-particle threshold. The relation holds up to neglected corrections of the form $e^{- m L}$, where $m$ is the physical mass of the lightest degree of freedom in the theory.

The precise definition of $\mathcal M (P^2)$ is given by
\begin{equation}
\mathcal M_{  \ell m ;   \ell' m'}(P^2) = \delta_{\ell \ell'} \delta_{m m'} \mathcal M^{(\ell)}(P^2) \,, \ \ \ \text{with} \ \ \  \mathcal M^{(\ell)}(s) = \frac{8 \pi E^\star}{\xi} \frac{1}{q^\star \cot \delta^{(\ell)}(q^\star) - i q^\star} \bigg \vert_{ E^{\star 2} = s} \,,
\end{equation}
where $\delta^{(\ell)}(q^\star)$ is the scattering phase shift, and $\xi = 1/2$ if the particles are identical and $\xi=1$ otherwise. 

The remaining ingredient is $F(P,L)$, a matrix of finite-volume geometric functions defined by
\begin{equation}
\label{eq:Fscdef}
F_{  \ell m ;   \ell' m'}(P,L)  \equiv 
\xi 
  \bigg [\frac{1}{L^3}\sum_{\mathbf{k}} - \int \! \! \frac{d^3 \textbf k}{(2\pi)^3}\, \bigg]
 \frac{\mathcal Y_{\ell m}(\textbf k^\star ) \,  \,   \mathcal Y^*_{\ell' m'}(\textbf k^\star)   
  }{2 \omega_{Pk1} 2 \omega_{k2}(E-\omega_{k2} -\omega_{Pk1} + i \epsilon )} ,
 \end{equation}
 where 
 \begin{align}
 \label{eq:Ylm}
\mathcal Y_{\ell m}(\textbf k^\star) & \equiv { \sqrt{4 \pi}  Y_{\ell m}(\hat {\textbf k}^\star)}
  \bigg (\frac{k^{\star}}{q^\star} \bigg)^{\ell} \,.
 \end{align}
In Appendix~\ref{sec:Ffunc} we review an efficient method to evaluate this, based on analytic expressions for the integrals defined using the cutoff functions introduced in Ref.~\cite{Kim:2005gf}. 

To close this subsection, we introduce one additional finite-volume matrix
\begin{equation}
\label{eq:Rdef}
\mathcal R(E_{n}, \textbf P) \equiv  \lim_{E \rightarrow E_{n}} \left[   \frac{(E - E_{n})}{F^{-1}(P,L) + \mathcal M(P^2)}\right] \,.
\end{equation}
This object, introduced in this form in Ref.~\cite{Briceno:2014uqa}, is the generalization of the Lellouch-L\"uscher factor \cite{Lellouch:2000pv} that relates finite-volume matrix elements of two-particle states to the corresponding infinite-volume decay and transition amplitudes. 

\subsection{Relating finite-volume matrix elements with $\textbf 2 + \mathcal J \to \textbf 2 $ transition amplitudes\label{sec:cov2to2}}

We are now ready to present our improved finite-volume formalism. The approach that we advocate here differs from that of Ref.~\cite{\origTwoTwo} in two key ways. 

First, the separation of singularities, required to disentangle finite-volume effects in the $\textbf 2 + \mathcal{J} \to \textbf 2$ amplitude, is done here using Lorentz invariant poles of the form $1/(k^2 - m^2)$. In the previous work we instead used $1/[2 \omega_{k}(k^0 - \omega_k)]$. As long as we consistently modify the pole form everywhere, it turns out that either choice is valid. The advantage of the present approach is that it ensures all infinite-volume quantities are Lorentz covariant and also simplifies the form of the new finite-volume function, $G$, arising due to the triangle diagram shown in Fig.~\ref{fig:iW_Gfunc}(b) and defined explicitly in Fig.~\ref{fig:iW_Gfunc}(d) and in Eq.~(\ref{eq:Gmat_gen}) below.

Second, we treat the single-particle matrix elements in a simpler way here than we did in Ref.~\cite{\origTwoTwo}. Our approach requires isolating the $\textbf 1 + \mathcal J \to \textbf 1$ matrix element in order to express the finite-volume effects of the triangle diagram [Fig.~\ref{fig:iW_Gfunc}(b)]. In our previous publication, a complicated scheme was presented to project the matrix element on shell. Though correct, we have come to realize that this procedure is unnecessary. The reason, as we explain in more detail below, is that one can decompose the matrix elements into kinematically determined tensor structures and form factors. Projecting the kinematic factors on shell is unnecessary, and removing this step gives a simpler approach that leads to the same infinite-volume observables.

\bigskip

We begin by introducing notation for the physical $\textbf 2 + \mathcal J \to \textbf 2$ matrix element that we are after [see also Fig.~\ref{fig:iW_Gfunc}(c)]
\begin{equation}
\cW_{\mu_1 \cdots \mu_n}(P_f, k'; P_i, k) \equiv \langle   P_f, k'; \text{out}    \vert   \mathcal J_{\mu_1 \cdots \mu_n}(0)  \vert  P_i, k; \text{in}     \rangle_{\rm conn.} \,.
\end{equation}
Here the initial and final states are standard two-particle asymptotic states with the usual relativistic normalization convention and $\mathcal J_{\mu_1 \cdots \mu_n}(0)$ is a generic local current insertion. As is shown in Fig.~\ref{fig:iW_Gfunc}(a), the initial state is built from particle 1 [with mass $m_1$ and on-shell momentum $(P_i - k)^\mu$ satisfying $(P_i-k)^2 = m_1^2$] and particle 2 [mass $m_2$, momentum $k^2 = m_2^2$]. The final state is built from the same pair, now carrying momenta $P_f-k'$ and $k'$. Following the discussion of the previous sub-section, the c.m.~frame 3-momenta are denoted by $\textbf k_i^\star$ and $\textbf k_f^\star$ and have magnitudes equal to $q_i^\star$ and $q_f^\star$ respectively. The label ``conn.'' emphasizes that only fully connected diagrams, those shown in Fig.~\ref{fig:iW_Gfunc}(c), contribute to the definition of the $\textbf 2 + \mathcal J \to \textbf 2$ matrix element.

A consequence of the on-shell constraints is that, once total energy and momenta are fixed, the two-particle states only have directional degrees of freedom, $\hat {\textbf k}^\star_i$ and $\hat {\textbf k}^\star_f$. However, in contrast to the scattering amplitude, $\mathcal M$, for $\cW$ a decomposition in spherical harmonics is not useful. The directional degrees of freedom sweep across pole singularities due to the diagrams in the second line of Fig.~\ref{fig:iW_Gfunc}(c), implying that the decomposition is only defined in the sense of distributions. More importantly, these long-distance singularities guarantee that higher partial waves will not be suppressed.   

The issue is easily resolved by removing the singular terms before decomposing in harmonics. This was already discussed in detail in Ref.~\cite{\origTwoTwo} where the quantity $\cW_\df$ was first introduced. In this work we define an alternative, Lorentz-covariant version of the divergence-free amplitude with the same symbol [see again the second line of Fig.~\ref{fig:iW_Gfunc}(c)]\footnote{This subtraction assumes that the current couples only to particle 1. In the case that the current couple to both particles, two additional terms must be subtracted. In particular, if the particles are identical one must always subtract four terms in which the propagators carry the four possible values of external momenta. See Eq.~(\ref{eq:WdfBothCouple}) as well as Ref.~\cite{\origTwoTwo} for explicit expressions.}
\begin{align}
\label{eq:Wdf}
\W_{{\rm df};\mu_1 \cdots \mu_n} \equiv \W_{\mu_1 \cdots \mu_n} - 
i  \overline{\M}(P_f,k',k) \frac{i}{(P_f-k)^2 - m_1^2} \w_{\mu_1 \cdots \mu_n}
-
 \w_{\mu_1 \cdots \mu_n}
\frac{i}{(P_i - k')^2 - m_1^2}
i \overline{\M'}(P_i, k',k)
  \,,
\end{align}
where $\w_{\mu_1 \cdots \mu_n}$ is the single-particle matrix element of the external current, defined in detail below, and
\begin{align}
\overline{\M}(P_f,k',k) & \equiv   
 \sum_{\ell} \mathcal M^{(\ell)}(P_f^2) \,  \big ( 2 \ell + 1 \big ) P_\ell \big (\hat {\textbf k}'^\star_f \cdot \hat {\textbf k}_f^\star \big) \, \bigg ( \frac{k^\star_f}{q^\star_f}  \bigg)^{\ell}  \,, \\
\overline{\M'}(P_i, k',k) & \equiv 
 \sum_{\ell} \mathcal M^{(\ell)}(P_i^2) \, \bigg ( \frac{k'^\star_i}{q^\star_i}  \bigg)^{\ell} \,  \big ( 2 \ell + 1 \big ) P_\ell \big (\hat {\textbf k}'^\star_i \cdot \hat {\textbf k}_i^\star \big)  \,.
\end{align}
Here $P_\ell (\cos \theta)$ are the standard Legendre polynomials, satisfying $\sum_{m=-\ell}^\ell 4 \pi Y^*_{\ell m}(\hat {\textbf x}) Y_{\ell m}(\hat {\textbf y}) = (2 \ell + 1) P_{\ell}(\hat {\textbf x} \cdot \hat {\textbf y})$. 
Unlike $\W$, $\W_{{\rm df}}$ admits a uniformly convergent decomposition in spherical harmonics
\begin{equation}
\W_{{\rm df};\mu_1 \cdots \mu_n}(P_f, k'; P_i, k) \equiv 4\pi Y^*_{\ell' m'} \big( \hat {\textbf k}'^\star_f  \big )   \    \W_{{\rm df};\mu_1 \cdots \mu_n; \ell' m'; \ell m}( s_f, s_i, Q^2)       \  Y_{\ell m} \big( \hat {\textbf k}^\star_i  \big )     \,,
\end{equation}
 where $s_i \equiv P_i^2 = E_i^{\star 2}$, $s_f \equiv P_f^2 =  E_f^{\star 2}$, and the repeated harmonic indices on the right-hand side are summed.

The subscript ``df'', short for divergence-free, refers only to the lack of kinematic singularities arising from a long lived state between the $\textbf 2 \to \textbf 2$ and $\textbf 1 + \mathcal J \to \textbf 1$ transitions. In particular, as we show in Sec.~\ref{sec:IAeval} below, $\W_{{\rm df}}$ does have so-called triangle singularities as a function of $E_i^\star$ and $E_f^\star$ associated with diagram of the kind depicted in Fig.~\ref{fig:iW_Gfunc}(b). These pose no problem to the harmonic decomposition but must, of course, be understood in order to successfully extract and interpret both $\W_{{\rm df}}$ and $\W$.

The scattering amplitude, $\overline {\mathcal M}$, is defined with powers of $(k^\star/q^\star)$, referred to as barrier factors. These must be included due to the manner in which the factors of $w_{\mu_1 \cdots \mu_n}$ and ${\mathcal M}$ arise in the triangle diagram of Fig.~\ref{fig:iW_Gfunc}(b). In particular, the loop is summed over all finite-volume momenta in such a way that the current insertion, $w_{\mu_1 \cdots \mu_n}$, as well as the external factors of $\mathcal M$, are sampled at off-shell values of momenta (i.e.~values for which $k^2 \neq  m_i^2$). Nonetheless, the power-law finite-volume effects are governed by the on-shell values of $w_{\mu_1 \cdots \mu_n}$ and ${\mathcal M}$ only, and the off-shell contributions can be absorbed into other infinite-volume quantities. The catch here is that the on-shell projection, effected via $k^\star \to q^\star$, amounts to replacing $k^{\star \ell}Y_{\ell m}(\hat {{\textbf k}^\star})$ with $q^{\star \ell}Y_{\ell m}(\hat {{\textbf k}^\star})$. The latter has spurious singularities near $\textbf k^\star = \textbf 0$ and thus more care is needed. The inclusion of barrier factors resolves the issue.

\bigskip

The single-particle matrix element, $\w_{\mu_1 \cdots \mu_n}$, appearing in Eq.~(\ref{eq:Wdf}), is a function of $(P_f - k, P_i - k)$ in the first term and $(P_f - k', P_i - k')$ in the second. Using the first term as a reference, the explicit definition is given in a three step processes. First,
the fully on-shell version is defined via a single-particle matrix element
\begin{equation}
\w^{\text{on}}_{\mu_1 \cdots \mu_n} (P_f-k,P_i-k) \equiv \langle P_f-k, m_1 \vert \mathcal J_{\mu_1 \cdots \mu_n}(0) \vert P_i - k, m_1 \rangle \,.
\end{equation}
Second, this is formally continued to an off-shell quantity in the context of some generic effective field-theory. The latter object is then decomposed into kinematic tensors and form factors
\begin{align}
\label{eq:w_off}
\w^{\text{off}}_{\mu_1 \cdots \mu_n} (P_f-k,P_i-k)
&=\sum_{j}\,\textbf{K}^{(j)}_{\mu_1 \cdots \mu_n}(k , P_f,P_i)\,f^{(j)}[Q^2; (P_f - k)^2; (P_i-k)^2] \,,
\end{align}
where the sum runs over all possible tensor structures for the given current.
Third, and finally, a partial on-shell projection is performed to define the version of $\w_{\mu_1 \cdots \mu_n}$ that appears in Eq.~(\ref{eq:Wdf}). In this step we set the virtualities within the form factors to be on shell [$(P_f - k)^2, (P_i-k)^2 \longrightarrow m_1^2$] and also set $k^0 = \omega_{k2}$ everywhere. We reach
\begin{align}
\label{eq:w_on}
\w_{\mu_1 \cdots \mu_n} (P_f-k,P_i-k) & \equiv \sum_{j}\,\textbf{K}^{(j)}_{\mu_1 \cdots \mu_n}(m,\textbf k,P_f,P_i) \bigg  |_{k^0=\omega_{k2}} \,f^{(j)}(Q^2)\,.
\end{align}
This definition is not completely on shell because, within %
$\textbf{K}^{(j)}_{\mu_1 \cdots \mu_n}$ only, the virtualities $(P_i-k)^2$ and $(P_f-k)^2$ may differ from $m_1^2$.

In what follows we will consider sums and integrals over the spectator momentum, $\textbf{k}$. With this in mind, it is convenient to rewrite $\textbf{K}^{(j)}_{\mu_1 \cdots \mu_n}$ as a sum of terms that isolate the dependence on this quantity
\begin{equation}
\label{eq:KtoCKomega}
\textbf{K}^{(j)}_{\mu_1 \cdots \mu_n}(m,\textbf k,P_f,P_i) \equiv \sum_{n'=0}^{n}   K^{\omega}_{\mu_1\cdots \mu_{n'}}(m,\textbf k) \   C^{(j)}_{\mu_{n'+1}\ldots \mu_{n}}(P_f,P_i) \,,
\end{equation}
where $C^{(j)}_{\mu_{n'+1}\ldots \mu_{n}}(P_f,P_i)$ has no indices for $n'=n$ and 
\begin{equation}
K^{\omega}_{\mu_1\cdots \mu_n}(m,\textbf k)  \equiv k_{\mu_1}\cdots k_{\mu_n}  \bigg  |_{k_0=\omega_{k2}} \,.
\end{equation}
Here the superscript $\omega$ indicates that the 4-momenta in $K^\omega$ are on shell. This will be important in Sec.~\ref{sec:G} below, where we introduce various off-shell versions of $K$ in our formulas for evaluating $G$.

 \bigskip
 
Having defined all of the infinite-volume quantities that enter our formalism, we now turn our attention to the finite volume. As mentioned in the introduction, we restrict attention to a finite cubic spatial volume, with periodicity $L$ applied to the fields in each of the three spatial directions. %
In this set-up, we consider a matrix element in which the local current $\mathcal J$ is sandwiched between two finite-volume states, each of which has the quantum numbers of the two-particle system. 
As we demonstrate in Appendix \ref{sec:newder}, following the derivation of Ref.~\cite{\origTwoTwo}, this LQCD observable is related to the infinite-volume $\textbf 2 + \mathcal J \to \textbf 2$ transition amplitude via
\begin{equation}
\Big |  \langle P_f, L \vert  \mathcal J^{\mu_1 \cdots \mu_n}(0)  \vert  P_i, L \rangle \Big |^2 
=\frac{1}{L^6} {\rm{Tr}}\left[
 \mathcal R( P_i) 
\Wtildf^{\mu_1 \cdots \mu_n}(P_i,P_f,L)
\mathcal R( P_f)  
\Wtildf^{\mu_1 \cdots \mu_n}(P_f,P_i,L)
\right] \,,
  \label{eq:2to2}
\end{equation}
 where $\Wtildf$ directly determines $\Wdf$ via
\begin{multline}
\label{eq:WtiltoWdf}
 \Wtildf^{\mu_1 \cdots \mu_n}(P_f,P_i,L)  - \Wdf^{\mu_1 \cdots \mu_n}(s_f,s_i,Q^2) \equiv  
  \sum_j
  \sum_{n'=0}^{n} 
  C^{(j), \mu_{n'+1}\ldots \mu_{n}}(P_f,P_i) f^{(j)}(Q^2)
\\
   \times
 \mathcal  M(s_f) \ G^{\mu_1 \cdots \mu_{n'}}(P_f,P_i,L) \ \mathcal M(s_i)  \,.
 \end{multline} 
 In the case of distinct particles that both couple to the current one must subtract two terms. In the second the particle labels $1$ and $2$ are swapped everywhere in the definition of $G$. In addition, the $f^{(j)}(Q^2)$ will take on different values if the particle species differs. These should be given a species label in the case that the current couples to both. See Eq.~(\ref{eq:WtiltoWdfAPP}) as well as Ref.~\cite{\origTwoTwo} for explicit expressions.
  
 In these expressions we have suppressed angular-momentum indices on $\mathcal R$, $ \Wtildf^{\mu_1 \cdots \mu_n}$, $\Wdf^{\mu_1 \cdots \mu_n}$, $\mathcal  M$ and $G^{\mu_1 \cdots \mu_{n'}}$. Each object carries the set $\ell' m'; \ell m$ and these are contracted between adjacent factors in the usual matrix multiplication. The trace in Eq.~(\ref{eq:2to2}) is also over this angular-momentum space. Note, by contrast, that the index set $\mu_1 \cdots \mu_n$ is not summed but rather fixed to common values for all objects appearing in these equations.
 
Finally, the matrix $G^{}_{\mu_1 \cdots \mu_{n'}}(P_f,P_i,L) $ is defined diagrammatically in 
 Fig.~\ref{fig:iW_Gfunc}(d) and has the explicit form
\begin{align}
\label{eq:Gmat_gen}
G_{\mu_1 \cdots \mu_n; \ell_f m_{f} ;     \ell_i m_{i}} (P_f,P_i,L)
&\equiv 
  \bigg [\frac{1}{L^3}\sum_{\mathbf{k}} - \int \! \! \frac{d^3 \textbf k}{(2\pi)^3}\, \bigg]
\mathcal Y_{\ell_f m_f}(\textbf k^\star_{f}) \,
 D(m,\textbf{k}) K^{ \omega}_{\mu_1\cdots \mu_n}(m,\textbf k)  \,   \mathcal Y^*_{\ell_i m_i}(\textbf k^\star_{i})   
  \,,
\end{align}
where
\begin{align}
D(m,\textbf{k})& \equiv \frac{1}{2\omega_{k2}}
\frac{1}
  {  (P_f-k)^2 - m_1^2 + i \epsilon  } 
\frac{1}
{  (P_i-k)^2 - m_1^2 + i \epsilon  }\bigg|_{k_0=\omega_{k2}}    \,.
\label{eq:D_def}
\end{align}
This differs from the form presented in Ref.~\cite{Briceno:2015tza} due to the aforementioned modifications: The poles are Lorentz invariant and the $\textbf 1 + \mathcal J \to \textbf 1$ matrix element is expressed in terms of tensor structures leading to $K^\omega$. Note that the modifications to $G$ are directly connected to those in the definition of $\Wdf$, Eq.~(\ref{eq:Wdf}). We have altered these two intermediate objects in such a way that $\cW$ is unchanged.

\subsection{Examples\label{sec:2to2ex}}

In this final subsection we show how the construction outlined above may be applied to specific, phenomenologically well-motivated examples.

\subsubsection{$(\pi^+ \pi^0)_{J=1} + j_\mu \to (\pi^+ \pi^0)_{J=1}$ \label{sec:rho2to2}}

We begin with the electro-magnetic form factors of a charged $\rho$ meson. The $\rho$ decays predominantly to the vector-isovector $\pi\pi$ state. Indeed for heavier than physical light-quark masses (such that $4 M_\pi > M_\rho$) and in the iso-spin symmetric theory, this is the only possible QCD decay channel. If the light-quark mass is further increased, the two-pion threshold eventually exceeds the $\rho$ mass ($2 M_\pi > M_\rho$) and the latter becomes a stable particle.
In this case, one can extract the form factors of the $\rho$ directly from finite-volume matrix elements. 
See, for example, Ref.~\cite{Shultz:2015pfa}. 

To determine the analogous observable at quark masses for which the $\rho \to \pi\pi$ decay occurs, it is necessary to first extract the $(\pi^+ \pi^0)_{J=1} + j_\mu \to (\pi^+ \pi^0)_{J=1}$ amplitude for a wide range of kinematic points. As depicted in Fig.~\ref{fig:road_map}, by fitting the dependence of the initial- and final-state energies to a functional ansatz, one may analytically continue these amplitudes to the complex-valued pole to obtain the $\rho$ form factors. Detailing the steps of this continuation will be the focus of future work. Here we focus on the extraction of the $(\pi^+ \pi^0)_{J=1} + j_\mu \to (\pi^+ \pi^0)_{J=1}$ amplitude for real $\pi \pi$ energies.

By interpolating isospin-one initial and final states ($I=1,m_I = \pm 1$), we project the system to a sector where all even angular momenta vanish. Thus, regardless of the values of $\textbf P_f$ and $\textbf P_i$, we will always have a finite-volume irrep that contains $J^P = 1^-$, with the next contamination coming from $J \geq 3$. Taking the latter to be negligible, we can approximate total angular momentum as a good quantum number. However, even in this simple case, the azimuthal component, $m_J$, is not a good quantum number in the finite volume. In general the positive and negative helicity states mix, but one can readily construct linear combinations of these that are invariant under transformations of the cubic group or its little groups~\cite{Thomas:2011rh}. 

Considering first the incoming state, we restrict attention to a specified set of finite-volume quantum numbers: $\Lambda_i$ and $\mu_i$, labeling the irrep and row respectively, of the little group defined by $\textbf{P}_i$. In addition we assume that, within this irrep, the interactions are dominated by $\ell_i = 1$ and a particular $m_i$ value so that the matrices can be truncated to single entries.  
Doing the same for the final-state, Eq.~(\ref{eq:2to2}) becomes%
\footnote{Note, the procedure for subducing the matrix elements onto the appropriate symmetry group is discussed in detail in Ref.~\cite{Briceno:2015tza}. Although some of the details of the formalism has changed, this aspect remains the same.}
\begin{equation}
\mathcal{W}_{L,{\rm df};\Lambda_f\mu_f ; \Lambda_i\mu_i}^{\mu}(P_f,P_i,L) =  e^{ i \delta^{I=1}_{\pi \pi}(q^\star_f)}     \frac{ \langle E_{f,n'}, \textbf P_f, L ;\Lambda_f\mu_f\vert    j^{\mu}(0)  \vert  E_{i,n}, \textbf P_i, L;\Lambda_i\mu_i \rangle}{ \vert \mathcal R^{I=1}_{\Lambda_f\mu_f}(P_f, L)  \mathcal R^{I=1}_{\Lambda_i\mu_i}(P_i, L) \vert^{1/2} }  e^{   i \delta^{I=1}_{\pi \pi}(q^\star_i)}   
 \,,
  \label{eq:rho2to2}
\end{equation}
where $\delta^{I=1}_{\pi \pi}$ is the elastic scattering phase shift for $\pi \pi \to \pi \pi$ in the $\rho$ channel and
\begin{equation}
\vert \mathcal R^{I=1}_{\Lambda\mu}(P, L) \vert^{1/2} \equiv  \frac{1}{L^3} \sqrt{\frac{q^\star}{16 \pi E^\star}}  \left [ \frac{E^\star}{E} \frac{\partial }{\partial E^\star} \Big ( \phi_{\Lambda\mu}^{\textbf d}(E^\star, L) + \delta^{I=1}_{\pi \pi}(q^\star) \Big ) \right ]^{-1/2} \,,
\end{equation}
with $\phi_{\Lambda\mu}^{\textbf d}(E^\star, L) $ defined for the irreps that couple to $\ell=1$ $\pi\pi$ in, for example, Ref.~\cite{Briceno:2014uqa}. We stress that all instances of $E_f$ and $E_i$ in these expressions are to be evaluated at any pair of finite-volume energies, $E_{f,n'}(L)$ and $E_{i,n}(L)$ respectively, satisfying Eq.~(\ref{eq:QC}). 
In Eq.~(\ref{eq:rho2to2}), $\mathcal{W}_{L,{\rm df};\Lambda_f\mu_f ; \Lambda_i\mu_i}^{\mu} $ is the subduced  version of $\mathcal{W}_{L,{\rm df}}^{\mu}$. As discussed in Ref.~\cite{Briceno:2015tza}, this can be obtained from $\mathcal{W}_{L,{\rm df};\ell_fm_{f} ; \ell_im_{i}}^{\mu}$ by rotating these into the helicity basis and then using the subduction matrices~\cite{Thomas:2011rh}. 
This subduction procedure requires no approximation. However each irrep couples to an infinite tower of partial waves, and only by neglecting these above a certain maximum value does one reach useful expressions.

Compared to Eq.~(\ref{eq:2to2}), in Eq.~(\ref{eq:rho2to2}) we have dropped the trace, since we are ignoring all but one partial wave, and have solved for $\Wtildf^{\mu}$. For the latter step there is a potential sign ambiguity that one must address. Note that $\mathcal R^{I=1}_{\Lambda\mu}(P, L) = \vert \mathcal R^{I=1}_{\Lambda\mu}(P, L) \vert e^{- 2 i \delta^{I=1}_{\pi \pi}(q^\star)}$ as is shown, for example, in Eqs.~(132)-(134) of Ref.~\cite{Briceno:2015csa}. The phases in $\mathcal R^{I=1}_{\Lambda\mu}(P_f, L)$ and $\mathcal R^{I=1}_{\Lambda\mu}(P_i, L)$ precisely generate the Watson phases within $ \Wtildf^{\mu}(P_f,P_i,L)$ as they must, since the finite-volume matrix element is real. The remaining $\pm$ ambiguity is constrained by the known value in the $Q^2 \to 0$ limit, but in certain cases a remaining ambiguity may survive.

The final step is to convert this to the infinite-volume quantity $\Wdf^{\mu }$ via
\begin{align}
\label{eq:WtiltoWdfrho}
\mathcal{W}^{\mu}_{{\rm df};\Lambda_f\mu_f ; \Lambda_i\mu_i}(s_f,s_i,Q^2) 
&=  
\mathcal{W}^{\mu}_{L,{\rm df};\Lambda_f\mu_f ; \Lambda_i\mu_i}(P_f,P_i,L)  \nn\\
&\hspace{-2cm}
-    f_{\pi^+}(Q^2)  \mathcal  M^{I=1}(s_f)  \Big [ 
  (P_f + P_i)^\mu  G_{\Lambda_f\mu_f ; \Lambda_i\mu_i}(P_f,P_i,L)  - 2   G^\mu_{\Lambda_f\mu_f ; \Lambda_i\mu_i}(P_f,P_i,L)    \Big ]  \mathcal M^{I=1}(s_i) \,.
 \end{align}
The general form of Eq.~(\ref{eq:WtiltoWdf}) is overly complicated for this application but still applies with $f^{(1)}(Q^2) = f^{(2)}(Q^2) = f_{\pi^+}(Q^2)$ corresponding to the usual (spacelike) pion form factor. Here we have also used the standard relation
\begin{equation}
\langle P_f - k; M_\pi \vert j_\mu(0) \vert P_i - k; M_\pi \rangle = (P_f + P_i - 2 k)_\mu \ f_{\pi^+}(Q^2) \,.
\end{equation}
In these expressions we are neglecting the electro-magnetic form factor of the neutral pion which is expected to be small but non-zero for $Q^2 \neq 0$. Finally we remark that a factor of $i$ may appear in this relation depending on the conventions for Euclidean or Minkowski gamma matrices. As the same gamma factors appear in all terms of our formalism changing conventions just amounts to multiplying both sides of $\mu$-indexed equations by a common factor.

\subsubsection{$(\pi^+ \pi^0)_{I_i} + j_\mu \to (\pi^+ \pi^0)_{I_f}$ \label{sec:pipi2to2}}

The electromagnetic current has both $I=0$ and $I=1$ components, but G parity guarantees that all matrix elements of the iso-scalar part between two $\pi \pi$ states must vanish. If we take the angular momentum to be unconstrained then the incoming $\pi \pi$ state may, in general carry isospin $I_i=0,1,2$. The iso-vector current then couples the fixed incoming isospin as follows:
$0 \to 1$, $1 \to 0,1,2$ and $2 \to 1,2$. If we further restrict attention to $\vert m_I \vert = 1$ states (i.e.~$\pi^+ \pi^0$ states) then this reduces to $1 \to 1,2$ and $2 \to 1,2$. An alternative way to distinguish these possibilities is by fixing orbital angular momentum: $J=0 \Leftrightarrow I=2$ and $J=1 \Leftrightarrow I=1$. 

In this way we identify four possible transitions involving $\pi^+ \pi^0$ states, the $p$-wave to $p$-wave matrix element considered above as well as $s$ to $s$, $s$ to $p$ and $p$ to $s$. It turns out that all four of these transitions are described by Eqs.~(\ref{eq:rho2to2}) to (\ref{eq:WtiltoWdfrho}) provided that we can neglect the effects of finite-volume mixing with $J \geq 2$ states. The only modifications are that one must generally project to different irreps $\Lambda$ to access the $J=0$ components and that the scattering amplitudes must correspond to the isospin of the state
\begin{equation}
\mathcal{W}_{L,{\rm df};\Lambda_f\mu_f,\Lambda_i\mu_i}^{\mu}(P_f,P_i,L) =  e^{ i \delta^{I_f}_{\pi \pi}(q^\star_f)}     \frac{ \langle E_{f,n'}, \textbf P_f, L ;\Lambda_f\mu_f\vert    j^{\mu}(0)  \vert  E_{i,n}, \textbf P_i, L;\Lambda_i\mu_i \rangle}{ \vert \mathcal R^{I_f}_{\Lambda_f\mu_f}(P_f, L)  \mathcal R^{I_i}_{\Lambda_i\mu_i}(P_i, L) \vert^{1/2} }  e^{   i \delta^{I_i}_{\pi \pi}(q^\star_i)}   
 \,,
\end{equation}
with
\begin{multline}
\mathcal{W}^{\mu}_{{\rm df};\Lambda_f\mu_f,\Lambda_i\mu_i}(s_f,s_i,Q^2) 
=  
\mathcal{W}^{\mu}_{L,{\rm df};\Lambda_f\mu_f,\Lambda_i\mu_i}(P_f,P_i,L)  \\
-    f_{\pi^+}(Q^2)  \mathcal  M^{I_f}(s_f)  \Big [ 
  (P_f + P_i)^\mu  G_{\Lambda_f\mu_f,\Lambda_i\mu_i}(P_f,P_i,L)  - 2   G^\mu_{\Lambda_f\mu_f,\Lambda_i\mu_i}(P_f,P_i,L)    \Big ]  \mathcal M^{I_i}(s_i) \,.
 \end{multline}

\subsubsection{Gluonic structure}

Thirty years ago, Jaffe and Manohar identified a structure function that provides a measure of the gluon distribution within hadrons~\cite{Jaffe:1989xy}. This has since lead to lattice calculations of the leading moments of these distributions, for example within the $\phi$-meson. Thus far, the calculations are restricted to heavy quark masses where the $\phi$ is stable within QCD~\cite{Detmold:2017oqb, Detmold:2016gpy}. Similarly, calculations of gluonic moments for light nuclei are already underway, again for values of the light quark masses that lead to the nuclei being deeply bound~\cite{Winter:2017bfs}.  The formalism presented here will allow for future calculations closer to the physical point by accommodating the finite-volume effects of loosely-bound as well as resonant states. %

In order to extract the leading moment of the gluon structure function, one must evaluate the traceless part of the product of two gluon energy-momentum tensors [$\mathcal{O}_{\mu\nu}\sim \mathcal G_{\mu\beta}\,{ \mathcal G}_{\beta \nu} $]. As one might expect, this is more complicated than the case considered above in part because it is a rank-two tensor. 
A starting point in extracting gluonic moments of resonances from LQCD would likely be to consider the $\rho$, discussed in Sec.~\ref{sec:rho2to2} above.
In this case, the relation between the finite-volume matrix elements and the transition amplitude %
is very similar to Eq.~(\ref{eq:rho2to2}). The only distinction arises in relating $\mathcal{W}^{\mu\nu}_{{\rm df}}$ and $\mathcal{W}^{\mu\nu}_{L,{\rm df}}$. To do so one must determine the scalar ($G$), vector ($G^\mu$), and tensor ($G^{\mu\nu}$) contributions to the finite-volume $G$-function and combine these with the relevant gluonic form-factors of the single-pion state. %
 
\bigskip

This concludes our discussion of the covariant formalism. %
The aim of the section was to provide a procedure by which three inputs: \emph{(i)} single-particle form factors, \emph{(ii)} $\textbf 2 \to \textbf 2$ scattering amplitudes and \emph{(iii)} finite-volume kinematic functions, can be combined with finite-volume two-particle matrix elements to extract the infinite-volume $\textbf 2 + \mathcal J \to \textbf 2$ transition amplitudes. In this recipe the ingredient that remains most obscure is the new finite-volume function $G^{\mu_1 \cdots \mu_n}(P_f,P_i,L)$, defined in Eq.~(\ref{eq:Gmat_gen}). Thus, in the next section, we give a detailed description of how this can be efficiently evaluated numerically.

\section{Evaluating $G(P_f,P_i,L)$\label{sec:G}}

Our aim is to evaluate 
\begin{align}
\label{eq:Gmat_PfnPi}
G_{\sigma} (P_f,P_i,L)
&\equiv 
  \bigg [\frac{1}{L^3}\sum_{\mathbf{k}} - \int \! \! \frac{d^3 \textbf k}{(2\pi)^3}\, \bigg]
\mathcal Y_{\ell_f m_f}(\textbf k^\star_{f}) \,
 D(m,\textbf{k}) K^{\omega}_{\mu_1\cdots \mu_n}(m,\textbf k)  \,   \mathcal Y^*_{\ell_i m_i}(\textbf k^\star_{i})   
  \,,
\end{align}
where we have introduced the collective index $\sigma \equiv [\mu_1 \cdots \mu_n; \ \ell_f m_{f} ;  \   \ell_i m_{i}]$. We also take the convention that if $\sigma$ is written as a low (high) index, then all of the Lorentz indices it contains are also understood to be low (high). The sum is straightforward to calculate numerically once a cutoff function has been included. %
We comment here that the ultraviolet divergences match between the sum and the integral, meaning that the difference has an unambiguous limit as the cutoff is removed and thus that $G^\sigma$ is a universal quantity with no scheme dependence. 

Evaluating the integral part of $G^\sigma$ turns out to be significantly more challenging. 
The integrand contains singularities associated with on-shell intermediate states and, although these are perfectly integrable given the $i \epsilon$ pole prescription,
numerical evaluations converge very slowly for standard numerical techniques. Thus it is highly advantageous to find analytical representations to the extent possible. 

For the case of $P_i = P_f$, it turns out that one can provide exact analytical expressions for the integral, as discussed in the following subsection. For the generic case, with $P_i \neq P_f$, we have not managed to obtain fully analytic results. Instead, we write the integral as the sum of two terms. The first includes all singularities and can be evaluated analytically to the level of Feynman parameters. The second is defined with a smooth integrand such that standard numerical integration is effective. Our approach for evaluating $G^\sigma$ in the case of $P_i \neq P_f$ case is detailed in Sec.~\ref{sec:GPfnPi} below.

\subsection{$P_i = P_f$\label{sec:GPfPi}}

The function $G^\sigma$ simplifies considerably when the initial and final momenta coincide, $P_i = P_f \equiv P$, i.e.~when the momentum transfer vanishes, $P_f^\mu - P_i^\mu = 0$. A particularly helpful feature of these kinematics is that a natural preferred frame emerges: the simultaneous rest frame of the incoming and outgoing particles, in which the spatial part of $P$ vanishes. 

Another consequence of $P_i = P_f$ is that the product of poles within $D(m,\textbf k)$ becomes a double pole of the form
\begin{equation}
D(m,\textbf{k}) = 
\frac{1}{2\omega_{k2}}
\left(\frac{1}
  {  (P-k)^2 - m_1^2 + i \epsilon  } \right)^2_{k^0 = \omega_{k2}} \,.
\end{equation}
Focusing on the factor within parenthesis, note that this can be rewritten as
\begin{equation}
\frac{1}
  {  (P-k)^2 - m_1^2 + i \epsilon  }  =   \frac{\omega_{q2}^{\star}}{  E^{\star}\, }
   \frac{1}{(q^{\star2}-k^{\star2}+ i \epsilon)   }
   - \frac{1}{  4 \omega_{q2}^{\star}  E^{\star} }+ \mathcal O(q^{\star 2}-k^{\star 2})  \,,
\end{equation}
where $k^\star$ and $q^\star$ are defined with respect to the $P$ rest frame.%

If $D(m, \textbf k)$ were to only contain a single pole, then, after acting with the sum-integral difference, only the leading-order, singular term would be relevant. This is because the sum-integral difference of smooth functions %
leads only to exponentially suppressed volume dependence that we neglect. However, in this case the first two terms in the expansion are important as they generate double and single poles upon squaring. This leads to
\begin{equation}
G_{\sigma} (P,P,L)
 = 
  \bigg [\frac{1}{L^3}\sum_{\mathbf{k}} - \int \! \! \frac{d^3 \textbf k}{(2\pi)^3} \bigg]  \frac{1}{2\omega_{k2}}
\mathcal Y_{\ell_f m_f}(\textbf k^\star) 
  K^{\omega}_{\mu_1\cdots \mu_n}(m,\textbf k)     \mathcal Y^*_{\ell_i m_i}(\textbf k^\star)      \overline D(m, k^\star, q^\star) \,,
  \label{eq:GwithDbar}
\end{equation}
where
\begin{equation}
\overline D(m, k^\star, q^\star) \equiv  \frac{\omega_{q2}^{\star 2}}{  E^{\star 2}}
   \frac{1}{(q^{\star2}-k^{\star2}+ i \epsilon)^2   }
   -
     \frac{1}{ 2 E^{\star2} }
   \frac{1}{(q^{\star2}-k^{\star2}+ i \epsilon)}  \,.
\end{equation}

At this stage it is useful to decompose the angular dependence within the tensors into a single set of spherical harmonics
\begin{equation}
\label{eq:decom}
\mathcal Y_{\ell_f m_f}(\textbf k^\star) 
  K^{\omega}_{\mu_1\cdots \mu_n}(m,\textbf k)     \mathcal Y^*_{\ell_i m_i}(\textbf k^\star) 
  \equiv \frac{\sqrt{4 \pi}}{(q^\star )^{\ell_i + \ell_f}}
  \sum_{J M}  \mathcal{C}_{\sigma, JM}(\pmb \beta , k^{\star}) \,   k^{\star J} {Y}_{J M}(\hat {\textbf k}^\star) \,.
  \end{equation}
As we explain in Appendix~\ref{sec:kin}, $\mathcal C_{\sigma, J M} (\pmb \beta, k^{\star })$ can be efficiently calculated by writing the factors within $K^\omega$ as boosted c.m.~frame vectors, $k^{ \mu} = {[\Lambda_{-\pmb \beta}]^{\mu}}_\nu k^{\star \nu}$. Such factors can then be written as spherical harmonics and, using Clebsch-Gordon coefficients, these can be combined with the external factors of $\mathcal Y_{\ell_f m_f}(\textbf k^\star)$ and  $ \mathcal Y^*_{\ell_i m_i}(\textbf k^\star)$ to identify the a final harmonic basis. As a final note, we stress that it is possible to unambiguously separate the dependence on $\pmb \beta = \textbf P/E$ and $k^{\star}$ within $\mathcal C_\sigma$, i.e.~one can vary $k^\star$ while holding $\pmb \beta$ constant. This will be important for the manipulations performed below.

The construction of  $\mathcal C_{\sigma, JM} (\pmb \beta, k^{\star})$ is discussed in detail in Appendix~\ref{sec:kin}. As a specific simple example, here we consider the case $\sigma = [\mu;\,10;\,10]$. The numerator within $G_\sigma$ is then
\begin{equation}
(q^{\star})^{2}  \mathcal Y_{10}(\textbf k^\star) 
k_\mu     \mathcal Y^*_{10}(\textbf k^\star) = 3 (k^{z \star } )^2 { [\Lambda_{-\pmb \beta}]_\mu}^{\nu} k_\nu^\star  \,.
\end{equation}
The current insertion, $k_\mu$, can be written as a combination of $\ell=0$ and $\ell=1$ spherical harmonics. Combining this with the two $\ell=1$ harmonics from the external states, one finds that $\mathcal C_{\sigma, JM}$ is zero for $J>3$. 

The $J M = 0 0$ component only has a non-zero contribution in the $k_0^\star$ component. Isolating this contribution and substituting the definition of the boost matrix, we reach
\begin{equation}
\label{eq:Cex00}
\mathcal C^\sigma_{00}(\pmb \beta, k^{\star }) =
k^{\star2} \omega_{k2}^\star  \frac{P^\mu}{E^\star} \,.
\end{equation}
The remaining nonzero coefficients, arising for $J \leq 3$, take a more complicated form in general, but simplify for $\textbf P \propto \hat {\textbf z}$. If we additionally focus on the $\mu=z$ component, then only three coefficients survive
\begin{equation}
\mathcal C^\sigma_{10}(\pmb \beta, k^{\star })  =\frac{3\sqrt{3} }{5} \frac{E}{E^\star}  k^{\star 2}  \,,  \ \ \ \ \ 
\mathcal C^\sigma_{20}(\pmb \beta, k^{\star })  =\frac{2\sqrt{5}}{5}   \frac{P^z}{E^\star}    \omega _{k2}^\star \,,  \ \ \ \ \ 
\mathcal C^\sigma_{30}(\pmb \beta, k^{\star })  =\frac{6 \sqrt{7} }{35}   \frac{E}{E^\star}     \,.
\end{equation}
We revisit this case below to show how it enters our final construction for $G^\mu_{  10 ;10}$.  

Returning to the main line of argument, we now substitute Eq.~(\ref{eq:decom}) into the expression for $G_{\sigma} (P,P,L) $, Eq.~(\ref{eq:GwithDbar}), to reach
\begin{equation}
\label{eq:GwithCdecom}
G_{\sigma} (P,P,L) =  \frac{\sqrt{4 \pi}}{(q^\star )^{\ell_i + \ell_f}}   \sum_{J M}  \bigg [\frac{1}{L^3}\sum_{\mathbf{k}}  \frac{\omega_{k2}^\star}{ \omega_{k2}} - \int \! \! \frac{d^3 \textbf k^\star}{(2\pi)^3} \bigg]  \frac{ \mathcal{C}_{\sigma, JM}(\pmb \beta , k^{\star})}{2\omega_{k2}^\star}
 \,   k^{\star J} {Y}_{J M}(\hat {\textbf k}^\star)   \overline D(m, k^\star, q^\star)      \,.
\end{equation}

The next step is to expand the $k^{\star}$ dependence in the first factor within the summand about the pole location 
\begin{equation}
\frac{ \mathcal{C}_{\sigma, JM}(\pmb \beta , k^{\star})}{2\omega_{k2}^\star} = \frac{ \mathcal{C}_{\sigma, JM}(\pmb \beta , q^{\star})}{2\omega_{q2}^\star}   - (q^{\star 2} - k^{\star 2}) \frac{\partial}{\partial q^{\star 2}}  \frac{ \mathcal{C}_{\sigma, JM}(\pmb \beta , q^{\star})}{2\omega_{q2}^\star}  + \mathcal O \big [(q^{\star 2} - k^{\star 2})^2 \big ] \,.
\end{equation}
In the second term we have rewritten the derivative with respect to $k^{\star 2}$ (to be evaluated afterwards at $k^{\star 2} = q^{\star 2}$) directly as a derivative with respect to $q^{\star 2}$. This is possible because $\mathcal{C}_{\sigma, JM}(\pmb \beta , k^{\star})$ has no implicit $q^{\star 2}$ dependence and it is formally possible to vary $q^{\star}$, and thus $E^\star$, while holding $\pmb \beta$ constant. 

Combining this with the definition of $\overline D(m, k^\star, q^\star) $ one can show
\begin{multline}
\frac{ \mathcal{C}_{\sigma, JM}(\pmb \beta , k^{\star})}{2\omega_{k2}^\star} \overline D(m, k^\star, q^\star) =  \frac{\omega_{q2}^{\star }}{ 2 E^{\star 2}}   \frac{\mathcal{C}_{\sigma, JM}(\pmb \beta , q^{\star})}{(q^{\star2}-k^{\star2}+ i \epsilon)^2   }
- \bigg [ \bigg ( \frac{1}{ 2 E^{\star2} }  + \frac{\omega_{q2}^{\star 2}}{  E^{\star 2}}
  \frac{\partial}{\partial q^{\star 2}}    \bigg )  \frac{ \mathcal{C}_{\sigma, JM}(\pmb \beta , q^{\star})}{2\omega_{q2}^\star}     
    \bigg ]  \frac{1}{q^{\star2}-k^{\star2}+ i \epsilon   }  \\
    + \mathcal O \big [(q^{\star 2} - k^{\star 2})^0 \big ]\,.
\end{multline}
Remarkably, the operator in parenthesis vanishes when acting on $1/\omega_{q2}^\star$ so that this reduces to
\begin{align}
\frac{ \mathcal{C}_{\sigma, JM}(\pmb \beta , k^{\star})}{2\omega_{k2}^\star} \overline D(m, k^\star, q^\star) & =  \frac{\omega_{q2}^{\star }}{ 2 E^{\star 2}}   \frac{\mathcal{C}_{\sigma, JM}(\pmb \beta , q^{\star})}{(q^{\star2}-k^{\star2}+ i \epsilon)^2   }
-   \frac{\omega_{q2}^{\star }}{ 2 E^{\star 2}}  
       \frac{\partial_{q^{\star 2}} \mathcal{C}_{\sigma, JM}(\pmb \beta , q^{\star})  }{q^{\star2}-k^{\star2}+ i \epsilon   }  
    + \mathcal O \big [(q^{\star 2} - k^{\star 2})^0 \big ] \,, \\
    & =   \frac{\omega_{q2}^{\star }}{ 2 E^{\star 2}}  \sum_{n=1}^2    \frac{ (- \partial_{q^{\star 2}})^{2-n }  \mathcal C _{\sigma, JM}(\pmb \beta , q^{\star})}{(q^{\star2}-k^{\star2}+ i \epsilon)^n   }  + \mathcal O \big [(q^{\star 2} - k^{\star 2})^0 \big ] \,.
\end{align}
It follows that Eq.~(\ref{eq:GwithCdecom}) can be rewritten as
\begin{equation}
G_{\sigma} (P,P,L) =  \frac{1}{(q^\star )^{\ell_i + \ell_f}}  \frac{\omega_{q2}^{\star }}{ 2 E^{\star 2}}  \sum_{n=1}^2    \sum_{J M} \big [(- \partial_{q^{\star 2}})^{2-n }  \mathcal C _{\sigma, JM}(\pmb \beta , q^{\star}) \big]  \bigg [\frac{1}{L^3}\sum_{\mathbf{k}}  \frac{\omega_{k2}^\star}{ \omega_{k2}} - \int \! \! \frac{d^3 \textbf k^\star}{(2\pi)^3} \bigg]    \frac{ \sqrt{4 \pi}  k^{\star J} {Y}_{J M}(\hat {\textbf k}^\star)  }{(q^{\star2}-k^{\star2}+ i \epsilon)^n   } 
      \,.
\end{equation}

\begin{figure}
\begin{center}
\includegraphics[width=\textwidth]{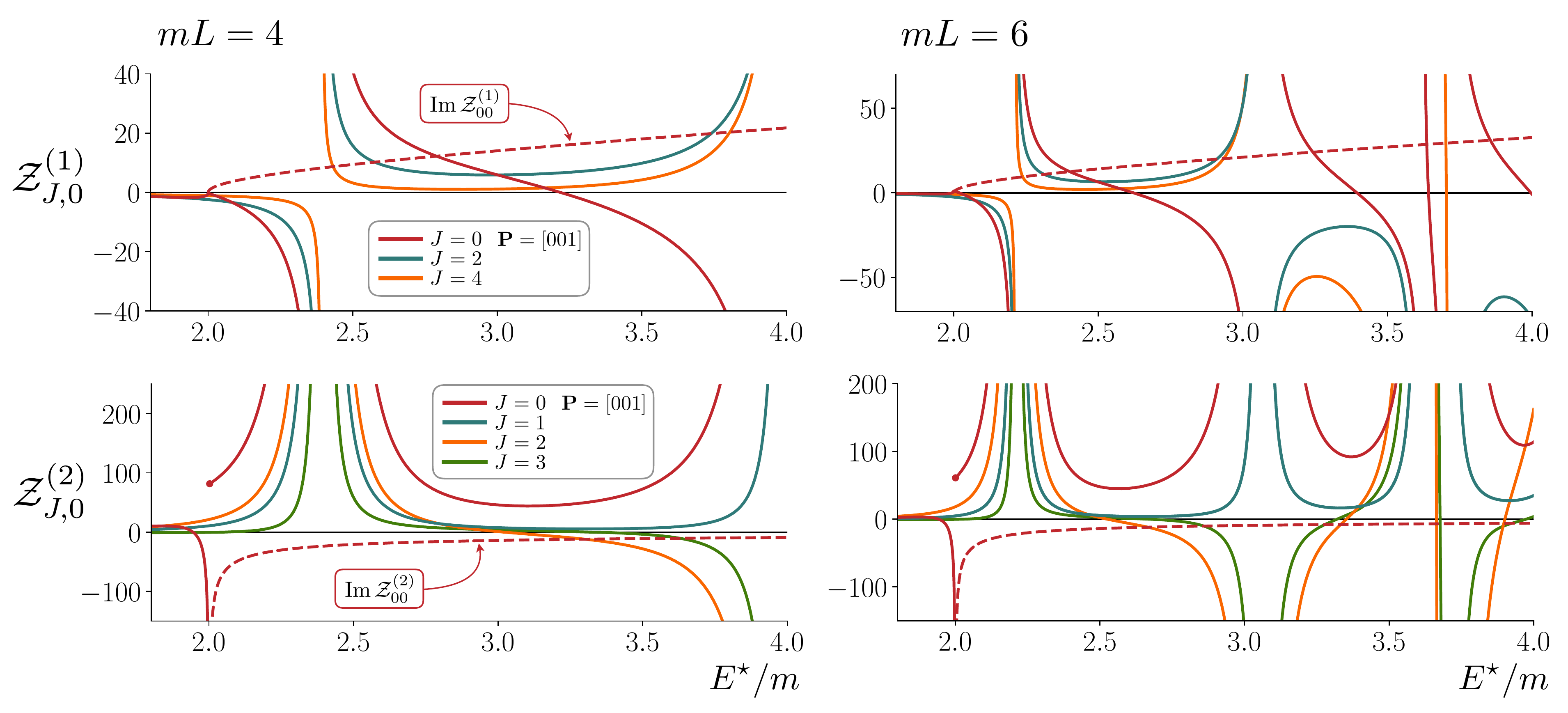}
\caption{Example plots for $\mathcal{Z}^{(1)}_{J M}$ (top two panels) and $\mathcal{Z}^{(2)}_{J M}$ (bottom two). All curves show the function plotted versus $E^\star$ for fixed spatial momentum, $\textbf P = (2 \pi / L) [001]$, for which only the $M=0$ components are non-zero (up to $M=4$). The real parts are shown as solid curves and, for $J=0$, the imaginary part is indicated with the dashed curve. As discussed in detail in Appendix \ref{app:whenZisZero}, for $n=1$ the odd $J$ are exponentially suppressed and indistinguishable from zero on the plotted scale.}
\label{fig:Zsummary}
\end{center}
\end{figure}

To reach our final form we make two additional modifications. First we introduce a cutoff function on the sum-integral difference to enable effective numerical evaluation. Second, we re-express our results in terms of dimensionless quantities $\mathbf{r}^\star=\mathbf{k}^\star L/(2\pi)$ and $x=q^\star L/(2\pi)$. Then, shuffling around terms and introducing a new geometric function, we conclude
\begin{equation}
G_{\sigma} (P,P,L) =  \frac{1}{(q^\star )^{\ell_i + \ell_f}}  \frac{\omega_{q2}^{\star }}{ 2 E^{\star 2}}  \sum_{n=1}^2 \frac{L^{2n-3}}{(2 \pi)^{2n}}   \sum_{J M}   \frac{(2 \pi)^J}{L^J}    \lim_{\alpha \to 0} \mathcal{Z}^{(n)}_{J M}(P,L,\alpha) \  (- \partial_{q^{\star 2}})^{2-n }   \mathcal C _{\sigma, JM}(\pmb \beta , q^{\star})      \,,
\end{equation}
where
\begin{equation}
\label{eq:Zdef}
\mathcal{Z}^{(n)}_{J M}(P,L,\alpha) =    \bigg [ \sum_{\mathbf{r}}  \frac{\omega_{k2}^\star}{ \omega_{k2}} - \int \!   d^3 \textbf r^\star   \bigg]    
 \,  \frac{  \sqrt{4 \pi}  \, r^{\star J} {Y}_{J M}(\hat{\textbf r}^\star)     }{ (x^{2}-r^{\star2}+ i \epsilon)^n }  e^{- \alpha (r^{\star2} - x^{2})^n} \,.
\end{equation}
These two equations give the main result of this sub-section. In Appendix~\ref{sec:Zcal} we give some details about the evaluation of $\mathcal{Z}^{(n)}_{J M}(P,L,\alpha)$. 
 In Fig.~\ref{fig:Zsummary} we plot $\mathcal{Z}^{(n)}_{J M}(P,L,\alpha)$ for $\textbf P  = (2 \pi / L) [001]$, for various values of $J$ and two different volumes. 

We give two final comments concerning the new kinematic function, $\mathcal Z^{(n)}$. First we note an advantage of the decomposition over a single spherical harmonic performed in Eq.~(\ref{eq:decom}). It is now  straightforward to use the symmetries of the finite-volume system to identify for which values of $J M$, $\mathcal Z^{(n)}_{JM}$ will be nonzero. This is discussed in detail in Appendix \ref{app:whenZisZero}.

Second, we remark that the cutoff function used here is designed with the property that the $\mathcal O(\alpha)$ correction cancels the pole, and thus generates a smooth quantity within the sum-integral difference. If, for example, one were to instead use $e^{- \alpha (r^{\star2} - x^{2})}$ for all $n$ values, this would still be formally valid, but would lead to corrections of the form $\alpha/L^{k}$, making it more difficult to estimate the $\alpha \to 0$ limit. In fact these can be systematically subtracted and, as we shown in Appendix~\ref{sec:alpha_dep},
 this proves to be an efficient alternative approach for evaluating $\lim_{\alpha \to 0} \mathcal{Z}^{(n)}_{J M}$.

\begin{figure}
\begin{center}
\includegraphics[width=\textwidth]{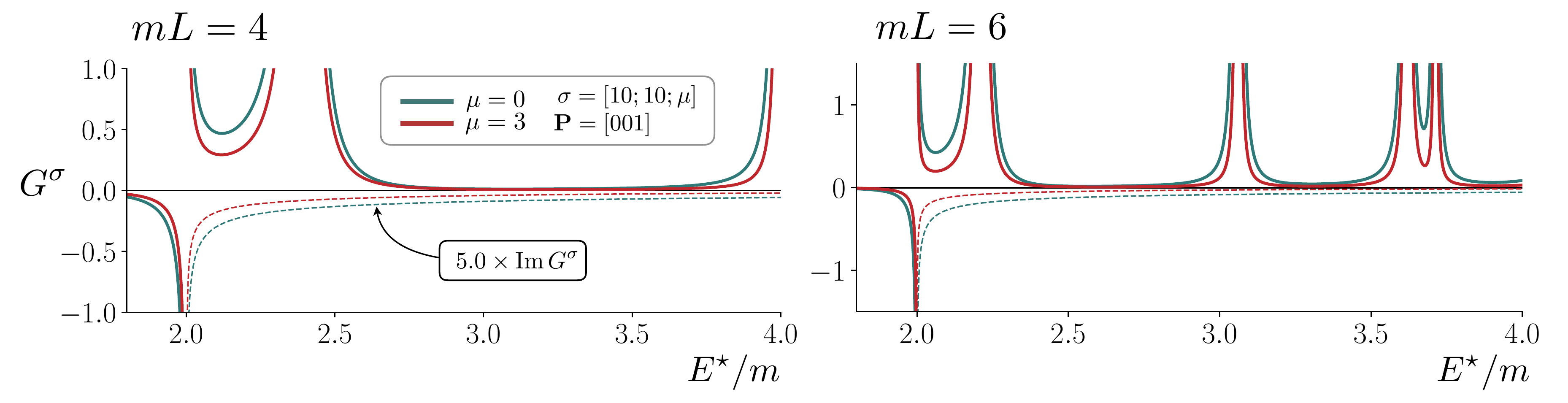}
\caption{Example plots for $G^\sigma\!(P_f,P_i,L)$ for the case of $P_i = P_f = (E, \textbf P)$ with $\textbf P=(2 \pi/L)[001]$. The real parts are shown as solid curves and the imaginary parts (multiplied by a factor to make the functional form visible) are dashed. At all non-interacting energy levels the function diverges as a double pole with a positive coefficient. }
\label{fig:GPiEQPf}
\end{center}
\end{figure}

We close this sub-section by returning to the specific case discussed above, $\sigma = [\mu=z;\,10;\,10]$ and $\textbf P = (2 \pi/L) [00d_z]$. Suppressing the arguments of $\mathcal Z^{(n)}$, the final result for $G^\sigma$ can be written as
\begin{multline}
G^{\sigma} \!(P,P,L) =   - \frac{1}{ q^{\star 2}}  \frac{\omega_{q2}^{\star }}{ 2 E^{\star 2}}   \frac{1}{ 4 \pi^2 L}   \left [       \mathcal{Z}^{(1)}_{00} \frac{P^z}{E^\star}  \bigg (   \omega_{q2}^\star  + \frac{q^{\star2}}{2 \omega_{q2}^\star}  \bigg )+     \frac{2 \pi}{L}  \mathcal Z_{10}^{(1)}  \frac{3\sqrt{3} }{5} \frac{E}{E^\star}  +    \frac{(2 \pi)^2}{L^2}  \mathcal{Z}^{(1)}_{20}    \frac{2\sqrt{5}}{5}   \frac{P^z}{E^\star} \frac{1}{2 \omega _{q2}^\star}      \right ]  \\
+
 \frac{1}{ q^{\star 2} }  \frac{\omega_{q2}^{\star }}{ 2 E^{\star 2}}   \frac{L}{(2 \pi)^{4}}  \left[     \mathcal{Z}^{(2)}_{0 0}   \frac{P^z}{E^\star}  q^{\star2} \omega_{q2}^\star     +    \frac{2 \pi}{L}  \mathcal{Z}^{(2)}_{1 0} \frac{3\sqrt{3} }{5} \frac{E}{E^\star}  q^{\star 2} +    \frac{(2 \pi)^2}{L^2}  \mathcal{Z}^{(2)}_{2 0}\frac{2\sqrt{5}}{5}   \frac{P^z}{E^\star}    \omega _{q2}^\star  +     \frac{(2 \pi)^3}{L^3}  \mathcal{Z}^{(2)}_{3 0}  \frac{6 \sqrt{7} }{35}   \frac{E}{E^\star}    \right ]    \,.
\end{multline}
Note that, in the case of mass-degenerate particles, $\mathcal Z^{(1)}_{JM}=0$ for all odd $J$. If the particles are at rest in the finite-volume frame then all odd-$J$ functions vanish as does $J=2$. This holds for both $n=1$ and $n=2$ for both degenerate and non-degenerate particles. [See again Appendix \ref{app:whenZisZero}.]
In Fig.~\ref{fig:GPiEQPf} we plot the real and imaginary parts of $G^{\mu}_{10; 10} (P,P,L)$ for $\textbf P  = (2 \pi / L) [001]$. 

\subsection{$P_i \neq P_f$\label{sec:GPfnPi}}

We now turn our attention to the more challenging general case of $P_i \neq P_f$. Note that this is realized if any of the four components of the 4-vectors differ, in particular also for $\textbf P_i = \textbf P_f$ but $E_i \neq E_f$.
As with $P_i = P_f$, here the evaluation of the sum is straightforward, while the integral is significantly more challenging. One of the major complications is that the two poles do not coincide in general as one varies the integration variable, $\textbf k$, but may overlap on a two-dimensional subspace for certain choices of external momenta. The contribution of this double-pole submanifold to the integral must be treated with care.

Though we have not found a fully analytic determination of the integral entering $G^\sigma$, we do have a recipe that gives the desired quantity accurately and with high efficiency. The approach is to rewrite the three-dimensional integral in terms of a $d^4 k$ integral plus a second $d^3 \textbf k$ integral with a smooth, singularity-free integrand. The four-dimensional integral, which carries all of the singularity structure, can then be reduced %
to a one-dimensional integral over a Feynman parameter. The second, smooth term can directly be evaluated using standard numerical integrators. 

To give the relevant expressions, we first introduce an extension of the cutoff function entering the definition of $\mathcal Z_{JM}^{(n)}$ in the previous subsection
\begin{align}
H( \bar \alpha, \textbf{k}) \equiv e^{- \bar \alpha (k^{\star 2}_i - q^{\star 2}_i) (k^{\star 2}_f - q^{\star 2}_f)} = e^{-   \alpha   (r^{\star 2}_i - x^{ 2}_i) (r^{\star 2}_f - x^{ 2}_f) } \,,
\end{align}
where $\bar \alpha = L^4/(2 \pi)^4 \alpha$ and $r^{\star 2}_i(2\pi/L)^2=k^{\star 2}_i$, $x^{ 2}_i(2\pi/L)^2=q^{\star 2}_i$, etc. %
We then write
\begin{align}
 {G}_{ \sigma} (P_f,P_i,L) = \lim_{\bar{\alpha} \to 0} \Big [  \mathcal {S}_{   \sigma }(\bar{\alpha},P_f,P_i,L)
-
 \mathcal {I}_{  \sigma }(\bar{\alpha},P_f,P_i)  \Big ]\,,
 \end{align}
where
\begin{align}
 \mathcal {S}_{   \sigma }(\bar{\alpha},P_f,P_i,L)
 & \equiv  \frac{1}{L^3}\sum_{\mathbf{k}} 
H(\bar{\alpha},\textbf{k}) \,
\mathcal Y_{\ell_f m_f}(\textbf k^\star_{f}) \,
 D(m,\textbf{k}) K^{\omega}_{\mu_1\cdots \mu_n}(m,\textbf k)  \,   \mathcal Y^*_{\ell_i m_i}(\textbf k^\star_{i})   
 \,, \\
 \mathcal {I}_{ \sigma }(\bar{\alpha},P_f,P_i)   
&\equiv  
 \int \! \frac{d^3 \textbf k}{(2\pi)^3} \,   H(\bar{\alpha},\textbf{k}) 
 \mathcal Y_{\ell_f m_f}(\textbf k^\star_{f}) \,
 D(m,\textbf{k}) K^{\omega}_{\mu_1\cdots \mu_n}(m,\textbf k)  \,   \mathcal Y^*_{\ell_i m_i}(\textbf k^\star_{i})   
 \,.
 \label{eq:Idef}
 \end{align}
The sum can be evaluated directly as written, %
and thus we make no further modifications to $\mathcal {S}_{   \sigma }$. The remainder of this section is dedicated to $\mathcal {I}_{   \sigma }$.

\subsubsection{Separation into $\mathcal I_{\mathcal A; \sigma}(  P_f, P_i)$ and $\mathcal I_{\mathcal N; \sigma}( \bar \alpha, P_f, P_i)$}

\label{sec:sep}

As summarized above, our approach is to split the integral into a singular part that can be evaluated semi-analytically, denoted $\mathcal I_{\mathcal A; \sigma}(  P_f, P_i)$, and a smooth remainder that is well-suited to numerical evaluation, denoted $\mathcal I_{\mathcal N; \sigma}(\bar \alpha,  P_f, P_i)$. To proceed we define
\begin{equation}
\label{eq:Dc}
D_{c}(m,  k) \equiv \frac{i}{k^2-m_2^2+i\epsilon}
\frac{1}
  {  (P_f-k)^2 - m_1^2 + i \epsilon  } 
\frac{ 1}
{  (P_i-k)^2 - m_1^2 + i \epsilon  } \,,
\end{equation}
and then introduce $D_r(m, \textbf k)$ via
\begin{equation}
\int \frac{d k^0}{2 \pi} D_c(m,k) \equiv D(m, \textbf k) + D_r(m, \textbf k) \,.
\end{equation}
In Appendix~\ref{app:INeval} we give an explicit expression for $D_r$. This term mops up the contributions from the two $P$-dependent poles in $D_c$. Here the subscripts $c$ and $r$ stand for {\em covariant} and {\em remainder}, respectively. The idea is to perform the integral of $D_c$ semi-analytically and that of $D_r$ numerically, and then to take the difference.

In these relations we have neglected the possible factors of $k_\mu$ and the spherical harmonics multiplying $D(m, \textbf k)$. To include these, we first introduce a tensor, $M$, defined such that
\begin{align}
\label{eq:tensorMdef}
 M^{\nu_1 \cdots \nu_N}_{ [\mu_1 \cdots \mu_n;  \ell_f m_{f} ;     \ell_i m_{i} ]}  K^{\omega}_{\nu_1\cdots \nu_N}(m,\textbf k) \equiv  \mathcal Y_{\ell_f m_f}(\textbf k^\star_{f}) \, K^{\omega}_{\mu_1\cdots \mu_n}(m,\textbf k)  \,   \mathcal Y^*_{\ell_i m_i}(\textbf k^\star_{i})    \,,
\end{align}
where $N = n+\ell_f+\ell_i$. A more explicit definition, together with various examples, is given in Appendix~\ref{sec:kin}. This simply amounts to recasting the factors of $\textbf k^\star$ within the harmonics as boosted factors of $k^\mu$.

To incorporate $K^{\omega}_{\nu_1 \cdots \nu_N}$, note that $D_r$ receives two contributions, one each from the poles at $k^0 = E_f + \omega_{ P_f  k1} - i \epsilon$ and $k^0 = E_i + \omega_{ P_i  k1}- i \epsilon$. (The third pole, at $k^0 = \omega_{k2} - i \epsilon$, generates the term we are after, $D(m,\textbf k)$.) We thus define
\begin{align}
\label{eq:DrfImpDef}
D_{rf}(m, \textbf k) & \equiv  \ointclockwise_{E_f + \omega_{ P_f  k1}} \frac{d k^0}{2 \pi} D_c(m,k) \,, \\
K^f_{\nu_1\cdots \nu_N}(m, \textbf k) & \equiv  k_{\nu_1}\cdots k_{\nu_N} \bigg \vert_{k^0 = E_f + \omega_{ P_f  k1}} \,,
\end{align}
where the integral here is a closed clockwise contour encircling the pole indicated. The definitions with an $i$ subscript are given by making the replacement $f \to i$ everywhere, and explicit expressions for $D_{ri}(m, \textbf k)$ and $D_{rf}(m, \textbf k)$ are given in App.~\ref{app:INeval}. With these quantities in hand, Eq.~(\ref{eq:Dc}) generalizes to
\begin{equation}
\label{eq:DcIndices}
\ointclockwise \frac{d k^0}{2 \pi} D_c(m,k) K_{\nu_1 \cdots \nu_N}(k) = D(m, \textbf k) K^{\omega}_{\nu_1 \cdots \nu_N}(m, \textbf k) +\mathcal K_{r; \nu_1 \cdots \nu_N}(m, \textbf k) \,.
\end{equation}
Here  the integral on the left-hand side is a closed clockwise contour encircling the three poles below the real axis and we have also introduced
\begin{equation}
\mathcal K_{r; \nu_1 \cdots \nu_N}(m, \textbf k) \equiv  D_{rf}(m, \textbf k)K^{f}_{\nu_1 \cdots \nu_N}(m, \textbf k) + D_{ri}(m, \textbf k) K^{i}_{\nu_1 \cdots \nu_N}(m, \textbf k)\,.
\end{equation}

The next step is to address the issue of ultraviolet convergence for these integrals. Equation (\ref{eq:Idef}) is manifestly convergent, due to the inclusion of the cutoff function $H(\bar{\alpha} , \textbf k)$. 
But to reach an integral that can be evaluated analytically it is convenient to introduce a second form of ultraviolet regularization. We explain the approach first for the special case of two indices, $N=2$. Here the integral has a logarithmic divergence that can be removed by taking Eq.~(\ref{eq:DcIndices}) and subtracting from this the same equation defined with $m_{1,2} \to  \Lambda$
\begin{multline}
\label{eq:n2}
\int \! \frac{d k^0}{2 \pi} \big [ D_c(m,k) - D_c(\Lambda,k) \big ] K_{\nu_1  \nu_2}(k) = \big [ D(m, \textbf k) K^{\omega}_{\nu_1  \nu_2}(m, \textbf k) -   D(\Lambda, \textbf k) K^{\omega}_{\nu_1  \nu_2}(\Lambda, \textbf k) \big ]  \\
+  \big [ \mathcal K_{r; \nu_1  \nu_2}(m, \textbf k)  -  \mathcal K_{r; \nu_1  \nu_2}(\Lambda, \textbf k)   \big ] \,.
\end{multline}
The regularization scale, $\Lambda$, is chosen so that the integrands that depend on it are smooth functions of $k$ with no need for an $i \epsilon$ prescription.  This holds for any $\Lambda$ satisfying $2 \Lambda >   \text{max}[E^{\star}_f,E^{\star}_i]$ though in practice it is useful to take the cutoff well above this minimum value.  %
On the left-hand side of Eq.~(\ref{eq:n2}) we have used that, for $N=2$, the $k^0$ integral can be extended to the entire real axis, with a vanishing arc at negative complex infinity. We additionally note that, as a result of the subtraction, the left-hand side and also both square bracketed terms on the right-hand side vanish as $1/\vert \textbf k \vert^5$ in the limit $\vert \textbf k \vert \to \infty$. These thus give convergent integrals with respect to $d^3 \textbf k$.

This approach can be extended to any number of $k_\mu$ factors, simply by forming more complicated linear combinations to cancel all divergent powers
\begin{equation}
\label{eq:fullLC}
\int \! \frac{d k^0}{2 \pi}  \sum_{j=0}^{n_j} c_j D_c(\Lambda_j,k)   K_{\nu_1 \cdots \nu_N}(k) =   \sum_{j=0}^{n_j} c_j D(\Lambda_j, \textbf k) K^{\omega}_{\nu_1 \cdots \nu_N}(\Lambda_j, \textbf k) +   \sum_{j=0}^{n_j} c_j \mathcal K_{r; \nu_1 \cdots \nu_N}(\Lambda_j, \textbf k)    \,,
\end{equation}
where we have introduced $c_0=1$ and $\Lambda_0= \{m_1, m_2\}$. As above, for $j>0$ we take $\Lambda_j$ such that the corresponding integrands are smooth functions of $k$ ($\Lambda_{j>0} >  \text{max}[E^{\star}_f,E^{\star}_i]$). In all cases, the linear combinations are constructed such that \emph{(i)} the $k^0$ integral extends to the entire real axis with a vanishing contribution from the arc at negative complex infinity, and \emph{(ii)} the left-hand side and each of the two sums on the right-hand side give convergent integrals with respect to $d^3 \textbf k$. In the following we sometimes refer to this as a Pauli-Villars-like regulator. We give a general algorithm for forming these linear combinations in Appendix~\ref{sec:detcjLambdaj}.

The final step is to multiply both sides of Eq.~(\ref{eq:fullLC}) by the cutoff function, and solve for the desired integral, defined in Eq.~(\ref{eq:Idef}). We deduce
\begin{equation}
\label{eq:Isep}
 \mathcal {I}_{ \sigma}(\bar{\alpha},P_f,P_i)   = \mathcal I_{\mathcal A; \sigma}(  P_f, P_i)  + \mathcal I_{\mathcal N;\sigma}(\bar{\alpha}, P_f, P_i)  \,,
\end{equation}
where
\begin{align}
\label{eq:IAdef}
\mathcal I_{\mathcal A; \sigma}(  P_f, P_i)  & \equiv  M^{\nu_1 \cdots \nu_N}_{\sigma}   \int \! \! \frac{d^4 k}{(2 \pi)^4}
\,K_{\nu_1 \cdots \nu_N}(k)
\, \sum_{j=0}^{n_j} c_j D_c(\Lambda_j,k)   
\,, \\[5pt]
\begin{split}
\mathcal I_{\mathcal N; \sigma}(\bar{\alpha}, P_f, P_i) &\equiv  M^{\nu_1 \cdots \nu_N}_{\sigma}  
\bigg [   \int \! \frac{d^3 \textbf k}{(2\pi)^3} \, 
\,
 [H(\bar{\alpha}, \textbf k) - 1]
 \,
\sum_{j=0}^{n_j} c_j \Big [ D(\Lambda_j, \textbf k) K^{\omega}_{\nu_1 \cdots \nu_N}(\Lambda_j, \textbf k) +\mathcal K_{r; \nu_1 \cdots \nu_N}
(\Lambda_j, \textbf k) \Big ]  
 \\
&\hspace{2cm}
 -  \int \! \frac{d^3 \textbf k}{(2\pi)^3}
 \,
     \, H(\bar{\alpha}, \textbf k)  \sum_{j=1}^{n_j}  c_j D(\Lambda_j, \textbf k) K^{\omega}_{\nu_1 \cdots \nu_N}(\Lambda_j, \textbf k)
 \\
&\hspace{2cm}
 -  \int \! \frac{d^3 \textbf k}{(2\pi)^3} \, 
 \, H(\bar{\alpha}, \textbf k)   \sum_{j=0}^{n_j} c_j   \mathcal K_{r; \nu_1 \cdots \nu_N}(\Lambda_j, \textbf k)    
 \bigg ] \,.
 \end{split}
\label{eq:INdef}
\end{align}
Equations (\ref{eq:Isep})-(\ref{eq:INdef}) are the main results of this subsection.

We emphasize here that the integrands in the definition of $\mathcal I_{\mathcal N}$ are smooth for all real $\textbf k$, and the integrals are ultraviolet convergent. For the second and third terms this follows from the fact that $H(\bar{\alpha}, \textbf k)$ decays exponentially, together with the observation that $\mathcal K_{r}$ is a smooth function, as is $D(\Lambda_j, \textbf k)$ for $j>0$. For the first term, smoothness is guaranteed because $[H(\bar{\alpha}, \textbf k) - 1] $ vanishes at the pole and ultraviolet convergence follows from the careful construction of the linear combination. As the integrands are smooth and the integrals are convergent, $\mathcal I_{\mathcal N; \sigma}(\bar{\alpha}, P_f, P_i) $ is well suited to numerical evaluation. %

\begin{figure}
\begin{center}
\includegraphics[width=\textwidth]{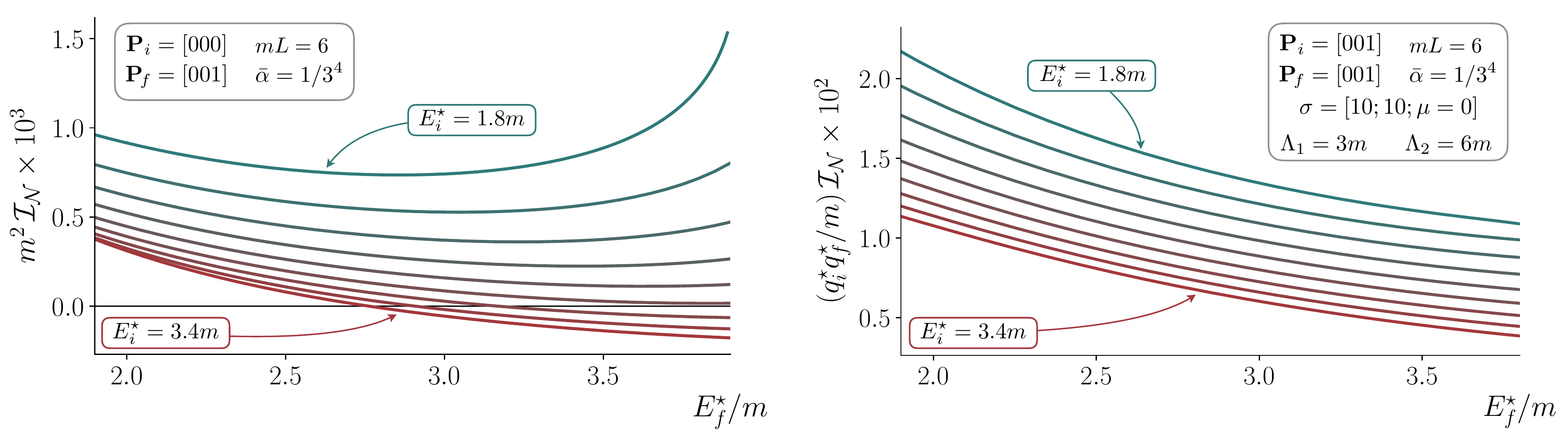}
\caption{Examples of the numerical integral $\mathcal I_{\mathcal N}^{\sigma}(\bar \alpha, P_f,P_i)$, plotted as a function of $E_f^\star$ with all other arguments fixed. In the left panel we consider the case of $\ell_f, m_f = \ell_i, m_i = 00$ with no factors of $k_\mu$ in the numerator. For these kinematics the function requires no subtractions and we directly evaluate $\mathcal I_{\mathcal N}$ at $\bar \alpha =1/3^4$, for fixed external 3-momenta and evenly spaced values of $E_i^\star$, as indicated. In the right panel we plot the function with a numerator factor of $3 (k^{z\star})^2 k_0$. In this case one requires Pauli-Villars-like subtractions, as described in the main text and summarized in the caption.
\label{fig:I_N}}
\end{center}
\end{figure}

It is instructive to consider a few specific examples of this construction, beginning with $n=0$, $\ell_i=\ell_f=0$, in which no factors of $k_\mu$ appear in the numerator of $\mathcal I_\sigma$. In this case the integrals are directly convergent, without any need for additional subtraction terms (i.e.~the sum over $j$ in Eq.~(\ref{eq:fullLC}) reduces to the $j=0$ term). Equations (\ref{eq:IAdef}) and (\ref{eq:INdef}) then reduce to 
\begin{align}
\mathcal I_{\mathcal A }(  P_f, P_i)  & \equiv    \int \! \! \frac{d^4 k}{(2 \pi)^4}   D_c(m,k)   \,, \\[5pt]
\begin{split}
\mathcal I_{\mathcal N}(\bar{\alpha}, P_f, P_i) & \equiv  \int \! \frac{d^3 \textbf k}{(2\pi)^3}   \Big [ D(m, \textbf k)   +  D_{rf}(m, \textbf k)   + D_{ri}(m, \textbf k) \Big ]   [H(\bar{\alpha}, \textbf k) - 1]  \\
& \hspace{200pt} -  \int \! \frac{d^3 \textbf k}{(2\pi)^3} \, H(\bar{\alpha}, \textbf k)   \Big [ D_{rf}(m, \textbf k)   + D_{ri}(m, \textbf k) \Big ]   \,.
\end{split}
\end{align}
This no-index version of $\mathcal I_{\mathcal N}$ is plotted in the left panel of Fig.~\ref{fig:I_N}. 

 We emphasize here that $\mathcal I_{\mathcal N}$, and thus also the original integral, diverges in the $\bar \alpha \to 0$ limit. This is because in the original integral the covariant propagators are evaluated at on-shell $k$ ($k^2 = m^2$), so the propagators scale as $1/\vert \textbf k \vert$ and the integrand as $d^3 \textbf k / \vert \textbf k \vert^3$. In other words the convergence of $\mathcal I_{\mathcal A}$ is always better than that of the original integral by two powers of $k$, resulting from the off-shell integration of $k^\mu$. For fixing the subtraction scheme in Eqs.~(\ref{eq:IAdef}) and (\ref{eq:INdef}), it is only necessary that $\mathcal I_{\mathcal A}$ be rendered finite. [See also Appendix \ref{sec:detcjLambdaj}.]

We close with one final example:~$\sigma = [\mu =0; 10; 10]$, corresponding to a factor of $\mathcal Y_{10} k_0 \mathcal Y^*_{10}$ in the numerator. This leads to an $\mathcal I_{\mathcal A;\sigma}$ integral with an integrand scaling as $d^4k k^2/k^6$, i.e.~diverging as $\log \Lambda$. Performing a single subtraction of the same integral with $m \to \Lambda$ is therefore sufficient to render the result finite. In fact, to improve the numerical evaluation of $\mathcal I_{\mathcal N;\sigma}$, and to test our general method, here we choose to make two subtractions. As explained in Appendix \ref{sec:detcjLambdaj}, one possible choice is to add an integral evaluated at $\Lambda = 3m$ (with coefficient $-35/27$) and a second at $\Lambda = 6m$ (with coefficient $8/27$). Implementing this in Eqs.~(\ref{eq:IAdef}) and (\ref{eq:INdef}) leads to convergent forms of $\mathcal I_{\mathcal A;\sigma}$ and $\mathcal I_{\mathcal N;\sigma}$ respectively, with integrands scaling as $d k/k^5$. $\mathcal I_{\mathcal N;\sigma}$ in this scheme is plotted in the right panel of Fig.~\ref{fig:I_N}.

As we include additional factors of $k_\mu$ the expressions complicate, first because we need additional terms in the sum over $j$ (to maintain convergent integrals) and second because the numerical integrals depend on multiple vector components. However, we find that no conceptual issues arise and the task amounts to coding Eq.~(\ref{eq:INdef}) with an efficient numerical integrator. We give some details on our approach in Appendix~\ref{app:INeval}, but consider $\mathcal I_{\mathcal N;\sigma}(\bar{\alpha}, P_f, P_i) $ as a numerically known function for the remainder of the main text. 

Thus it remains only to evaluate $\mathcal I_{\mathcal A; \nu_1 \cdots \nu_{N}}(  P_f, P_i) $, to which we now turn.%

\subsubsection{Evaluating $\mathcal I_{\mathcal A; \sigma}(  P_f, P_i)$\label{sec:IAeval}}

We first use the tensor $M^{\nu_1 \cdots \nu_N}_{ \sigma}$, introduced in Eq.~(\ref{eq:tensorMdef}) above, to define a version of $\mathcal I_{\mathcal A}$ with no spherical-harmonic indices
\begin{align}
\label{eq:Gmat_PfnPi}
\mathcal I_{\mathcal A; \sigma}(  P_f, P_i) \equiv M^{\nu_1 \cdots \nu_N}_{ \sigma}  \mathcal I_{\mathcal A; \nu_1 \cdots \nu_N }(P_f,P_i) \,.
\end{align}
The integrals on the right-hand side can then be written as
\begin{equation}
\label{eq:IAsumoverDR}
\mathcal I_{\mathcal A; \nu_1 \cdots \nu_{N}}(  P_f, P_i)   \equiv  \lim_{\delta \to 0} \sum_{j=0}^{n_j} c_j \mathcal I_{\mathcal A; \nu_1 \cdots \nu_{N}}(  P_f, P_i,  \Lambda_j, \delta)  \,,
\end{equation}
where
\begin{equation}
\label{eq:IAdimreg}
\mathcal I_{\mathcal A; \nu_1 \cdots \nu_{N}}(  P_f, P_i, m, \delta)   \equiv    \int \! \! \frac{d^{4 - \delta} k}{(2 \pi)^{4 - \delta}}   \frac{i}{k^2-m_2^2+i\epsilon}
\frac{1}
  { (P_f-k)^2 - m_1^2 + i \epsilon  } 
\frac{ 1}
{  (P_i-k)^2 - m_1^2 + i \epsilon  }  k_{\nu_1} \cdots k_{\nu_N}   \,.
\end{equation}
Here we have used the fact that the sum over $j$ gives a convergent integral, and is thus equal to the $\delta \to 0$ limit of the integral in $4 - \delta$ dimensions. Then, at fixed delta, one can exchange the orders of summation and integration, leading to Eq.~(\ref{eq:IAsumoverDR}).

To evaluate the integral in Eq.~(\ref{eq:IAdimreg}), we first perform the standard Wick rotation on the $k^0$ integration contour (counter-clockwise to the imaginary axis), and similarly rotate the time components $P_{i,0}$ and $P_{f,0}$. We then make the variable redefinitions
\begin{equation}
k_0 \equiv ik_{E,0},\ P_{f,0} \equiv i P_{E,f,0}  ,\ P_{i,0} = i P_{E,i,0}   \,,
\end{equation}
to reach
\begin{align}
\mathcal I_{\mathcal A; \nu_1 \cdots \nu_{N}}( P_f, P_i, m, \delta) 
&=
\xi_{\nu_1\cdots \nu_N}
\int\frac{d^{4 - \delta} k_E}{(2\pi)^{4 - \delta}}
\frac{1}{k^2_E+m_2^2}
\frac{1}
  {  (P_{E,f}-k_E)^2 + m_1^2   } 
\frac{ 1
 }{  (P_{E,i}-k_E)^2 + m_1^2   } k_{E,\nu_1}\,\cdots  k_{E,\nu_{N}} \,, \\
&=
\xi_{\nu_1\cdots \nu_N}
\frac{\partial}{i\partial \chi_E^{\nu_1}}
\cdots 
\frac{\partial}{i\partial \chi_E^{\nu_N}}
\mathcal I^{\chi}(P_f, P_i, m,\delta)
\bigg|_{\chi=0} \,,
\label{eq:DerWRTalpha}
\end{align}
where $\xi_{\nu_1\cdots \nu_N}\equiv(i)^{\delta_{{\nu_1},0}+\cdots+ \delta_{{\nu_N},0}}$. {Note here that the indices are not contracted between $\xi_{\nu_1\cdots \nu_N}$ and the momenta but rather are held fixed on both sides of the equation.}

In the second step we have introduced the generating functional
\begin{align}
\label{eq:Ialphadef}
\mathcal I^{\chi}(P_f, P_i, m, \delta)
& \equiv
\int\frac{d^{4 - \delta} k_E}{(2\pi)^{4 - \delta}}
\frac{e^{i\chi_E \cdot k_E} }{k^2_E+m_2^2}
\frac{1}
{  (P_{E,f}-k_E)^2 + m_1^2  } 
\frac{ 1}
{  (P_{E,i}-k_E)^2 + m_1^2   } \,.
\end{align}
As we show in Appendix~\ref{app:Jeval}, this reduces to
\begin{equation}
\mathcal I^{\chi}(P_f, P_i, m, \delta)
 = 
 \int_0^1dx\int_0^{1-x}dy
\, e^{- i\chi\cdot (x P_{f}+y P_{i})}
   \sum_{n=0}^\infty   \frac{ (\chi^2 )^{n}}{ 4^n n!}     
  \frac{\Gamma(1-n + \delta/2)}{(4\pi)^{2 - \delta/2}  }  M(m,x,y)^{2n - 2 - \delta} \,,
  \label{eq:ExpAlpha}
 \end{equation}
 where
  \begin{align}
\label{eq:Mdef}
M(m,x,y)^2 & =  (1-x-y)m_2^2+ (x+y)m_1^2-x(1-x) s_{f}-y(1-y) s_{i} + x y\, (Q^2 + s_{f} + s_{i})-i\epsilon  \,.
 \end{align}
In these results we have analytically continued back to real $P_{i,0}$ and $P_{f,0}$ and expressed all quantities in terms of the 4-vectors $P_f$ amd $P_i$ as well as the Lorentz invariants $Q^2= - (P_i-P_f)^2$, $s_i=P_i^2,$ and $s_f=P_f^2$.
The corresponding analytic continuation of Eq.~(\ref{eq:DerWRTalpha}) is given by
\begin{align}
\mathcal I_{\mathcal A}^{\nu_1 \cdots \nu_{N}}( P_f, P_i, m, \delta)   &=
\frac{\partial}{-i\partial \chi_{\nu_1}}
\cdots 
\frac{\partial}{-i\partial \chi_{\nu_N}}
\mathcal I^{\chi}(P_f, P_i, m,\delta)
\bigg|_{\chi=0} \,.
\label{eq:DerWRTalphaMINK} 
\end{align}
Taken together, Eqs.~(\ref{eq:ExpAlpha})-(\ref{eq:DerWRTalphaMINK}) %
give the main result of this subsection.

As above, it is instructive to consider $\mathcal I_{\mathcal A}$ for $N=0$, i.e.~with no factors of $k_\mu$ in the numerator. In this case Eq.~(\ref{eq:DerWRTalphaMINK}) is evaluated with no $\chi$-derivatives and gives
  \begin{align}
\mathcal I_{\mathcal A}( P_f, P_i, m, \delta)  %
 & =   \frac{\Gamma(1 + \delta/2)}{(4\pi)^{2 - \delta/2}}     \int_0^1dx\int_0^{1-x}dy \,
  \frac{1}{M(m,x,y)^{ 2 + \delta}  } \,.
 \end{align}
 
 As noted at the end of the previous subsection, this integral is convergent in the $\delta \to 0$ limit, and simplifies to %
  \begin{align}
\mathcal I_{\mathcal A}( P_f, P_i, m, 0) %
&=
 \frac{1}{(4\pi)^{2}}
\frac{1}{s_{i}}
\int_0^1dx\int_0^{1-x}dy
\,
\frac{
1}
{[y-y_+(x)][y-y_-(x)]} \,,
\label{eq:IAypoles}
 \end{align}
where we have substituted 
 \begin{equation}
 M(m,x,y)^2 = s_i  [y-y_+(x)][y-y_-(x)] \,,
 \label{eq:Mdef_ys}
 \end{equation}
 with
\begin{align}
y_\pm(x) & \equiv \frac{1}{2} \left( A \pm  \sqrt{A^2 + B + i\epsilon } \right) \,, \\[5pt]
\label{eq:Axdef}
A & \equiv 1+\frac{m_2^2-m_1^2+x(q^2- s_{f}- s_{i})}{s_i} \,, \\
 B & \equiv - 4\frac{m^2_2 -x (m^2_2-m^2_1) -  x(1-x)\,s_f}{s_i}  \,.
\label{eq:Bxdef}
\end{align}

The final analytic step is to evaluate the integral with respect to $y$.
We do so via the identity
\begin{equation}
\int_0^{1-x}\frac{dy}{y-(f(x) \pm i \epsilon)} =
 \log \left \lvert \frac{1 - x -f(x)}{f(x)} \right \rvert + i \arctan \! \left( \frac{1 - x-\text{Re} f(x) }{ \text{Im} f(x)  \pm \epsilon} \right )  +  i \arctan \! \left( \frac{\text{Re} f(x) }{\text{Im}f(x)  \pm \epsilon} \right )   \equiv \mathcal L_{\pm \epsilon}[f(x)]
 \label{eq:calLdef}
\,,
\end{equation}
where in the second equality we have introduced a short-hand for the result of the integral. Note that, as long as $\text{Im}f(x)$ is nonzero, we can safely set $\epsilon = 0$ in these expressions.  In the case that the imaginary part does vanish, then we use the relation $\lim_{\epsilon \to 0} \arctan(a/\epsilon) = \text{sign}[a] \pi/2$.

Applying this to Eq.~(\ref{eq:IAypoles}) we deduce
  \begin{equation}
\mathcal I_{\mathcal A}( P_f, P_i, m, 0)  =
\int_0^1dx   F^{(1)}(x) \,,
\label{eq:IA_00}
 \end{equation}
where $F^{(1)}(x)$ [also given in Eq.~(\ref{eq:f1})] is
 \begin{equation}
F^{(1)}(x) \equiv \frac{1}{(4\pi)^{2} s_i}   \frac{\mathcal L_{+ \epsilon}[y_+(x)] - \mathcal L_{- \epsilon}[y_-(x)]}{y_+(x)-y_-(x)} \,.
\end{equation} 
In this case $\text{Im} y_{+}(x) = - \text{Im} y_{-}(x) = \text{Im} \sqrt{A^2 + B}/2$. Thus the substitution $\lim_{\epsilon \to 0}\arctan(a/\epsilon) = \text{sign}[a] \pi/2$ is only needed when the argument of the square root is positive. As we explain in detail in Appendix \ref{app:triangle}, evaluating the remaining integral over $x$ reveals that $\mathcal I_{\mathcal A}(P_f,P_i)$ has triangle singularities that arise whenever $P_f$ and $P_i$ take on values for which all three particles in the triangle of Fig.~\ref{fig:iW_Gfunc}(b) can go on shell.

\begin{figure}
\begin{center}
\includegraphics[width=0.7\textwidth]{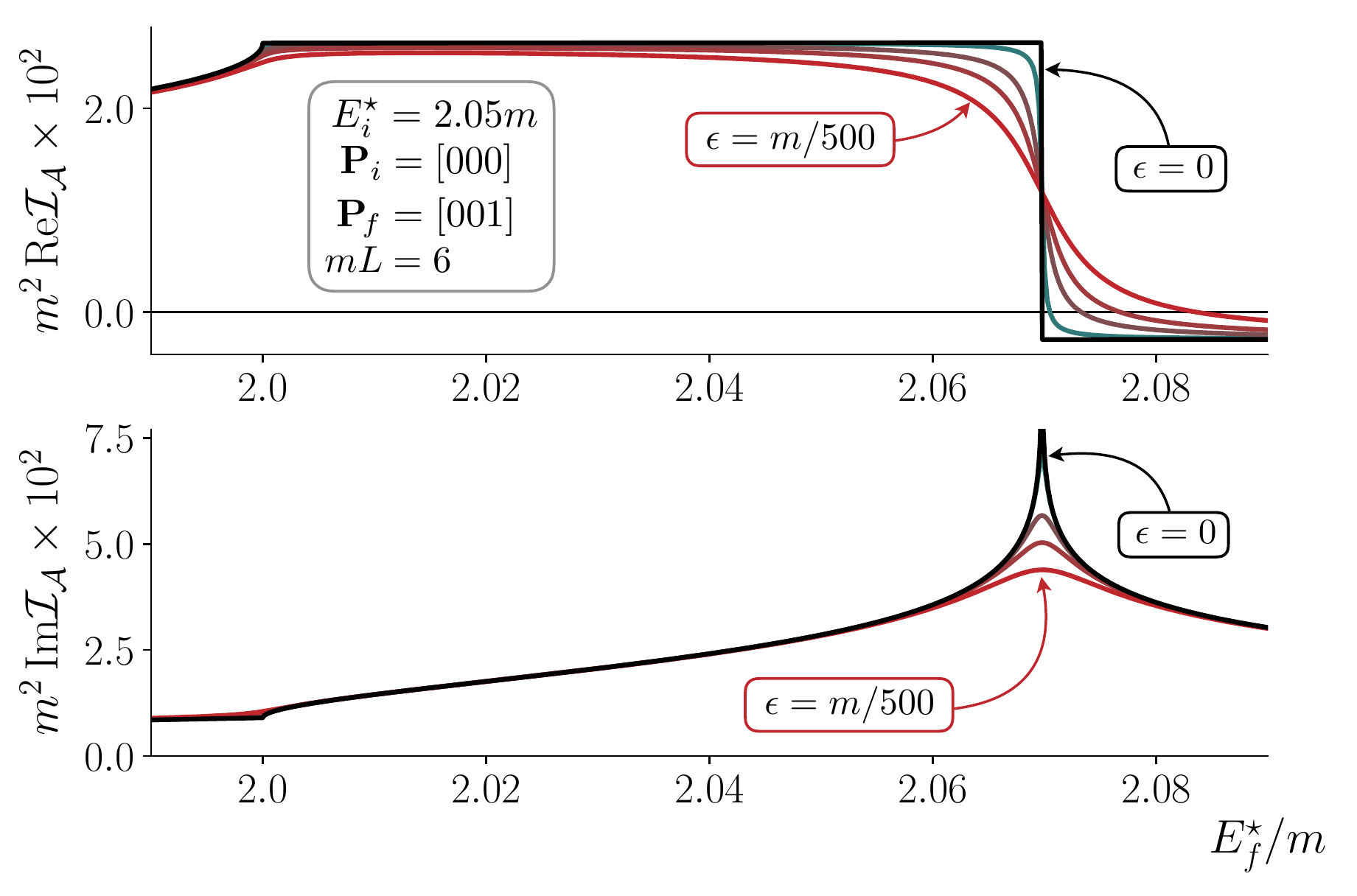}
\caption{The real and imaginary parts of $\mathcal I_\mathcal{A}$, generated using the single-parameter integral of Eq.~(\ref{eq:IA_00}), with $m_1 = m_2$ and all other kinematics as indicated in the legend. The fixed external coordinates are chosen such that a singularity arises at $E_{f,c}^\star \approx 2.07$. As explained in the text, the discontinuity is induced when a pole in the Feynman parameter, $x$, crosses into the integrated region. We vary the value of $\epsilon$ (in the $i \epsilon$ pole prescription) to illustrate how the singularity arises as $\epsilon \to 0$.%
}
\label{fig:eps_dep}
\end{center}
\end{figure}

More precisely, we show in Appendix \ref{app:triangle} that the singularity locations are governed by the discriminant of the polynomial $A^2 + B \equiv a x^2 + b x + c$, given by
\begin{equation}
\label{eq:Xdef}
X (s_f,s_i,Q^2) \equiv b^2 - 4 a c = m_1^2 \left((s_f-s_i)^2+Q^2(2 m_2^2 -m_1^2 + s_f+s_i)\right)-Q^2 \left(m_2^4-m_2^2 \left(Q^2+s_f+s_i\right)+s_f s_i\right)  \,.
\end{equation}
Critical kinematics are realized whenever $X (s_f,s_i,Q^2) = 0$, so that $A^2 + B = (x-x_c)^2$, and in addition, $x_c$ and $y_c \equiv y_+(x_c) $ fall in the integrated range. It can be shown that these conditions are equivalent to the ones found using Landau's
singularity classification \cite{Landau:1959fi}.

\begin{figure}
\begin{center}
\includegraphics[width=0.7\textwidth]{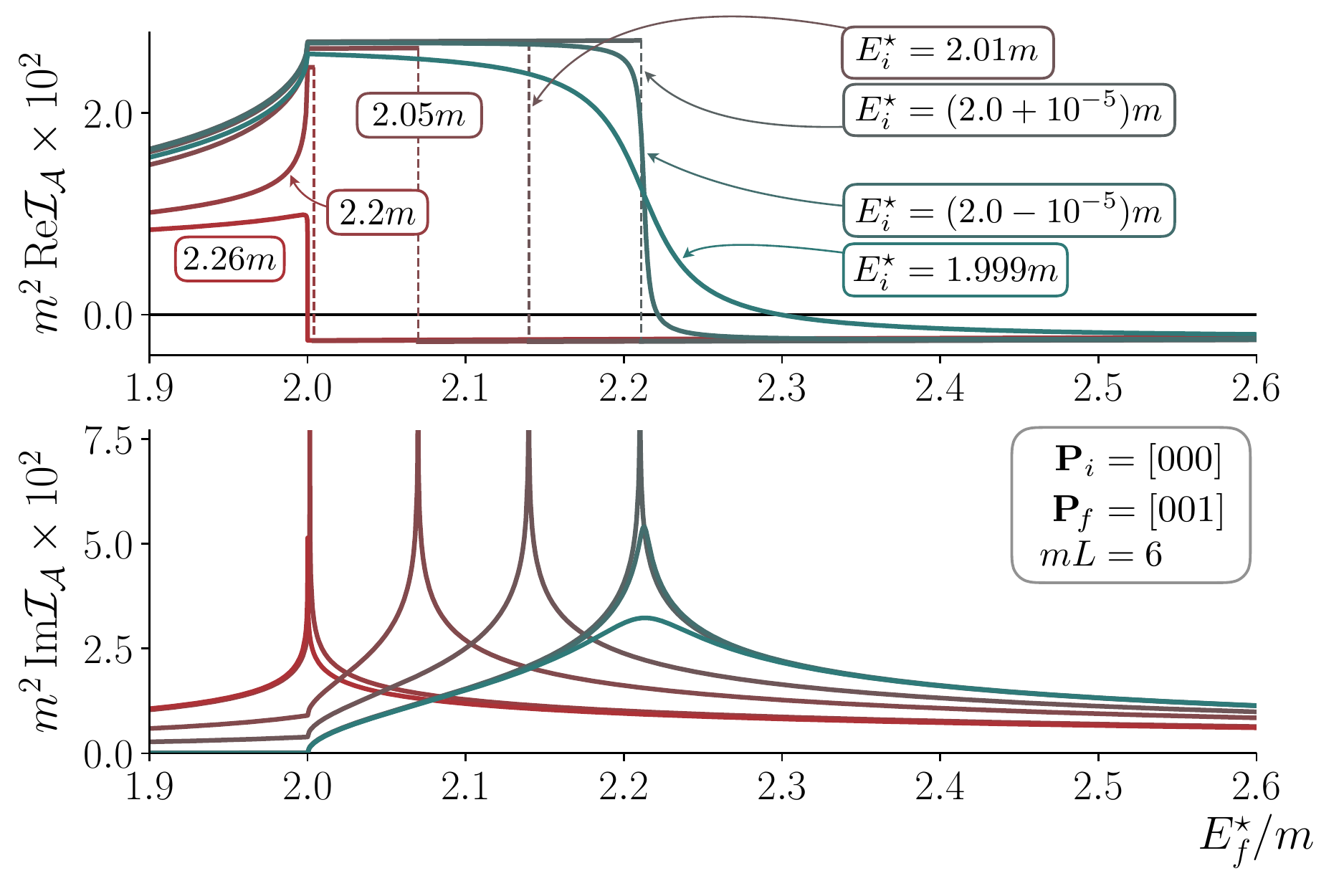}
\caption{The real and imaginary parts of $\mathcal I_\mathcal{A}$, generated using the single-parameter integral of Eq.~(\ref{eq:IA_00}), evaluated piecewise in order to work exactly at $\epsilon=0$ as described in the paragraph containing Eq.~(\ref{eq:calLdef}). As indicated, the various curves correspond to fixed values of $E_i^\star$, chosen to illustrate the behavior of the triangle singularity.  For $E_i^\star < 2m$, $\mathcal I_\mathcal{A}$ is a smooth function of $E_f^\star$ away from threshold. As $E_i^\star$ approaches threshold from below, $\text{Re} \mathcal I_\mathcal{A}$ forms a step-function singularity and $\text{Im} \mathcal I_\mathcal{A}$ a logarithmic divergence. When $E_i^\star$ is then further increased, this singularity moves to lower values of $E_f^\star$, eventually colliding with the threshold cusp.
}
\label{fig:diff_Eis}
\end{center}
\end{figure}

At values of $P_f$ and $P_i$ satisfying these conditions, the real part of $\mathcal I_{\mathcal A}$ has a step-function discontinuity of height
\begin{align}
\text{Disc}(\mathcal I_{\mathcal A})
&= \frac{1}{16\sqrt{(P_i\cdot P_f)^2 - \, s_f s_i}}\,,
\label{eq:DiscVal}
\end{align}
and the imaginary part shows a logarithmic divergence.
In Fig.~\ref{fig:eps_dep} we illustrate how these singularities form in the $\epsilon \to 0$ limit of the $i \epsilon$ pole prescription. 
In Fig.~\ref{fig:diff_Eis} we show the singularity structure as a function of $E_f^\star$ for various fixed values of $E_i^\star$. In particular one sees that, for sub-thresold $E_i^\star$, $\mathcal I_{\mathcal A}$ is a smooth function away from the two-particle production threshold. As $E_i^\star$ approaches $2m$ a step forms in $\text{Re} \mathcal I_{\mathcal A}$ and $\text{Im} \mathcal I_{\mathcal A}$ develops a log divergence. Then, as $E_i^\star$ is further increased, the location of the singularity in $E_f^\star$ moves towards and eventually collides with the two-particle threshold. 

To complete this subsection we would like to comment on the behavior of this singularity for some special set of kinematics. First, in the case of identical initial and final 3-momenta, i.e.~$\textbf{P}_i=\textbf{P}_f$, $\mathcal I_{\mathcal A}$ does not have any other singularities apart from those arising at threshold, and therefore the $G$-function will not exhibit a triangle singularity. This is consistent with our analysis of the $P_f=P_i$ case, and with the numerical example shown in Fig.~\ref{fig:Pi001Pf001}. In other words, given that all the external momenta in the triangle diagram are time-like, the condition of all three internal propagators to be on shell cannot be realized. A second example is the special case of $m_1 = m_2$ and $s_i = s_f = s$. Solving $X(s,s,Q^2)=0$ in this simple case leads to a singular manifold given by
\begin{align}
Q^2 = -4s \left(1-\frac{s}{4 m^2}\right) \,.
\end{align}
In this case it is easier to visualize that this condition is equivalent to that of all three intermediate particles in the triangle diagram, Fig~\ref{fig:iW_Gfunc}(b), going on-shell. (See also Refs.~\cite{Bayar:2016ftu, Coleman:1965xm}.)

This concludes our discussion of $\mathcal I_{\mathcal A}$ within the main text.  In Appendix~\ref{sec:IAsigmas} we extend the results here by explicitly evaluating $\mathcal I_{\mathcal A; \nu_1}$, $\mathcal I_{\mathcal A; \nu_1  \nu_2}$ and $\mathcal I_{\mathcal A; \nu_1  \nu_3  \nu_3}$. For these integrals we find that the above-threshold discontinuities persist, but are milder when factors of $k_\mu$ appear in the numerator.

 \subsubsection{Examples of $G^\sigma\!(P_f,P_i,L)$}

\begin{figure}
\begin{center}
\includegraphics[width=\textwidth]{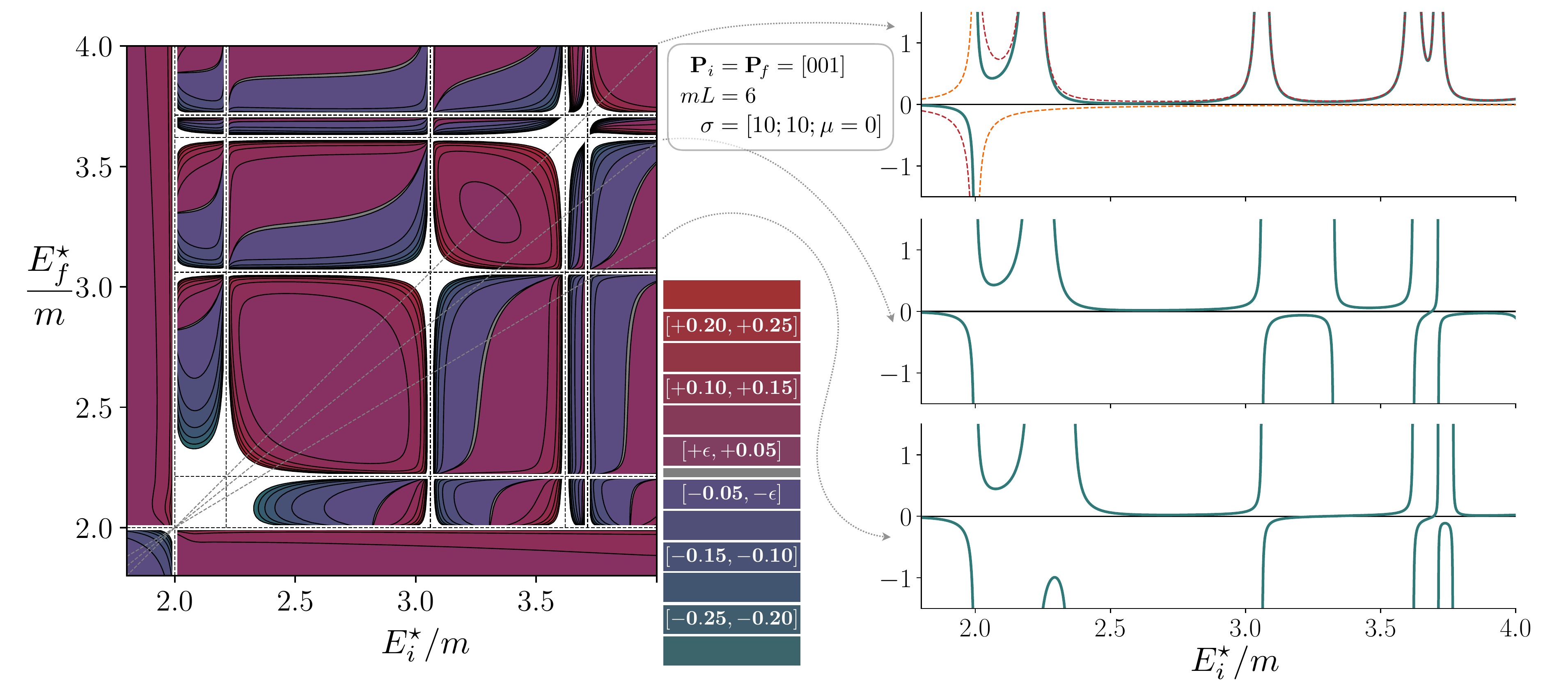}
\caption{Contour plot representation of $\text{Re}\,G^{\mu=0}_{10;10}(P_f,P_i,L)$ for $\textbf P_{\!f} = \textbf P_i = [001]$ and $m L=6$. The grey diagonal dashed lines in the left panel indicate slices defined by $E_f^\star = s (E_i^\star - 2 m) + 2m$, for $s=1.0,0.8,0.6$. We plot $\text{Re}\,G^{\sigma}$ along these slices in the right panel as indicated. The top right panel corresponds to $P_f = P_f$ and the plot matches Fig.~\ref{fig:GPiEQPf} well within the expected $e^{- m L}$ discrepancies.  
Here we also separately show the contributions from $\mathcal I_{\mathcal A}$ (dashed orange) and the remaining contribution to $\text{Re}\,G^{\sigma}$ (dashed red).  \label{fig:Pi001Pf001}}
\end{center}
\end{figure}
\begin{figure}
\begin{center}
\includegraphics[width=\textwidth]{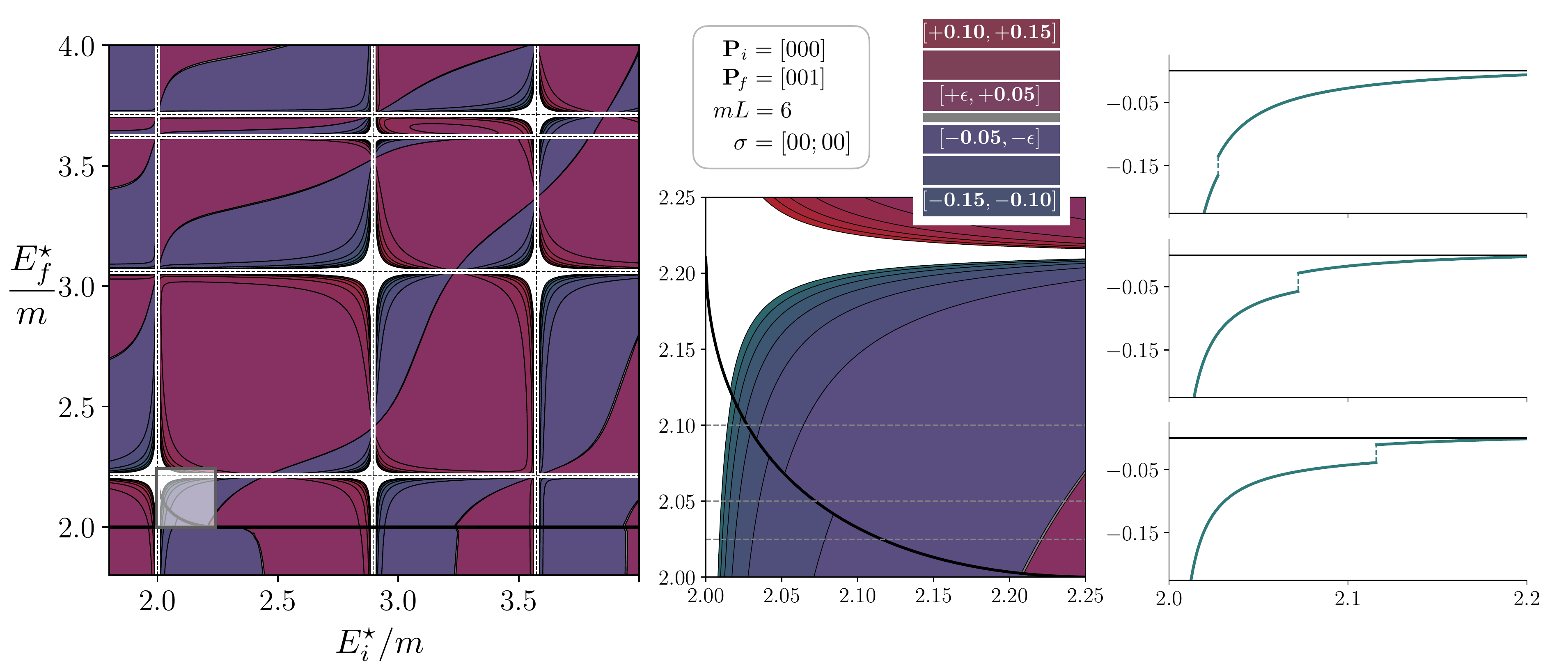}
\caption{Contour plot representation of $\text{Re}\,G_{00;00}(P_f,P_i,L)$ for  $\textbf P_i = [000]$, $\textbf P_f=[001]$ and $m L=6$.
Allowing the spatial momenta to differ means that the finite-volume spectrum must be different for the incoming and outgoing states, as is apparent from the different positions of the poles, corresponding to non-interacting levels (vertical and horizontal dashed lines in the left panel). Another feature of differing spatial momenta is the appearance of triangular singularities, emphasized in the middle and right panels. On the far right we plot three slices, running over the step discontinuity in $\text{Re}\,G^{\sigma}$.}
\label{fig:Pi000Pf001}
\end{center}
\end{figure}

 Having discussed the integral entering $G^\sigma\!(P_f,P_i,L)$ in great detail we are now ready to put everything together and evaluate the complete function. We do so for two different examples of external kinematics. First, in Fig.~\ref{fig:Pi001Pf001}, we consider the case of $\textbf P_i = \textbf P_f = (2 \pi/L)[001]$ with $\sigma = [\mu=0; 10; 10]$. Then, in Fig.~\ref{fig:Pi000Pf001} we take $\textbf P_i = [000] $ and $ \textbf P_f=(2 \pi/L)[001]$ with $\sigma = [00;00]$, i.e.~with no numerator factors. In both cases we fix $mL=6$ and plot $\text{Re}\,G^\sigma$ for all values of $E_i^\star$ and $E_f^\star$ in the region of interest. 
 
 As explained in the figure captions, each example illustrates important issues and features that arise. In Fig.~\ref{fig:Pi001Pf001} we consider various diagonal slices of the $E_i^\star$, $E_f^\star$ plane. We find that the result for $E_i^\star = E_f^\star$ is in perfect agreement with the $P_i = P_f$ result determined by combining the various $\mathcal Z_{JM}$ functions. This provides a strong check on the two different methods.  
 Figure \ref{fig:Pi001Pf001} also illustrates that double poles arise along the $E_i^\star = E_f^\star$ line, but these split to single poles as the slice is rotated away from this singular choice.
 
 In addition, Fig.~\ref{fig:Pi001Pf001} illustrates the results of using the Pauli-Villars-like regulator to separate the integral into $\mathcal I_{\mathcal A}$ and $\mathcal I_{\mathcal N}$. As discussed towards the end of Sec.~\ref{sec:sep}, the original integral contains a factor scaling as $\mathcal Y_{10} k^\mu \mathcal Y_{10} \sim \vert \textbf k \vert^3$ in the numerator leading to a $\log \Lambda$ divergence in $\mathcal I_{\mathcal A}$. Following Appendix \ref{sec:detcjLambdaj} we handle this by evaluating Eqs.~(\ref{eq:IAdef}) and (\ref{eq:INdef}) with $\{\Lambda_1, \Lambda_2\} = \{3m, 6m\}$ and $\{c_1, c_2\} = \{-35/27,8/27\}$. This removes not only the divergence but also a $dk/k^3$ term in the integrand to further optimize the numerical convergence of $\mathcal I_{\mathcal N}$. We already gave the result for $\mathcal I_{\mathcal N}$ in this prescription in the right panel of Fig.~\ref{fig:I_N}. Here the result is combined with the sum and $\mathcal I_{\mathcal A}$ to reach $G^\sigma$. With all building blocks summed together, the dependence on the Pauli-Villars parameters, $\Lambda_i$, cancels (between $\mathcal I_{\mathcal A}$ and $\mathcal I_{\mathcal N}$) as does the dependence on the smooth cutoff parameter, $\bar \alpha$, (between $\mathcal I_{\mathcal N}$ and the sum). 
 
Turning now to Fig.~\ref{fig:Pi000Pf001}, this result displays two additional features of $G^\sigma$. First we see that, when $\textbf P_i$ and $\textbf P_f$ differ, the non-interacting two-particle poles appear in different locations for $E_i^\star$ and $E_f^\star$. Interactions shift the finite-volume energies away from these singularities so that $G^\sigma$, like the Lellouch-L\"uscher factors, will generally be evaluated away from the divergent locations. However, as with all finite-volume kinematic functions, this implicit knowledge of the non-interacting spectrum is a key ingredient in the all-orders correction of the scattering-state volume effects.

Second, we see the appearance of triangle singularities inherited through $\mathcal I_{\mathcal A}$. Such features are simply part of the correct definition of $G^\sigma$. Indeed, because the singularity structure is directly induced by the infinite-volume diagram of Fig.~\ref{fig:iW_Gfunc}(b), it also appears within the infinite-volume $\textbf 2 + \mathcal J \to \textbf 2$ transition amplitude itself. The steps, cusps and log divergences of $G^\sigma$ are present in both $\mathcal W$ and $\mathcal W_{\text{df}}$ for exactly the same kinematics. Thus, understanding the features is crucial to extracting and interpreting the infinite-volume observables that we are after. We discuss the role of discontinuities within the transition amplitude in more detail in future work.

\section{Conclusion \label{sec:conclusion}}

In this work we have presented a modified version of the finite-volume formalism for studying $\textbf 2 + \mathcal J \to \textbf 2$ transition amplitudes. This is closely related to the approach of Ref.~\cite{Briceno:2015tza}, but differs in that all infinite-volume quantities are Lorentz covariant and the $\textbf 1 + \mathcal J \to \textbf 1$ matrix elements have been reformulated in terms of standard form factors. As explained in Sec.~\ref{sec:cov2to2} and Appendix \ref{sec:newder}, the new result is reached by making minor adjustments to the derivation presented in Ref.~\cite{Briceno:2015tza}. For example, in that work finite-volume effects are expressed as sums over poles of the form $1/[2 \omega_k (k^0 - \omega_k)]$, and here the same effects are expressed via invariant poles, $1/(k^2 - m^2)$. 

These changes lead to a modified form of the finite-volume function, denoted $G$, with an added benefit that the new form is easier to evaluate numerically. As described in Sec.~\ref{sec:IAeval}, the Lorentz covariant structure allow us to write the integral appearing in $G$ in terms of Feynman parameters. This reduction is also crucial to revealing the analytic structure of $G$, including the triangle singularities described in great detail Sec.~\ref{sec:IAeval} and Appendix \ref{app:triangle}, and illustrated in Figs.~\ref{fig:eps_dep}, \ref{fig:diff_Eis} and \ref{fig:Pi000Pf001}. We also recall that the form of $G$ presented in Ref.~\cite{Briceno:2015tza} carries four sets of spherical harmonic indices, resulting from a cumbersome description of the $\textbf 1 + \mathcal J \to \textbf 1$ matrix elements. By contrast, our improved expression carries only the angular momentum indices of the external states, together with Lorentz indices to describe the current insertion.

To avoid proliferation of flavor and channel indices, in this work we restricted attention to kinematics for which a single channel of two scalar particles is open.  
Accommodating multiple channels is straightforward given the results of Ref.~\cite{Briceno:2015tza}. %
Incorporating particles with spin has yet to be considered for these types of observables and is the subject of future work. 
This final generalization is of great importance given the phenomenological interest in two-nucleon matrix elements~\cite{Chang:2017eiq, Savage:2016kon}. From
 our previous experience with spinning particles~\cite{Briceno:2014oea, Briceno:2015tza}, we expect that the extension will be relatively straightforward. 

Looking to less trivial extensions, it would be of great interest to extended these ideas in order to develop an approach for extracting non-local matrix elements of two-particle systems. This would make it possible to extract distribution functions of resonant states, following the methods of Refs.~\cite{Ji:2013dva}, and would open the possibility for lattice QCD calculations of two-body contributions to double-beta decays~\cite{Tiburzi:2017iux, Nicholson:2018mwc}. Matrix elements of non-local operators suffer from different types of finite-volume artifacts. These depend crucially on whether the operators are displaced in Euclidean time, as in Ref.~\cite{Christ:2015pwa}, or in a spatial direction, as considered for example in Ref.~\cite{Briceno:2018lfj}. %
Finally, extensions of this work to energies for which three or more particles can go on-shell should be feasible in the future, especially given the recent progress in understanding the finite-volume spectrum of three-particle states~\cite{Briceno:2012rv, Hansen:2014eka, Mai:2017bge, Hammer:2017kms, Briceno:2017tce,  Briceno:2018aml}. 

Returning to the present formalism, several open questions persist that we plan to address in future work. For example, $\Wdf$ is defined using the partial-wave basis, while the form factors of resonances or bound states are more naturally described using Lorentz decomposition. It is always possible to relate the partial-wave and Lorentz bases. However, due to the reduction of rotational symmetry in the finite-volume, we do not expect a one-to-one correspondence between the finite-volume matrix elements and the different Lorentz components of $\Wdf$. For example, in the $\rho$ channel with nonzero spatial momenta in the finite-volume frame, the different helicity components mix to different finite-volume irreps. This means that the components are sampled by different finite-volume quantization conditions and thus at different energies. As a result, just as is done in the analysis of coupled-channel scattering~\cite{Wilson:2014cna, Dudek:2016cru, Moir:2016srx, Woss:2018irj, Dudek:2014qha, Briceno:2017qmb}, it will be necessary to perform global fits of the matrix elements using Eq.~(\ref{eq:2to2}). This requires a detailed understanding of the analytic structure of these amplitudes in which triangle singularities play a crucial role.

\section{Acknowledgements}

RAB acknowledges support from U.S. Department of Energy contract DE-AC05-06OR23177, under which Jefferson Science Associates, LLC, manages and operates Jefferson Lab, and the U.S. Department of Energy Early Career award contract DE-SC0019229. AB acknowledges support from U.S. Department of Energy, Office of Science, Office of Nuclear Physics, under Award No. DE-SC0010300.
The authors would like to thank A. Jackura, as well as J. Dudek, R. Edwards, D. Wilson, and the rest of the Hadron Spectrum Collaboration for useful discussions.

\appendix

\section{Derivation of the covariant formalism\label{sec:newder}}

\begin{figure}
\begin{center}
\includegraphics[width=.8\textwidth]{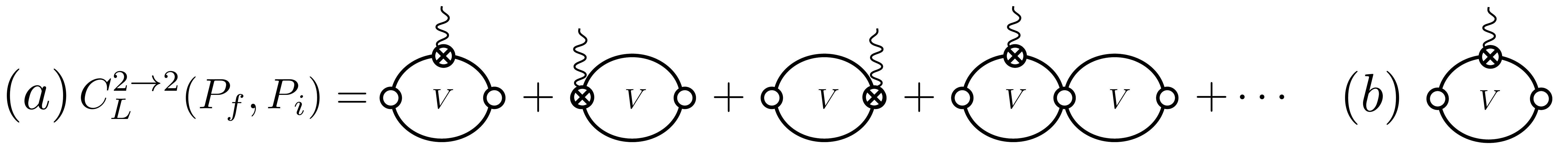}
\caption{
 (a)
 The finite-volume three-point correlator used to derive the $\textbf 2 + \mathcal J \to \textbf 2$ formalism. All symbols common to Fig.~\ref{fig:iW_Gfunc} have the same definition. The open circles, new to this figure, denote Bethe-Salpeter kernels, defined to include all diagrams besides the $s$-channel two-particle reducible set, shown explicitly. The two-particle loops shown explicitly are evaluated in a finite volume as indicated with the $V$ label.
 (b) The finite-volume loop within the correlator that leads to the appearance of the $G$-function. 
 }
\label{fig:C3pt}
\end{center}
\end{figure}

In this appendix we describe how the derivation of Ref.~\cite{\origTwoTwo} must be modified to give the covariant version of the formalism, presented in Sec.~\ref{sec:cov2to2} above. As in our previous derivation, we restrict attention to a finite cubic spatial volume, implemented by requiring all fields to have periodicity $L$ in the three spatial directions. 

Within this set-up we introduce a finite-volume three-point function, $C_L^{2 \to 2}(P_f,P_i)$, defined as the sum over all possible diagrams connecting the initial and final states to the inserted current. See also Fig.~\ref{fig:C3pt}(a). For $E_i^\star, E_f^\star $ below the next multi-hadron threshold (such that only a single two-particle channel can propagate), all volume effects scaling as a power of $1/L$ are captured by the skeleton expansion shown in the figure. Here the label $V$ within the loops stands for volume and denotes that the diagrams are defined with the spatial momenta summed over the discrete set allowed by the periodic boundary conditions, $\textbf k = 2 \pi \textbf n/L$ with $\textbf n$ a 3-vector of integers. The corresponding diagrams in an infinite volume are given by replacing these sums with integrals and are represented by a loop with no label.

The power-like volume effects of $C_L^{2 \to 2}(P_f,P_i)$ are encoded in the skeleton expansion of Fig.~\ref{fig:C3pt}(a), built from fully dressed hadron propagators (indicated by the simple black lines) and two-particle Bethe-Salpeter kernels (indicated by open circles). The vertices with a current insertion are given by the same diagrammatic set defining propagators and kernels, but with the current attached at all possible locations. In the kinematic window of interest, the difference between the finite- and infinite-volume definitions of propagators and kernels as $e^{- m L}$.

\bigskip

The set-up here is identical to that of Ref.~\cite{\origTwoTwo}. Indeed the only modifications required are in the evaluation of the two-particle loop, in which the current couples to one of the two particles. The relevant diagram is shown in Fig.~\ref{fig:C3pt}(b). To explain how the analysis is altered, we begin by recalling the finite-volume reside of this diagram, given in Eq.~(27) of Ref.~\cite{\origTwoTwo}
\begin{align}
\mathcal G_{L,\mu_1 \cdots \mu_n} & \equiv  
  \bigg [\frac{1}{L^3}\sum_{\mathbf{k}} - \int \! \! \frac{d^3 \textbf k}{(2\pi)^3}\, \bigg]
\frac{1}{2\omega_{k2}}   i\mathcal L(P_f, k)  \,  \Delta(P_f - k) \, \w^{\text{off}}_{\mu_1 \cdots \mu_n} (P_f-k,P_i-k) \, \Delta(P_i - k) \, i\mathcal R^\dagger(P_i, k) \bigg \vert_{k^0=\omega_{k2}} \,.
\label{eq:GL1}
\end{align}

Here $\mathcal L(P_f, k) $ and $ \mathcal R^\dagger(P_f, k)$ are generic endcap functions to be replaced with Bethe-Salpeter kernels or the overlap to the interpolators in the final derivation. 
In contrast to Ref.~\cite{\origTwoTwo}, we define $\mathcal G_{L,\mu_1 \cdots \mu_n}$ with these endcaps accompanied by factors of $i$. The difference arises because we formulate the derivation here with Minkowski momenta (in contrast to the Euclidean conventions of the previous publication). In this set-up the $i$ factors give a more natural extension to the Bethe-Salpeter kernels, multiplied by this factor due to the weight, $e^{iS}$, in the quantum path integral. Our second notational modification is to represent the $\textbf 1 + \mathcal J \to \textbf 1$ insertion with a set of Lorentz indices, and to make explicit that the quantity is off-shell, i.e. that $(P_f - k)^2, (P_i - k)^2 \neq m_1^2$. [See also Eq.~(\ref{eq:w_off}) above.] Our third and final alteration is to restrict attention to a single channel, thus removing the $a$ and $b$ indices from Eq.~(27) or Ref.~\cite{\origTwoTwo}.

We now express $\mathcal L(P_f, k) $, $ \mathcal R^\dagger(P_f, k)$ and $\w^{\text{off}}_{\mu_1 \cdots \mu_n} (P_f-k,P_i-k)$ in terms of their on-shell counterparts, plus corrections. For the endcap functions this is done exactly as in Ref.~\cite{\origTwoTwo}, by first defining
\begin{align}
\label{eq:onshellcomp}
\mathcal L(P_f,  q^\star_f  \hat {\textbf{k}}^\star_f)=\sum_{\ell_f m_f }\sqrt{4 \pi} Y_{\ell_f m_f }(\hat {\textbf k}^\star_f)\mathcal L_{\ell_f m_f }(P_f),
\ \ \ \ \ 
\mathcal R^\dagger(P_i,  q^\star_i  \hat {\textbf{k}}^\star_i)
=\sum_{\ell_i m_i } \sqrt{4 \pi} Y^*_{\ell_i m_i }(\hat {\textbf k}^\star_i)\mathcal R^\dagger_{\ell_i m_i}(P_i) \,,
\end{align}
and then recombining the components with the $\mathcal{Y}_{\ell m}$ harmonics defined in Eq.~(\ref{eq:Ylm})
\begin{equation}
\mathcal L_{\mathrm{on}}(P_f, \textbf k^\star_f ) \equiv \sum_{\ell_f m_f }\mathcal{Y}_{\ell_f m_f }( {\textbf k}^\star_f)\mathcal L_{\ell_f m_f }(P_f) \,,
\ \ \ \ \ 
\mathcal R_{\mathrm{on}}^\dagger(P_i, \textbf k^\star_i )
\equiv \sum_{\ell_i m_i }  \mathcal{Y}^*_{\ell_i m_i }( {\textbf k}^\star_i)\mathcal R^\dagger_{\ell m}(P_i) \,.
\label{eq:finalonfunc}
\end{equation}
With these in hand we introduce the $\delta$ operator as follows
\begin{align}
\label{eq:Ltodelta}
\mathcal L(P_f,k)\bigg \vert_{k^0=\omega_{k2}} & = \mathcal L_{\mathrm{on}}(P_f, \textbf k^\star_f ) + [\mathcal L  \delta](P_f, k) \,, \\
\mathcal R^\dagger(P_i,k) \bigg \vert_{k^0=\omega_{k2}}  & = \mathcal R^\dagger_{\mathrm{on}}(P_i, \textbf k^\star_i ) + [\delta \mathcal R^\dagger](P_i,k) \,.
\label{eq:Rtodelta}
\end{align}
These results match Eqs.~(41) and (42) of Ref.~\cite{\origTwoTwo}, up to the minor notational differences discussed above. A key property that we will use below is that $[\mathcal L  \delta](P_f, k)$ vanishes like $(P_f - k)^2 - m^2_1$ in the on-shell limit, and $[\delta \mathcal R^\dagger](P_i,k)$ vanishes like $(P_i - k)^2 - m^2_1$.
  
We now imitate this separation with the $\textbf 1 + \mathcal J \to \textbf 1$ insertion, and in doing so introduce the first major difference as compared to our earlier work. Beginning with Eq.~(\ref{eq:w_off}) of the main text, above, we introduce the shorthand $k_f \equiv P_f-k$ and $k_i \equiv P_i-k$ to write
\begin{align}
\w^{\text{off}}_{\mu_1 \cdots \mu_n} (k_f,k_i) & = \sum_{j}\,\textbf{K}^{(j)}_{\mu_1 \cdots \mu_n}(k , P_f,P_i)\, f^{(j)}(Q^2, k_f^2, k_i^2) \,.
\end{align}
We now follow the approach introduced in the main text by only projecting the scalar form factors, $f^{(j)}$, to their on-shell values. To understand the idea note that
\begin{equation}
\label{eq:nokhat}
f^{(j)}(Q^2, k_f^2, k_i^2)=  f^{(j)} \Big [ Q^2; \  P_f^2 + k^2 - 2 E^\star_f \omega^{\star f}_{k2}     ; \  P_i^2 + k^2 - 2 E^\star_i \omega^{\star i}_{k2}   \Big   ]  \,,
\end{equation}
where $\omega^{\star f}_{k2} $ is the temporal component of $k^{\mu \star f}$, given by boosting $k^\mu = (\omega_{k2}, \textbf k)$ to the $\textbf P_f = \textbf 0$ frame, and $\omega^{\star i}_{k2} $ is the $f \to i $ analog. 

The key point is that, when expressed in this way, $f^{(j)}$ has no dependence on $\hat {\textbf k}_f^\star$ or $\hat {\textbf k}_i^\star$. Thus there is no need to decompose in harmonics, nor to include barrier factors. The on-shell projection is simply 
\begin{align}
 f^{(j)}(Q^2 ) & \equiv  f^{(j)}(Q^2, m_1^2, m_1^2)    = f^{(j)} \Big [ \cdots  \Big   ] \bigg \vert_{k^\star_f = q^\star_f, k^\star_i = q^\star_i} \,, \\
  f^{(j)}_{\text{off},\text{on}} ( Q^2 , k_f^2 ) & \equiv f^{(j)}(Q^2,k_f^2, m_1^2) =  f^{(j)} \Big [ \cdots  \Big   ] \bigg \vert_{  k^\star_i = q^\star_i} \,, \\
   f^{(j)}_{\text{on},\text{off}} ( Q^2 , k_i^2 ) & \equiv f^{(j)}(Q^2,m_1^2,k_i^2) =  f^{(j)} \Big [ \cdots  \Big   ] \bigg \vert_{k^\star_f = q^\star_f}  \,.
\end{align}
To avoid clutter we have suppressed the arguments on the right-hand side, identical to those of Eq.~(\ref{eq:nokhat}). We stress here that the on-shell projections are subtle in that $\textbf k$ is used to define two separate variables $k^\star_f$ and $k^\star_i$. The separation is unambiguously given by whether the original $\textbf k$ appears in $k_f = P_f-k$ or $k_i = P_i-k$. With the induced $k_f^\star$ and $k_i^\star$ dependence, it is possible to separately project the initial and final states on shell via $k^\star_i \to q^\star_i$ and $k^\star_f \to q^\star_f$ respectively.

We next form linear combinations of the on- and partially-off-shell form factors in direct analogy to Eq.~(37) of Ref.~\cite{\origTwoTwo} to reach
\begin{align}
\label{eq:ftodelta}
f^{(j)}(Q^2,k_f^2,k_i^2)
=f^{(j)}(Q^2)
+
\delta f^{(j)}(Q^2)
+
 f^{(j)}(Q^2)\delta
+
\delta f^{(j)}(Q^2)\delta \,.
\end{align}
The individual terms on the right-hand side are defined by analogy to Eqs.~(38)-(40) of Ref.~\cite{\origTwoTwo} and have the explicit form
\begin{align}
\delta f^{(j)}(Q^2)
&\equiv  f^{(j)}_{\text{off},\text{on}} ( Q^2 , k_f^2 ) - f^{(j)}(Q^2) \,,  \\
 f^{(j)}(Q^2)\delta
&\equiv     f^{(j)}_{\text{on},\text{off}}( Q^2 , k_i^2 )  - f^{(j)}(Q^2) \,,  \\
\delta f^{(j)}(Q^2)\delta
& \equiv 
f^{(j)}(Q^2, k_f^2,k_i^2) + f^{(j)}(Q^2)
-  f^{(j)}_{\text{off},\text{on}} ( Q^2 , k_f^2 )  
-  f^{(j)}_{\text{on},\text{off}}( Q^2 , k_i^2 )  - f^{(j)}(Q^2) \,.
\end{align}
Note that functions with a $\delta$ on the left side (right side) also depend on $k_f^2$ ($k_i^2$), but we keep this dependence implicit to avoid clutter. As in Ref.~\cite{\origTwoTwo}, the key point here is that $\delta$ appearing on either side of the function indicates that the latter vanishes in the on-shell limit, scaling as $k_f^2 - m_1^2$ (for $\delta$ on the left) and as $k_i^2 - m_1^2$ (for delta on the right). 
 
 It now remains only to separate the propagators by defining
 \begin{equation}
 \label{eq:propsep}
 \Delta(k^2) \equiv \mathcal D(k^2) + \mathcal S(k^2) \,,
 \end{equation}
 where
 \begin{equation}
  \mathcal D(k^2) \equiv \frac{i}{k^2 - m_1^2 + i \epsilon} \,,
 \end{equation}
 is the non-interacting covariant propagator. Our conventions are such that $ \Delta(k^2)$ has unit residue at the single particle pole, implying $\mathcal S(k^2)$ is smooth and finite near $k^2 = m_1^2$. This form of separation, in which $\mathcal D(k^2)$ remains Lorentz invariant, is the second key distinction relative to our earlier formalism.
 
At this stage we have separated the endcaps [Eqs.~(\ref{eq:Ltodelta}), (\ref{eq:Rtodelta})] the single-particle form factor [Eq.~(\ref{eq:ftodelta})] and the covariant propagator [Eq.~(\ref{eq:propsep})] into on-shell terms plus corrections. Substituting these four identities into Eq.~(\ref{eq:GL1}) then gives the analog of Eq.~(43) of Ref.~\cite{\origTwoTwo}. The final step is to rearrange terms according to singularity structure. Of the 64 terms (reach by multiplying 4 binomials as well as the form-term separation of $f^{(j)}$) all but 17 are smooth at both poles, and thus give only exponentially suppressed contributions to the sum-integral difference. Those that are singular break into three classes, 8 with only the $1/(k_f^2 + m^2 - i \epsilon)$ singularity, 8 with $1/(k_i^2 + m^2 - i \epsilon)$, and a single maximally singular term that leads to the appearance of $G^\sigma\!(P_f,P_i,L)$.

To give an explicit form for the single-pole terms we need to introduce one final piece of notation. We define the operator $\delta_{\text{df}}$ via
\begin{multline}
\Big [ \big [\mathcal L(P_f, k) \Delta(k_f^2) \w_{\mu_1 \cdots \mu_n}(k_f, k_i) \big ] \delta_{\mathrm{df}} \Big ] \equiv
\sum_{j}
\textbf{K}^{(j)}_{\mu_1 \cdots \mu_n}(m,\textbf k,P_f,P_i) \\ \times
 \bigg [\mathcal L(P_f,k)     \Delta(k_f^2)  f^{(j)}( Q^2 , k_f^2,k_i^2 ) 
  -   \mathcal L_{\mathrm{on}}(P_f, \textbf k^\star_f ) \mathcal D (k_f^2)  f^{(j)}_{\text{on},\text{off}}( Q^2 ,k_i^2 ) \bigg] \,,
 \end{multline}
 and similarly for $\delta_{\text{df}}$ acting on the left.
 
 We can make use of this to rewrite Eq.~(\ref{eq:GL1}) as 
 \begin{multline}
 \mathcal G_{L,\mu_1 \cdots \mu_n}   = 
 -\sum_{j}
 f^{(j)}(Q^2) 
 i \mathcal L_{\ell_f m_f } 
 \bigg (  \bigg [\frac{1}{L^3}\sum_{\mathbf{k}} - \int \! \! \frac{d^3 \textbf k}{(2\pi)^3}  \bigg]      
\mathcal Y_{\ell_f m_f}(\textbf k^\star_{f}) 
D(m,\textbf{k})    
 \textbf{K}^{(j)}_{\mu_1 \cdots \mu_n}(m,\textbf k,P_f,P_i) 
\mathcal Y^*_{\ell_i m_i}(\textbf k^\star_{i})    \bigg )
  i\mathcal R^\dagger_{\ell_i m_i }  
\\
 +     \Big [ \big [i \mathcal L(P_f, k) \Delta(k_f^2) \w_{\mu_1 \cdots \mu_n}(k_f, k_i)  {\xi}^{-1}\big ] \delta_{\mathrm{df}} \Big ]    
   iF(P_i,L)    i \mathcal R^\dagger 
\\ +
i \mathcal L
   iF(P_f,L)   
 \Big [ \delta_{\mathrm{df}} \big [  {\xi}^{-1} \w_{\mu_1 \cdots \mu_n}(k_f, k_i) \Delta(k_i^2) i \mathcal R^\dagger(P_i, k)  \big ]  \Big ]  \,,
\end{multline}
where $D(m,\textbf{k})$ is defined in Eq.~(\ref{eq:D_def}).
On the second line the square bracketed quantities have been decomposed in spherical harmonics and carry implicit indices that are contracted with $F(P,L)$, defined in Eq.~(\ref{eq:Fscdef}). We have included the inverted symmetry factor, $\xi^{-1} = 2$ for identical particles, to compensate the factor in the definition of $F$. Note that this arises naturally from the fact that the current couples to each of the two particles when these are identical. This implies that $\mathcal W_{\text{df}}$ is defined with four subtraction terms, given by coupling the current to each of the four external propagators. When $\mathcal W_{\text{df}}$ is then projected to definite angular momenta, these terms become pairwise redundant leading to the factors of $2$. In the case of non-identical particles that both couple one must sum over the two choices of species mass within $\Delta(k_f^2)$ as well as the alternative mass assignments within $D(m, \textbf k)$, i.e.~$\{m_1, m_2\} \to \{m_2, m_1\} $.

Using Eq.~(\ref{eq:KtoCKomega}), we can write the first term on the right-hand side in terms of $G_{\mu_1 \cdots \mu_n; \ell_f m_{f} ;     \ell_i m_{i}}(P_f,P_i,L)$, defined in Eq.~(\ref{eq:Gmat_gen}). Following the steps outlined in Ref.~\cite{Briceno:2015tza}, one finally arrives at Eq.~(\ref{eq:2to2}) -- the relationship between the finite-volume matrix element and $i\W_{{\rm df};\mu_1 \cdots \mu_n} $, defined in Eq.~(\ref{eq:Wdf}). The latter emerges through the identity $\delta_{\text{df}} \mathcal W \delta_{\text{df}} \equiv \mathcal W_{\text{df}}$.

{

We close by giving explicit expressions for the case that the current has a non-negligible coupling to both of the two particles. When two distinct species, $1 \neq 2$, each admit a $\textbf 1 + \mathcal J \rightarrow \textbf 1$ transition, then the definition of $\W_{{\rm df}}$, Eq.~(\ref{eq:Wdf}), is replaced by an expression with four subtractions
\begin{multline}
\label{eq:WdfBothCouple}
\W_{{\rm df};\mu_1 \cdots \mu_n} \equiv \W_{\mu_1 \cdots \mu_n} - 
i  \overline{\M}(P_f,k',k) \frac{i}{(P_f-k)^2 - m_1^2} \w_{1;\mu_1 \cdots \mu_n}
-
 \w_{1;\mu_1 \cdots \mu_n}
\frac{i}{(P_i - k')^2 - m_1^2}
i \overline{\M'}(P_i, k',k)  \\
- 
i  \overline{\M}(P_f,k',P_f - P_i + k) \frac{i}{(P_f-P_i+k)^2 - m_2^2} \w_{2; \mu_1 \cdots \mu_n}
-
 \w_{2; \mu_1 \cdots \mu_n}
\frac{i}{  (P_i - P_f + k')^2 - m_2^2}
i \overline{\M'}(P_i, P_i - P_f + k',k)  \,.
\end{multline}
Here the $ \w_{2; \mu_1 \cdots \mu_n}$ in the bottom line have arguments $(k', P_i - P_f + k') $ and $ (k', P_i - P_f + k') $ respectively. We have no freedom to choose these once the external momenta are fixed.  

The extra subtractions also lead to an additional $G$-dependent term in the relation between $ \Wtildf $   and $ \Wdf$. But in this expression, as a result of the sum, we do have the freedom to re-label the coordinates. With this in mind it is most convenient to rewrite $ \w_{2; \mu_1 \cdots \mu_n}$ as a function of $(P_f - k, P_i -  k)$ and decompose exactly as in the main text
\begin{align}
\w_{2;\mu_1 \cdots \mu_n} (P_f - k, P_i -  k) & \equiv \sum_{j}\,\textbf{K}^{(j)}_{2;\mu_1 \cdots \mu_n}(m,\textbf k,P_f,P_i) \bigg  |_{k^0=\omega_{k1}} \,f_2^{(j)}(Q^2) \,, \\
\textbf{K}^{(j)}_{2;\mu_1 \cdots \mu_n}(m,\textbf k,P_f,P_i) & \equiv \sum_{n'=0}^{n}   K^{\omega}_{2; \mu_1\cdots \mu_{n'}}(m,\textbf k) \   C^{(j)}_{2;\mu_{n'+1}\ldots \mu_{n}}(P_f,P_i) \,.
\end{align}
This then allows us to write 
\begin{multline}
\label{eq:WtiltoWdfAPP}
 \Wtildf^{\mu_1 \cdots \mu_n}(P_f,P_i,L)  - \Wdf^{\mu_1 \cdots \mu_n}(s_i,s_f,Q^2) \equiv   \\
  \sum_j
  \sum_{n'=0}^{n} 
  C_1^{(j), \mu_{n'+1}\ldots \mu_{n}}(P_f,P_i) f_1^{(j)}(Q^2)
\big [ \mathcal  M(s_f)  G_{12}^{\mu_1 \cdots \mu_{n'}}(P_f,P_i,L)  \mathcal M(s_i) \big ] \\
 +
 \sum_j
  \sum_{n'=0}^{n} 
  C_2^{(j), \mu_{n'+1}\ldots \mu_{n}}(P_f,P_i) f_2^{(j)}(Q^2)
\big [ \mathcal  M(s_f)  G_{21}^{\mu_1 \cdots \mu_{n'}}(P_f,P_i,L)  \mathcal M(s_i) \big]
   \,,
 \end{multline} 
 where $G_{12}$ is exactly equal to the quantity defined in Eqs.~(\ref{eq:Gmat_gen}) and (\ref{eq:D_def}) of the main text and $G_{21}$ is the same but with $m_1 \leftrightarrow m_2$ everywhere.
 
Turning to the case of identical particles here the relation between between $\W_{{\rm df}}$  and $ \W$ again has four subtractions, exactly as in Eq.~(\ref{eq:WdfBothCouple}). The four terms continue to be distinct due to the four different momentum assignments. However, the relation between $ \Wtildf $   and $ \Wdf$ is exactly as Eq.~($\ref{eq:WtiltoWdf}$) of the main text, a single $G$-dependent term with no symmetry factors. This follows from the fact that identical particles lead to a unique diagram of the form shown in Fig.~\ref{fig:C3pt}(b), and that this has no symmetry factor, even in the case that the three hadrons in the loop are identical.

}

\section{Details for evaluating the finite-volume functions \label{sec:appFG} }

In the following subsections we collect various details relevant for the evaluation of the two finite-volume functions that enter our formalism $F_{  \ell m ;   \ell' m'}(P,L) $ and $G^\sigma\!(P_f,P_i,L)$.

\subsection{Index gymnastics \label{sec:kin}}

We first discuss various identities for rearranging spherical-harmonic and Lorentz indices in the evaluation of $G_\sigma (P_f,P_i,L)$. Begin with the case of $P_i=P_f$, in particular with Eq.~(\ref{eq:decom}) of the main text. Multiplying both sides by $(q^\star )^{\ell_i + \ell_f}$ and substituting the definition of $\mathcal Y_{\ell m}(\textbf k^\star)$, this reduces to 
\begin{equation}
\label{eq:decomapp}
4 \pi (k^\star)^{\ell_i + \ell_f}  Y_{\ell_f m_f}(\hat {\textbf k}^\star) 
  k_{\mu_1} \cdots  k_{\mu_n}   Y^*_{\ell_i m_i}(\hat {\textbf k}^\star)    \bigg \vert_{k^0 = \omega_{k2}} 
  =  \sqrt{4 \pi} 
  \sum_{J M}  \mathcal{C}_{\sigma, JM}(\pmb \beta , k^{\star}) \,   k^{\star J} {Y}_{J M}(\hat {\textbf k}^\star) \,.
  \end{equation}
The first aim of this subsection is to use this result to derive a useful expression for evaluating $\mathcal{C}_{\sigma, JM}(\pmb \beta , k^{\star}) $.
  
Substituting the relation $k_\mu = {[\Lambda_{-\pmb \beta}]_\mu}^\nu  k^\star_\nu$, 
we reach
\begin{equation}
\label{eq:decomapp2}
{[\Lambda_{-\pmb \beta}]_{\mu_1}}^{\nu_1}
 \cdots 
 {[\Lambda_{-\pmb \beta}]_{\mu_n}}^{\nu_n}
   \Big [ 4 \pi (k^\star)^{\ell_i + \ell_f}   Y_{\ell_f m_f}(\hat {\textbf k}^\star) 
  k^\star_{\nu_1} \cdots  k_{\nu_n}^\star   Y^*_{\ell_i m_i}(\hat {\textbf k}^\star)  \big ]   \bigg \vert_{k^0 = \omega_{k2}} 
  =  \sqrt{4 \pi} 
  \sum_{J M}  \mathcal{C}_{\sigma, JM}(\pmb \beta , k^{\star}) \,   k^{\star J} {Y}_{J M}(\hat {\textbf k}^\star) \,.
  \end{equation}
At this stage, the factor in square brackets is a simple polynomial in the coordinates $ k_x^\star$, $ k_y^\star$ and $k_z^\star$ with additional dependence on the magnitude entering through $\omega_{k2}^\star$. Thus, for a given set of indices, one can readily determine an explicit expression, and then use the orthogonality of the $Y_{\ell m}$s, to deduce
\begin{equation}
\label{eq:decomapp3}
\mathcal{C}_{\sigma, JM}(\pmb \beta , k^{\star}) = 
{[\Lambda_{-\pmb \beta}]_{\mu_1}}^{\nu_1}
 \cdots 
  {[\Lambda_{-\pmb \beta}]_{\mu_n}}^{\nu_n}
 \  \sqrt{4 \pi} (k^\star)^{\ell_i + \ell_f - J}  \int d \Omega^\star {Y}^*_{J M}(\hat {\textbf k}^\star)  \Big [ Y_{\ell_f m_f}(\hat {\textbf k}^\star) 
  k^\star_{\nu_1} \cdots  k_{\nu_n}^\star  Y^*_{\ell_i m_i}(\hat {\textbf k}^\star)      \Big ] \bigg \vert_{k^0 = \omega_{k2}} \,.
\end{equation}

As a specific example, we return to the case of $\sigma = [\mu=z; 10; 10]$ and $\textbf P = [00d_z]$, already discussed in the main text.  For these indices, Eq.~(\ref{eq:decomapp3}) reduces to
\begin{equation}
\label{eq:decomapp3spec}
\mathcal{C}^\sigma_{JM}(\pmb \beta , k^{\star}) =    \frac{3}{ \sqrt{4 \pi}} (k^\star)^{2 - J} 2 \pi \delta_{M,0} \int_{-1}^1 d \cos \theta^\star \  {Y}^*_{J 0}(\theta^\star)  \Big [ \cos^2\! \theta^\star \big (
 \beta^z \gamma \,  \omega_{k2}^\star  + \gamma \,  k^\star \cos \theta^\star \big )    \Big ]  \,.
\end{equation}
Here we have used the fact that, with $\mu$ fixed to $z$, we only need to sum over one row of the boost matrix, $
{[\Lambda_{-\pmb \beta}]^{z}}_{\nu}
= (\beta^z \gamma, 0, 0, \gamma)$.  The integrals are now trivial to evaluate. For example, the $JM=00$ component reduces to
\begin{equation}
\mathcal{C}^\sigma_{00}(\pmb \beta , k^{\star}) = \beta^z \gamma  \omega_{k2}^\star  \frac{3}{  4 \pi } (k^\star)^{2 } 2 \pi   \int_{-1}^1 d \cos \theta^\star \    \cos^2 \theta^\star 
 = k^{\star2} \omega_{k2}^\star  \frac{P^z}{E^*}  \,,
\end{equation}
where we have used $\beta^z = P^z/E$ and $\gamma = E/E^\star$. This matches Eq.~(\ref{eq:Cex00}) of the main text.

 An alternative approach for determining $\mathcal{C}_{\sigma, JM}(\pmb \beta , k^{\star}) $ is to note that the product of two spherical harmonics can be written in terms of Clebsch-Gordon coefficients
 \begin{align}
{Y}_{\ell_1 m_1}(\textbf k^\star)\,
{Y}_{\ell_2 m_2}(\textbf k^\star)
&=
\sum_{\ell_3 m_3}
\sqrt{\frac{(2\ell_1+1)(2\ell_2+1)}{(2\ell_3+1)\,4\pi}}
\langle \ell_10\ell_20|\ell_30\rangle
\langle \ell_1m_1\ell_2m_2|\ell_3m_3\rangle
{Y}_{\ell_3 m_3}(\textbf k^\star) \,,
\label{eq:2Y_1Y}
\end{align}
and that each factor of $k^\star_\nu$ can be rewritten using the identity
    \begin{align}
k^{ \star \nu}    & = \sqrt{4\pi}
\left(\omega^\star_{k2}Y_{00}(\hat {\textbf k}^\star),
k^\star \frac{Y_{1-1}(\hat {\textbf k}^\star)-Y_{11}(\hat {\textbf k}^\star)}{\sqrt{6}},
k^\star \frac{ Y_{1-1}(\hat {\textbf k}^\star)+  Y_{11}(\hat {\textbf k}^\star)}{-i\sqrt{6}},
k^\star \frac{Y_{10}(\hat {\textbf k}^\star)}{\sqrt{3}}
\right)
\equiv 
\sum_{\ell m<2} {T}^{\nu}_{\ell m}(k^{\star}) \, Y_{\ell m}(\hat {\textbf k}^\star) \,,
\end{align}
where the last equality defines ${T}^{\nu}_{\ell m}(k^{\star})$.
Substituting this into the left-hand side of Eq.~(\ref{eq:decomapp2}) leads to sums over products of spherical harmonics on that side of the equation. These can be pairwise combined using Clebsch-Gordon coefficients, until the left-hand side is reduced to a sum over a single harmonic. Then one can use the orthogonality of spherical harmonics to match this, term by term, to the right-hand side and thereby determine the values of $\mathcal{C}_{\sigma, JM}(\pmb \beta, k^\star)$.

\bigskip

We next consider the case of $P_i \neq P_f$. As explained in the main text, here we find it more useful to convert indices in the other direction, i.e.~to trade all dependence on spherical harmonics for additional momentum 4-vectors. The key distinction between this case and that discussed above is that we no longer have a natural c.m.~frame. The rest frames of $P_i$ and $P_f$ differ, and expressing the integrand in either frame leads to ugly expressions. 
This, together with the need to reach covariant expressions that we can evaluate semi-analytically, led us to introduce $M_\sigma^{\nu_1 \cdots \nu_N}$, in Eq.~(\ref{eq:Gmat_PfnPi}) above. The definition can be re-expressed as
\begin{align}
\label{eq:Mdef2}
M^{\nu_1 \cdots \nu_N}_{ [\mu_1 \cdots \mu_n;  \ell_f m_{f} ;     \ell_i m_{i} ]}  k_{\nu_1}\cdots k_{\nu_N} & = { \frac{4 \pi}{(q^\star_{f})^{\ell_f}(q^\star_{i})^{\ell_i}} \big [ ( k^{\star}_{f}  )^{\ell_f}    Y_{\ell_fm_{ f}}(\hat {\textbf k}^\star_{f})} \big ]
      k_{\mu_1}\cdots k_{\mu_n}  \big [  ( k^{\star}_{i}  )^{\ell_i}     {   Y^*_{\ell_i m_{ i}}(\hat {\textbf k}_{i})} \big ] \,.
\end{align}
The second, and final aim of this subsection, is to derive a useful expression for $M_\sigma^{\nu_1 \cdots \nu_N}$.

First we introduce a new set of tensors, denoted $\mathcal T$, that allow us to express the right-hand side in terms of 4-vectors
\begin{align}
\label{eq:Mdef3}
    M^{\nu_1 \cdots \nu_N}_{ [\mu_1 \cdots \mu_n;  \ell_f m_{f} ;     \ell_i m_{i} ]}  k_{\nu_1}\cdots k_{\nu_N} & =  {  \mathcal T^{\alpha_1 \cdots \alpha_{\ell_f}}_{[\ell_f m_{f}  ]}  \mathcal T^{*\gamma_1 \cdots \gamma_{\ell_i}}_{[\ell_i m_{i}  ]}  \  k^{\star f}_{\alpha_1} \cdots  k^{\star f}_{\alpha_{\ell_f}}        k_{\mu_1}\cdots k_{\mu_n}    k^{\star i}_{\gamma_1} \cdots  k^{\star i}_{\alpha_{\ell_i}} } \,.
\end{align}
The exact definition of the  $\mathcal T^{\alpha_1 \cdots \alpha_{\ell_f}}_{[\ell m   ]}$ can be inferred by comparing Eqs.~(\ref{eq:Mdef2}) and (\ref{eq:Mdef3}).
 Note that they contain the $\sqrt{4 \pi}/(q^\star)^{\ell}$ prefactors and also encode the combinations of $\textbf k^\star$ components needed to form the various spherical harmonics. For example $\sqrt{4 \pi} k^\star Y^*_{10}(\hat {\textbf k}^\star)/q^\star = -\sqrt{3} k_{\mu=3}^\star/q^\star$ implies $\mathcal T^{\mu}_{[10]} = \sqrt{3} (0,0,0,-1)/q^\star$, since $k_\mu^\star =(\omega^\star_{k2}, -\textbf k^\star)$. The single-index tensor, $\mathcal T^{\alpha}_{[\ell  m   ]}$, is closely related to $T^\nu_{ \ell m}$, introduced above.  
 
 The final step 
  is to boost all 4-vectors to the finite-volume frame. We deduce  
\begin{multline}
M^{\nu_1 \cdots \nu_N}_{ [\mu_1 \cdots \mu_n;  \ell_f m_{f} ;     \ell_i m_{i} ]}   =     \mathcal T^{\alpha_1 \cdots \alpha_{\ell_f}}_{[\ell_f m_{f}  ]}  \mathcal T^{*\gamma_1 \cdots \gamma_{\ell_i}}_{[\ell_i m_{i}  ]} \\ \times   
 {[\Lambda_{\pmb \beta_f}]_{\alpha_1}}^{\nu_1}
  \cdots
  {[\Lambda_{\pmb \beta_f}]_{\alpha_{\ell_f}}}^{\nu_{\ell_f}}
 \times \delta_{\mu_1}^{\nu_{\ell_f+1}}  \cdots  \delta_{\mu_n}^{\nu_{\ell_f+n}} \times 
  {[\Lambda_{\pmb \beta_i}]_{\gamma_1}}^{\nu_{\ell_f+n+1}}
  \cdots
    {[\Lambda_{\pmb \beta_i}]_{\gamma_{\ell_i}}}^{\nu_{N}}
  \,.
\end{multline}

\subsection{Evaluating $F_{  \ell m ;   \ell' m'}(P,L)  $ \label{sec:Ffunc}}

We next turn to the finite-volume matrix, $F(P,L)$. For convenience we repeat the definition given in Eq.~(\ref{eq:Fscdef}) above 
\begin{align}
\label{eq:Fapp1}
F_{  \ell m ;   \ell' m'}(P,L)  
& \equiv
\xi 
  \bigg [\frac{1}{L^3}\sum_{\mathbf{k}} - \int \! \! \frac{d^3 \textbf k}{(2\pi)^3}\, \bigg]
 \frac{\mathcal Y_{\ell m}(\textbf k^\star )   \,   \mathcal Y^*_{\ell' m'}(\textbf k^\star)   
  }{2 \omega_{Pk1} 2 \omega_{k2}(E - \omega_{Pk1}-\omega_{k2}  + i \epsilon )} 
   \,.
\end{align}
Our aim here is to rewrite this in terms of $\mathcal Z^{(1)}$. With this in mind, we first observe
   \begin{align}
\label{eq:Fapp2}
 F_{  \ell m ;   \ell' m'}(P,L)   &=
\frac{\xi }{2  E^{\star}}   \sum_{JM} B_{JM}^{[\ell m; \ell' m']}     \frac{1}{(q^\star)^J}
  \bigg [\frac{1}{L^3}\sum_{\mathbf{k}} - \int \! \! \frac{d^3 \textbf k}{(2\pi)^3}\, \bigg] \bigg ( \frac{k^\star}{q^\star} \bigg )^{\!\!\ell + \ell' - J}
  \frac{2\omega_{k2}^\star}{2\omega_{k2}}
 \frac{  \sqrt{4 \pi}  (k^\star)^J Y_{JM}(\hat {\textbf k}^\star )   }{ q^{\star 2} - k^{\star 2} +i\epsilon} \,,
 \end{align}
 where we have used the fact that $2 \omega_{Pk1} (E - \omega_{Pk1}-\omega_{k2}  + i \epsilon )$ can be replaced with $2 E^\star (q^{\star 2} - k^{\star 2} +i\epsilon)/2 \omega_{k2}^\star$ up to smooth terms in the integrand that lead to exponentially suppressed volume dependence.   The definition of $ B_{JM}^{[\ell m; \ell' m']} $ can be inferred by comparing Eqs.~(\ref{eq:Fapp1}) and (\ref{eq:Fapp2}) and is given more explicitly by
 \begin{align}
 \label{eq:Bdef}
 B_{JM}^{[\ell m; \ell' m']}   
 &\equiv   \sqrt{4 \pi} \int   \! d \Omega^\star \,  Y^*_{JM}(\hat {\textbf k}^\star)   \big [ Y_{\ell m}(\hat {\textbf k}^\star)    \,     Y^*_{\ell' m'}(\hat {\textbf k}^\star) \big ]  \,,\nn\\
 &
 =
(-1)^{m'} \sqrt{\frac{(2\ell+1)(2\ell'+1)}{(2J+1)}}
\langle \ell 0\ell'0|J 0\rangle
\langle \ell m\ell -m'|JM\rangle,
 \end{align}
where we have used Eq.~(\ref{eq:2Y_1Y}) to reach the last equality. 
  
The final step here is to note that the additional barrier factor, $(k^\star/q^\star)^{\ell + \ell' - J}$, appearing in Eq.~(\ref{eq:Fapp2}), can in fact be set to 1. This is justified because $ (k^\star)^J Y_{JM}(\hat {\textbf k}^\star ) $ is an analytic function and because $\ell + \ell' - J \geq 0$ for all nonzero $B_{JM}^{[\ell m; \ell' m']} $. The latter point directly follows from the explicit expression of the $B_{JM}^{[\ell m; \ell' m']} $ in terms of the Clebsch-Gordon coefficients. As a result, the difference $(k^\star/q^\star)^{\ell + \ell' - J} - 1$ cancels the pole, leading to a smooth summand and a suppressed sum-integral difference.  

 Removing this extra factor, and re-expressing the sum and integral with dimensionless coordinates, we conclude
 \begin{align}
 \label{eq:Fapp3}
  F_{  \ell m ;   \ell' m'}(P,L)  &= 
\frac{\xi  }{  8 \pi^2 L E^{\star}}   \sum_{JM} B_{JM}^{[\ell m; \ell' m']} 
\frac{(2 \pi)^{J}}{ (q^\star L)^{J}}
\lim_{\alpha \to 0}  \mathcal{Z}^{(1)}_{J M}(P,L,\alpha) \,.
\end{align}

 \subsection{$\alpha$-dependence of  $\mathcal{Z}^{(n)}_{J M}$\label{sec:alpha_dep}}
Here we explain our choice of cut-off function used in the definition of $\mathcal{Z}^{(n)}_{J M}$, Eq.~(\ref{eq:Zdef}), recalled here for convenience
\begin{equation}
\label{eq:Zdef_app}
\mathcal{Z}^{(n)}_{J M}(P,L,\alpha) =    \bigg [ \sum_{\mathbf{r}}  \frac{\omega_{k2}^\star}{ \omega_{k2}} - \int \!   d^3 \textbf r^\star   \bigg]    
 \,  \frac{  \sqrt{4 \pi}  \, r^{\star J} {Y}_{J M}(\hat{\textbf r}^\star)     }{ (x^{2}-r^{\star2}+ i \epsilon)^n }  e^{- \alpha (r^{\star2} - x^{2})^n} \,.
\end{equation}
For $n \geq 2$ the integral and sum are individually convergent in the limit $\alpha \to 0$. Nonetheless, evaluating the sums for various non-zero $\alpha$ and extrapolating $\alpha \to 0$ turns out to be more efficient than saturating the $\alpha = 0$ expression directly.  

By including the power of $n$ in the cutoff function, $e^{- \alpha (r^{\star2} - x^{2})^n}$, we ensure that differentiating with respect to $\alpha$ gives a smooth summand. This, in turn, implies that $\partial_\alpha \mathcal Z^{(n)}$ vanishes up to terms that are exponentially suppressed in the volume
 \begin{align}
\frac{\partial      \mathcal{Z}^{(n)}_{J M}(P,L,\alpha)  }{\partial \alpha}
&=    
-    (-1)^n\bigg [ \sum_{\mathbf{r}}  \frac{\omega_{k2}^\star}{ \omega_{k2}} - \int \!   d^3 \textbf r^\star   \bigg]    
 \,  {  \sqrt{4 \pi}  \, r^{\star J} {Y}_{J M}(\hat{\textbf r}^\star)     }   \,,    \\
 &=    
-    (-1)^n
\left(\frac{L}{2\pi}\right)^J
\sum_{\mathbf{n}\neq0}
\int \!   \frac{d^3 \textbf k }{(2\pi)^3}
\frac{\omega_{k2}^\star}{ \omega_{k2}}
   e^{i L \textbf{n}\cdot \textbf{k} }
  {  \sqrt{4 \pi}    k^{\star J} {Y}_{J M}(\hat{\textbf k}^\star)     }  =   \mathcal{O}( e^{-mL}) \,.
\end{align}
Using this result in an expansion about $\alpha = 0$ then gives
 \begin{align}
\mathcal{Z}^{(n)}_{J M}(P,L,\alpha)  &=     \mathcal{Z}^{(n)}_{J M}(P,L,0) + \mathcal O(\alpha\,e^{-mL} ) \,.
\end{align}

Of course, it is possible to define the $\mathcal{Z}^{(n)}_{JM}$ functions with a milder cutoff function, for example
\begin{equation}
\widetilde{\mathcal{Z}}^{(n)}_{J M}(P,L,\alpha) \equiv    \bigg [ \sum_{\mathbf{r}}  \frac{\omega_{k2}^\star}{ \omega_{k2}} - \int \!   d^3 \textbf r^\star   \bigg]    
 \,  \frac{  \sqrt{4 \pi}  \, r^{\star J} {Y}_{J M}(\hat{\textbf r}^\star)     }{ (x^{2}-r^{\star2}+ i \epsilon)^n }  e^{- \alpha (r^{\star2} - x^{2})} \,.
 \label{eq:Ztilde}
\end{equation}
For $n=1$ this equivalent to ${\mathcal{Z}}^{(n)}_{J M}(P,L,\alpha)$, but for $n > 1$ it is a less useful prescription, due to an enhancement of the $\alpha$ corrections
 \begin{align}
\widetilde{\mathcal{Z}}^{(n)}_{J M}(P,L,\alpha) -{\mathcal{Z}}^{(n)}_{J M}(P,L,0) 
&=    
\alpha\,\bigg [ \sum_{\mathbf{r}}  \frac{\omega_{k2}^\star}{ \omega_{k2}} - \int \!   d^3 \textbf r^\star   \bigg]    
\frac{  \sqrt{4 \pi}  \, r^{\star J} {Y}_{J M}(\hat{\textbf r}^\star)     }{ (x^{2}-r^{\star2}+ i \epsilon)^{n-1} }
\,+\mathcal{O}(\alpha^2)\,,
 \nn\\
 &=    
\alpha
 \,{\mathcal{Z}}^{(n-1)}_{J M}(P,L,0) +\mathcal{O}(\alpha^2)\,.
 \label{eq:Ztilde_alphadep}
 \end{align}
As ${\mathcal{Z}}^{(n-1)}_{J M}(P,L,0)$ is itself a singular function for $n > 1$, we deduce that the difference between the optimal version, ${\mathcal{Z}}^{(n)}_{J M}(P,L,\alpha)$, and the alternative, $\widetilde{\mathcal{Z}}^{(n)}_{J M}(P,L,\alpha)$ can take on arbitrarily large values for any finite $\alpha$.

\begin{figure}
\begin{center}
\includegraphics[width=0.7\textwidth]{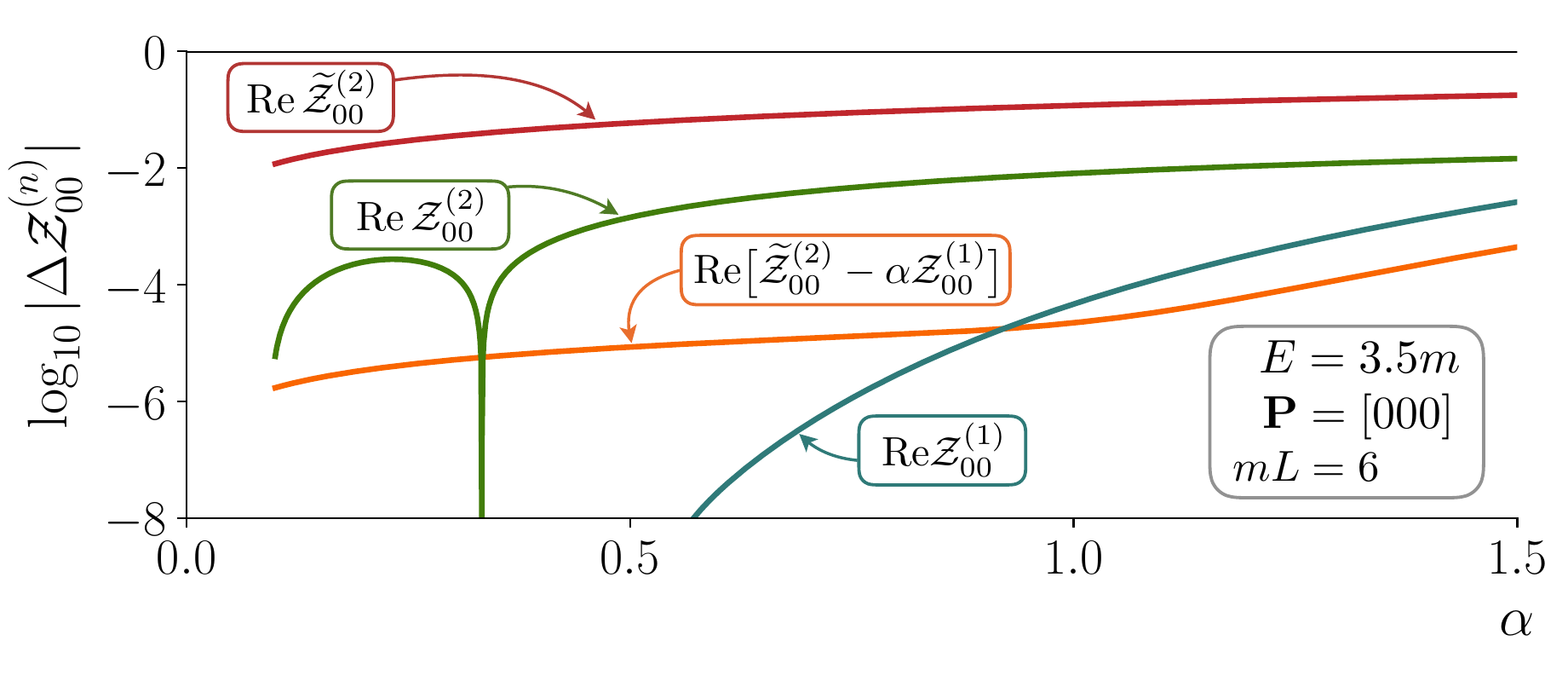}
\vspace{-10pt}
\caption{Dependence of $\mathcal{Z}^{(n)}_{00}$ on the cutoff parameter $\alpha$. Here we plot the log of the magnitude of the residual where $\Delta \mathcal Z(\alpha) \equiv [ \mathcal Z(\alpha) -  \mathcal Z(0)  ]/  \mathcal Z(0) $. Comparing the top two curves clearly shows that the cut-off function advocated in the main text, Eq.~(\ref{eq:Zdef}), has a milder $\alpha$ dependence than the alternative form, denoted $\widetilde{\mathcal{Z}}^{(2)}_{J M}$ and defined in Eq.~(\ref{eq:Ztilde}). The reason is that the latter has $\alpha$ depence with power-law $L$ scaling. In fact, the leading $\alpha$ behavior of $\widetilde{\mathcal{Z}}^{(2)}_{J M}$ is exactly given by ${\mathcal{Z}}^{(1)}_{J M}$ and subtracting this gives a highly improved result, as shown.
}
\label{fig:alpha_dep}
\end{center}
\end{figure}

 In Fig.~\ref{fig:alpha_dep} we compare the $\alpha$-dependence of ${\mathcal{Z}}^{(n)}_{J M}(P,L,\alpha) $ and $\widetilde{\mathcal{Z}}^{(n)}_{J M}(P,L,\alpha) $ for $n=1,2$, and show that the large $\alpha$-dependence of the latter is well described by Eq.~(\ref{eq:Ztilde_alphadep}). In the $\alpha \to 0$ limit the two prescriptions agree, but to optimize the numerical evaluation we advocate the form of Eq.~(\ref{eq:Zdef_app}) and use only this definition throughout the remainder of the text.

\subsection{Evaluating $\mathcal{Z}^{(n)}_{J M}(P,L,\alpha) $ \label{sec:Zcal}}
 
 In this section we discuss the evaluation of $\mathcal{Z}^{(n)}_{J M}(P,L,\alpha)$, in particular the integral part of this quantity. As already mentioned in Sec.~\ref{sec:GPfPi}, the integral entering $\mathcal Z^{(n)}$ vanishes for all entries besides $J M = 00$, implying
 \begin{align}
\mathcal{Z}^{(n)}_{J M}(P,L,\alpha) & =      \sum_{\mathbf{r}}  \frac{\omega_{k2}^\star}{ \omega_{k2}}  \frac{  \sqrt{4 \pi}    (r^{\star})^{ J} {Y}_{J M}(\hat{\textbf r}^\star)     }{ ( x^{2}-r^{\star2}+ i \epsilon)^n }  e^{- \alpha (r^{\star2} - x^{2})^n} -\delta_{J0} \delta_{M0} \Xi^{(n)}(x, \alpha)\,,
\end{align}
where
\begin{align}
 \Xi^{(n)}(x, \alpha) \equiv
4\pi \int^\infty_0  d r^\star r^{\star 2}    \frac{   e^{- \alpha ( r^{\star2}- x^2)^n   }}{ (x^{2}-r^{\star2}+ i \epsilon  )^n  } \,.
\end{align}
 
For the single-pole function, $\mathcal Z^{(1)}$, this can be evaluated analytically and takes the form
\begin{align}
 \Xi^{(1)}(x, \alpha) =
4\pi \left[ - \sqrt{\frac{\pi}{4 \alpha}}
e^{\alpha  x^2}
+
\frac{\pi  x}{2}
 {\rm Erfi}\left(\! \sqrt{\alpha  x^{ 2} }\right)
-i \frac{\pi x}{2} 
\right]\,,
\label{eq:Xi0}
\end{align}
 where ${\rm Erfi}(x)$ is the imaginary error function function, defined by $d {\rm Erfi}(z)/dz= 2  e^{z^2}/\sqrt{\pi}$ and ${\rm Erfi}(0) = 0$.

We now consider $n > 1$. Although we only require $n=2$ for the present work, we find it instructive to consider all values together. 
A consequence of the cutoff function, together with the higher pole factors, is that we are not able to evaluate  $ \Xi^{(n)}(x, \alpha) $ analytically for $n > 1$. Instead, following our usual trick, we separate the expression into two terms: one that can be evaluated analytically and another that is smooth and converges rapidly under numerical integration
\begin{align}
\Xi^{(n)}(x, \alpha)
&=
4\pi \int^\infty_0  d r^\star r^{\star 2}    \frac{   1   }{ (x^{2}-r^{\star2}+ i \epsilon    )^n}  
+
\delta \Xi^{(n)}(x, \alpha) \,, \\
&=
-i\frac{2\pi^2}{(n-1)!}\left(-\frac{\partial}{\partial x^2} \right)^{n-1} \sqrt{x^2}
+
\delta \Xi^{(n)}(x, \alpha) \,,
\label{eq:Xin}
\end{align}
where
\begin{equation}
\delta \Xi^{(n)}(x, \alpha) \equiv 4\pi \int^\infty_0  d r^\star r^{\star 2}    \frac{    e^{- \alpha ( r^{\star2}- x^2)^n   } - 1    }{ (x^{2}-r^{\star2}    )^n}      \,.
\end{equation}

We close with a final remark concerning the $n=2$ case, of direct relevance for $\textbf 2 + \mathcal J \to \textbf 2$ transition amplitudes. Here the relevant integral is
\begin{align}
\Xi^{(2)}(x, \alpha)
&=
\frac{i\pi^2}{x}
+
\delta \Xi^{(2)}(x, \alpha) \,.
\label{eq:Xi2}
\end{align}
Recalling $x \propto q^\star \propto \sqrt{s-s_{\rm th}}$, where $s = P^2$ is the c.m.~energy and $s_{\rm th} = (m_1 + m_2)^2$, we deduce that for $P_i = P_f$, $G_{\sigma}$ generically has an inverse square-root singularity at two-particle production threshold. This implies that $\mathcal W_{\text{df}}$, as well as $\mathcal W$, must have the same singularity. This behavior is visible in the values of $\mathcal{Z}_{JM}^{(2)}$ plotted in Fig.~\ref{fig:Zsummary}. In particular, we observe that $\mathcal{Z}_{JM}^{(1)}$ has milder behavior near threshold.

 \subsection{Symmetry contraints on $\mathcal{Z}^{(n)}_{J M}$\label{app:whenZisZero}}

\renewcommand{\arraystretch}{1.5}
\begin{table}
\begin{center}
\begin{tabular}{  c | c | c | c | c  }
$\textbf P$ & \ \  $(n)$ \ \  & $m_1,m_2$ &   $\mathcal Z^{(n)}_{JM} = 0$  & comments \\ \hline  \hline
$[000]$ & \ \ all \ \  & general & \ \ \  for all $M \not \in 4 \mathbb Z$, all $J \not \in 2 \mathbb Z$, $JM=20$ \ \ \  &   also $\mathcal Z^{(n)}_{J, M} =  \mathcal Z^{(n)}_{J, -M}$ \\ \hline
\ $[00d_z]$ \ &  \ \ all \ \   & general &  for all $M \not \in 4 \mathbb Z$  &  also $\mathcal Z^{(n)}_{J, M} =  \mathcal Z^{(n)}_{J, -M}$  \\ \hline 
\ $[0d_yd_z]$ \ & \ \ all \ \ & general & & $\mathcal Z^{(n)}_{J, M} =  \mathcal Z^{(n)}_{J, -M}$ \\ \hline
\ \ general \ \   &  \ \ 1 \ \   & \ \  $m_1 = m_2$ \ \ &  for all $J \not \in 2 \mathbb Z$ \ \ \ & \ \        only up to $  \mathcal O(e^{- mL})$  \ \   \\ \hline
\end{tabular}
\caption{Summary of the conditions under which $\mathcal Z^{(n)}_{JM} = 0$. \label{tab:ZJMzero}
}
\end{center}
\end{table}
\renewcommand{\arraystretch}{1.0}

To efficiently implement the formalism it is useful to identify, based on symmetry arguments, the values of $JM$ for which $\mathcal{Z}^{(n)}_{J M}(P,L,\alpha) = 0$. In this subsection we review these constraints and discuss subtleties that arise for $n > 1$. 
Our results are summarized in Table \ref{tab:ZJMzero}.

We begin with $\textbf P = [000]$. The properties of the zero-momentum zeta function are well-known \cite{Luscher:1990ux,Rummukainen:1995vs}, but we still think it useful to review the arguments here, in order to better understand the  generalization to $\textbf P \neq [000]$ and $n>1$.
Note that the zero-momentum zeta function must be unchanged if we flip $\textbf r$ everywhere in the summand. Taking the expression for $JM \neq 00$ we write
 \begin{align}
\mathcal{Z}^{(n)}_{J M}((E, \textbf 0),L,\alpha) & =      \sum_{\mathbf{r}}    \frac{  \sqrt{4 \pi}    (\vert \! - \! \textbf r \vert)^{ J} {Y}_{J M}(- \hat {\textbf r})     }{( x^{2}- (- \textbf r )^2  )^n}  e^{- \alpha ((- \textbf r )^2 - x^{2})^n}  \,.
\end{align}
Substituting ${Y}_{J M}(- \hat {\textbf r}) = (-1)^J {Y}_{J M}( \hat {\textbf r})  $ then gives $\mathcal{Z}^{(n)}_{J M}((E, \textbf 0),L,\alpha)  = (-1)^J \mathcal{Z}^{(n)}_{J M}((E, \textbf 0),L,\alpha) $ implying that the zeta function vanishes for all odd $J$. We can further rewrite the summand with $(r_x,r_y,r_z) \longrightarrow (-r_x,r_y,r_z)$ and use $Y_{JM}(\theta,  \pi - \phi) =  Y_{J,-M}(\theta,\phi)$ to show that $\mathcal{Z}^{(n)}_{J  M}((E, \textbf 0),L,\alpha)  =  \mathcal{Z}^{(n)}_{J , -M}((E, \textbf 0),L,\alpha) $. Similarly, a $\pi/2$ rotation about the $z$-axis, together with the identity $Y_{JM}(\theta,  \phi - \pi/2) = e^{- i M \pi/2} Y_{JM}(\theta,\phi)$ implies that $M$ must be divisible by $4$ for the zeta function to be non-zero. 

A final zero-momentum case worth mentioning is $JM=20$. To see that this vanishes as well, note that the corresponding spherical harmonic is proportional to $3 r_z^2 - r^2$. The sum of this structure (times a function of $\textbf r^2$) over the integer set, $\textbf r \in \mathbb Z^3$, is clearly identical to the same with $3 r_x^2 - r^2$ or with $3 r_y^2 - r^2$. Thus $\mathcal{Z}^{(n)}_{20}$ is equally well defined by averaging the three possibilities. But this gives a summand proportional to $3 ( r_x^2 + r_y^2 + r_z^2 ) - 3 r^2 = 0$, implying $\mathcal{Z}^{(n)}_{20} = 0$ as claimed.
The conditions under which $\mathcal{Z}^{(n)}_{J M} = 0$ for  $\textbf P = [000]$  are summarized in the first line of Table \ref{tab:ZJMzero}.

\bigskip

We now turn to non-zero momenta of type $\textbf P = [00 d_z]$. As this momentum type only breaks the symmetry in the $z$ direction, the invariance under $(r_x,r_y,r_z) \longrightarrow (-r_x,r_y,r_z)$, as well as the $\pi/2$ rotation around the $z$-axis, give the same constraints as for  $\textbf P = [000]$.\footnote{Of course, it is true in all cases that the sum over $\textbf r \in \mathbb Z^3$ is invariant under any octahedral transformation on $\textbf r$. The relevant question is whether this leads to a useful constraint on $\mathcal Z_{JM}^{(n)}$.} Similarly, for $\textbf P = [0 d_y d_z]$, flipping only $r_x$ gives the same relation as above. 

By contrast, parity is broken for any non-zero momentum so that the argument based on $\textbf r \to - \textbf r$ no longer holds. 
For example, for $m_1 = m_2$ and $\textbf P = [00d_z]$, the summand defining $\mathcal{Z}^{(n)}_{J M} $  now depends on 
\begin{equation}
\textbf r^\star = \bigg (r_x, \  r_y, \  -  \frac{2 \pi d_z }{E^\star \! L}  \bigg [ \frac{m^2 L^2}{4 \pi^2} + \textbf r^2 \bigg ]^{1/2}  + \frac{E}{E^\star} r_z \bigg) \,.
\end{equation}
This vector does not transform in a useful way under a flip of $\textbf r$.
Remarkably, in the degenerate-mass case, the single pole functions, $\mathcal Z^{(1)}_{JM},$ continue to vanish for all odd $J$.
More precisely, these are smooth functions with exponentially suppressed volume dependence and thus scale as terms that have been dropped in the derivation. As we now explain, this is due to an accidental symmetry  
 inherited from the non-relativistic  system.

The following argument holds for all values of $\textbf P = (2 \pi/L) \textbf d$ and so we present the results for the general case. The approach is based on the results of Ref.~\cite{\KSS} and we begin by recalling Eqs.~(62) and (66) from that work 
\begin{align}
\textbf r^\star_{\parallel}   & = \textbf R^\star_{\parallel} -  \frac{\textbf P}{E} \frac{L}{2 \pi} \frac{{k}^{\star2} - {q}^{\star2}}{\omega_{k}^\star + E^\star/2}    \,, \ \ \ \  \textbf r^\star_{\bot} =  \textbf R^\star_{\bot}  \,, \\
( x^{2}-r^{\star2}+ i \epsilon) & = \gamma \omega_{k2}^\star /\omega_{k2} (x^{2} - \textbf R^{\star2} + i \epsilon) + \mathcal O[(x^{2} - \textbf R^{\star2})^2] \,.
\end{align}
where $\textbf R^\star =  \hat \gamma^{-1}(\textbf r - \textbf d/2)$. Here $\textbf r^\star_{\parallel} $ and $\textbf r^\star_{\bot} $ are the vector components parallel and perpendicular to $\textbf d$. We have also introduced the operator $\hat \gamma^{-1} (\textbf r^\star_{\parallel} , \textbf r^\star_{\bot} ) = (   \gamma^{-1}\textbf r^\star_{\parallel} , \textbf r^\star_{\bot} ) $. Substituting these results into our definition of $\mathcal Z_{JM}^{(n)}$ we find
 \begin{align}
 \label{eq:ZnRG}
\mathcal{Z}^{(n)}_{J M}(P,L,\alpha) & =      \sum_{\mathbf{r}} \bigg [ \frac{\omega_{k2}^\star}{ \omega_{k2}}  \bigg (  \frac{\gamma \omega_{k2}^\star}{ \omega_{k2}} \bigg )^{-n}   \frac{  \sqrt{4 \pi}    (R^{\star})^{ J} {Y}_{J M}(\hat{\textbf R}^\star)     }{( x^{2}-  R^{\star 2}+ i \epsilon  )^n}  
 + \mathcal O((x^2 - r^{\star2} )^{1-n})  \bigg ]\,.
\end{align}

Now note that, for $n=1$, this function exhibits two special features, both unique to the single-pole case. First, the factors of $\omega_{k2}/\omega_{k2}^\star$ multiplying the pole exactly cancel; second, the subleading term becomes a smooth function of the summed coordinate, $\textbf r$. We thus reach
 \begin{align}
\mathcal{Z}^{(1)}_{J M}(P,L,\alpha) & =     \frac{1}{\gamma} \sum_{\mathbf{r}}     \frac{  \sqrt{4 \pi}    (R^{\star})^{ J} {Y}_{J M}(\hat{\textbf R}^\star)     }{( x^{2}-  R^{\star 2}+ i \epsilon  )} 
 + \mathcal O(e^{-m L})\,.
\end{align}
This simplified form, incidentally the form first derived for the moving frame quantization condition \cite{Rummukainen:1995vs}, makes the accidental symmetry that we are after manifest. In particular, we can now use that the sum over $\textbf r \in \mathbb Z^3$ is invariant under $\textbf r \longrightarrow \textbf d - \textbf r$. Under this transformation $\textbf R^\star \to - \textbf R^\star$ leaving $R^\star = \vert \textbf R^\star \vert$ unchanged. Thus every factor in the summand is invariant except for ${Y}_{J M}(- \hat{\textbf R}^\star) = (-1)^J {Y}_{J M}( \hat{\textbf R}^\star) $. We deduce that, for odd $J$ and degenerate masses, $\mathcal{Z}^{(1)}_{J M}=\mathcal O(e^{- m L})$. However, the remaining symmetries do not survive due to the factor of $\hat \gamma^{-1}$ in the definition of $\textbf R^\star$.

Finally we stress that the vanishing of odd $J$ due to the accidental symmetry only holds for $n=1$. As is clear from Eq.~(\ref{eq:ZnRG}), for all other $n$ values, the ratio of omegas does not cancel, leading to another factor that changes under the $\textbf r \to \textbf d - \textbf r$ transformation. Thus, while the odd-$J$ $\mathcal Z^{(1)}_{JM}$ have no poles, for $n=2$ the functions already exhibit the full double-pole behavior. [See also Fig.~\ref{fig:Zsummary}.] In addition, for $n>1$, the subleading term contains a $(n-1)$th order pole that also generates an important contribution to the zeta function.

At this stage we have completed all details relevant for $P_i = P_f$ and therefore turn to the more complicated case of $P_i \neq P_f$, beginning with the numerical integral denoted $\mathcal I_{\mathcal N; \sigma}(\alpha, P_f, P_i)$.

\subsection{Evaluating $\mathcal I_{\mathcal N; \sigma }(\alpha, P_f, P_i)$ \label{app:INeval}}

Here we give some further details on the evaluation of $\mathcal I_{\mathcal N; \sigma }(\alpha, P_f, P_i)$, defined in Eq.~(\ref{eq:INdef}) of the main text. Recall that this is a 3-dimensional integral with a smooth integrand, to be evaluated numerically. When added to the semi-analytic expression, $\mathcal I_{\mathcal A; \sigma}$, it gives the full integral entering the definition of $G_\sigma$.

The complication we wish to address here is that the definition in the main text requires evaluating a large number of integrals defined with factors of $k_\mu$ (but with no spherical harmonics) in the integrand. These are then contracted with the tensor $M^{\nu_1 \cdots \nu_N}_{\sigma}$ to reach the final expression. While highly advantageous for the analytic integral, the Lorentz-index-based expressions lead to a costly determination of $\mathcal I_{\mathcal N; \sigma}$. For example, for $\ell_f = \ell_i = 1$ and a single current index, the covariant form contains factors of $k_{\nu_1} k_{\nu_2} k_{\nu_3}$ that naively require the evaluation of $64$ terms.

To improve this situation, it is preferable to move $M^{\nu_1 \cdots \nu_N}_{\sigma}$ back inside the integrals defining $\mathcal I_{\mathcal N; \sigma}$, and thereby rewrite the \emph{integrands} to only carry the final $\sigma$ index. This procedure is a bit subtle, because the time components of the $k^\mu$ are evaluated at various values. To proceed we first recall that $\mathcal{I}_{\mathcal N, \sigma}$ generally involves terms evaluated at the physical masses, $m_1, m_2$, together with regulating integrals evaluated at some higher scale $\Lambda_j$. For the physical-mass terms, time components are evaluated at $k^0 \in \big \{ \omega_{k2}, E_f + \omega_{P_fk1}, E_i + \omega_{P_i k1} \big \}$, and for the regulating integrals at $k^0 \in \big \{ \sqrt{k^2 + \Lambda_j^2}, E_f + \sqrt{(\textbf{P}_f-\textbf{k})^2 + \Lambda_j^2}, E_i + \sqrt{(\textbf{P}_i-\textbf{k})^2 + \Lambda_j^2} \big \}$. Some of these four vectors are recombined into the harmonics, $\mathcal Y_{\ell m}(\textbf k^\star)$, and, because the time and space components mix upon boosting $k^{\star}_\mu ={{ [\Lambda_{ \pmb \beta}]_\mu}^\nu} k_\nu$, we end up with strange spatial components in some of the harmonics.

To give concrete expressions it is convenient to define
\begin{align}
\label{eq:OmegaK1}
\mathcal N_\sigma(\Omega_k,  \Lambda, \textbf k) \equiv M^{\nu_1 \cdots \nu_N}_{ \sigma} \Big [   k_{\nu_1}\cdots k_{\nu_N} \Big \vert_{k^0 = \Omega_k} \Big ]  = \mathcal   Y_{\ell_f m_{ f}}(  {\textbf k}^{\star, \Omega}_{f})  
     \Big [   k_{\mu_1}\cdots k_{\mu_n} \Big \vert_{k^0 = \Omega_k} \Big ]    \mathcal   Y^*_{\ell_i m_{ i}}(  {\textbf k}^{\star, \Omega}_{i})      \,.
\end{align}
Here $\Omega_k$ represents any of the possible choices made for the temporal component of the 4-vectors. 
In each of these three cases, an implicit mass dependence enters and the second argument, $\Lambda$, refers to this mass dependence. 
In the following we will use $m$ or $\Lambda_0$ to indicate that the $\omega$s are evaluated at their physical masses and $\Lambda_{j>0}$ to indicate evaluation at an unphysical value $m_1 = m_2 = \Lambda_{j>0}$. In short, the first two entries in $\mathcal N_\sigma$ simply serve to indicate the value at which all $k^0$ are evaluated.
We stress that the tensor $ M^{\nu_1 \cdots \nu_N}_{ \sigma}$ does not depend on these parameters but only on the c.m.~frame energies $E_i^\star$ and $E_f^\star$ as well as the boost velocities $\pmb \beta_i$ and $\pmb \beta_f$. Thus, the only modification to the spherical harmonics is that they now depend on ${\textbf k}^{\star, \Omega}$, defined via
\begin{equation}
\label{eq:OmegaK2}
(\Omega_k^\star, \ {\textbf k}^{\star, \Omega})^\mu \equiv { [\Lambda_{\pmb \beta}]^\mu}_\nu \ \big (\Omega_k, \  \textbf k \big )^\nu \,.
\end{equation}
If we set $\Omega_k = \omega_{k2}$ and $\Lambda = m$, then we exactly recover the spherical harmonic definitions used everywhere else in this work. 

With our new numerator function  in hand, we are ready to give our final form for $\mathcal I_{\mathcal N; \sigma}$
\begin{align}
\begin{split}
\mathcal I_{\mathcal N; \sigma}(\alpha, P_f, P_i) &\equiv  
   \int \! \frac{d^3 \textbf k}{(2\pi)^3} \, 
\,
 [H(\alpha, \textbf k) - 1]
 \,
\sum_{j=0}^{n_j} c_j \Big [ D(\Lambda_j, \textbf k) \mathcal  N_\sigma  (\omega_{k} , \Lambda_j, \textbf k) + \mathcal K_{r;\sigma}(\Lambda_j, \textbf k)  \Big ]  
 \\
& %
 -  \int \! \frac{d^3 \textbf k}{(2\pi)^3}
 \,
     \, H(\alpha, \textbf k)  \sum_{j=1}^{n_j}  c_j D(\Lambda_j, \textbf k)  \mathcal  N_\sigma  (\omega_{k} , \Lambda_j, \textbf k)
      \\
& %
 -  \int \! \frac{d^3 \textbf k}{(2\pi)^3} \, 
 \, H(\alpha, \textbf k)   \sum_{j=0}^{n_j} c_j     \mathcal K_{r;\sigma}(\Lambda_j, \textbf k)    
 \,,
 \end{split}
\end{align}
where
\begin{equation}
\label{eq:calKrsigma}
\mathcal K_{r;\sigma}(\Lambda, \textbf k) \equiv D_{rf}(\Lambda, \textbf k) \mathcal N_{\sigma}(E_f+\omega_{P_fk1}, \Lambda, \textbf k) +  D_{ri}(\Lambda, \textbf k) \mathcal N_{\sigma}(E_i+\omega_{P_i k1}, \Lambda, \textbf k)  \,.
\end{equation}
This is identically equal to the quantity defined in Eq.~(\ref{eq:INdef}) of the main text. The only difference is that we have absorbed $M^{\nu_1 \cdots \nu_N}_{ \sigma}$ inside the integrands, via the new function $\mathcal N_\sigma$.

To complete the specification we require explicit expressions for $D_{rf}(m, \textbf k) $ and $D_{ri}(m, \textbf k) $
\begin{align}
\label{eq:Drf}
D_{rf}(m, \textbf k) &=
\frac{1}{2 \omega_{P_fk1}}
\frac{1}
  { ( E_f+\omega_{P_fk1})^2 - \omega_{k2}^2  }
\frac{ 1} 
{  (E_i-E_f-\omega_{P_fk1})^2 - \omega_{P_ik1}^2   } \,, \\
\label{eq:Dri}
D_{ri}(m, \textbf k) &=
\frac{1}{2 \omega_{P_ik1}}
\frac{1}
  { ( E_i+\omega_{P_ik1})^2 - \omega_{k2}^2   } 
\frac{ 1} 
{  (E_f-E_i-\omega_{P_ik1})^2 - \omega_{P_fk1}^2   }  \,.
\end{align}
The motivation for these quantities is discussed in the text around Eq.~(\ref{eq:DrfImpDef}), where an implicit definition is also given.

We close this subsection with one final  simplification to $\mathcal{I}_{\mathcal{N}}$. We show that it is always possible to simplify the numerical evaluation from a three- down to a two-dimensional integral. To show this we must prove the following key identity
\begin{equation}
\label{eq:INcov}
\mathcal{I}^{ \mu_1 \cdots \mu_n}_{\mathcal N; \ell_f m_f;\ell_i m_i}(\alpha, P_f,P_i) =
 {[\Lambda_{-\pmb \beta}]^{\mu_1}}_{\nu_1}
 \cdots 
 {[\Lambda_{-\pmb \beta}]^{\mu_n}}_{\nu_n}\,
  \mathcal{I}^{ \nu_1 \cdots \nu_n}_{\mathcal{N};\ell_f m_f;\ell_i m_i}(\alpha,\Lambda_{\pmb{\beta}}P_f,\Lambda_{\pmb{\beta}}P_i ) \,.
\end{equation}

Begin the proof by considering the generic integral
\begin{equation}
\label{eq:calXdef}
\mathcal X^{ \mu_1 \cdots \mu_n}(P_f, P_i) \equiv \int \frac{d^4 k}{(2 \pi)^4} \mathcal F(P_f, P_i, k) \mathcal G(\textbf k_i^\star, \textbf k_f^\star) k^{\mu_1} \cdots k^{\mu_n} \,,
\end{equation}
assumed to be convergent. Here we have separated the integrand into a Lorentz-scalar function, $\mathcal F$, together with an arbitrary function of the c.m.~frame momenta, $\mathcal G$. The latter is also Lorentz invariant in the vacuous sense, i.e.~because its arguments carry a frame label. Now act on both sides with $\Lambda_{\pmb{\beta}}$, on each index
\begin{equation}
 {[\Lambda_{\pmb \beta}]^{\mu_1}}_{\nu_1}
 \cdots 
 {[\Lambda_{\pmb \beta}]^{\mu_n}}_{\nu_n}  \mathcal X^{ \nu_1 \cdots \nu_n}(P_f, P_i) \equiv \int \frac{d^4 k}{(2 \pi)^4} \mathcal F(P_f, P_i, k) \mathcal G(\textbf k_i^\star, \textbf k_f^\star) \  {[\Lambda_{\pmb \beta}]^{\mu_1}}_{\nu_1}
 \cdots 
 {[\Lambda_{\pmb \beta}]^{\mu_n}}_{\nu_n}  k^{\nu_1} \cdots k^{\nu_n} \,.
\end{equation}
To simplify the right-hand side we perform a change of integration variable $k'^\mu \equiv {[\Lambda_{\pmb \beta}]^{\mu}}_{\nu} k^\nu$ and also define $P_f'^\mu \equiv {[\Lambda_{\pmb \beta}]^{\mu}}_{\nu} P_f^\nu$ and $P_i'^\mu \equiv {[\Lambda_{\pmb \beta}]^{\mu}}_{\nu} P_i^\nu$ 
\begin{equation}
 {[\Lambda_{\pmb \beta}]^{\mu_1}}_{\nu_1}
 \cdots 
 {[\Lambda_{\pmb \beta}]^{\mu_n}}_{\nu_n}  \mathcal X^{ \nu_1 \cdots \nu_n}(P_f, P_i) \equiv \int \frac{d^4 k'}{(2 \pi)^4} \mathcal F(P_f', P_i', k') \mathcal G(\textbf k_{i}^\star, \textbf k_{f}^\star)  \,  k'^{\nu_1} \cdots k'^{\nu_n} \,.
\end{equation}
Here we have used the Lorentz-invariance of the various building blocks, including the fact that $\textbf k_i^\star$ and $\textbf k_i^\star$ must be unchanged if we replace $k$, $P_f$ and $P_i$ with their boosted counterparts. This result, which can be rewritten as
\begin{equation}
 {[\Lambda_{\pmb \beta}]^{\mu_1}}_{\nu_1}
 \cdots 
 {[\Lambda_{\pmb \beta}]^{\mu_n}}_{\nu_n}  \mathcal X^{ \nu_1 \cdots \nu_n}(P_f, P_i) \equiv  \mathcal X^{ \nu_1 \cdots \nu_n}(P'_f, P'_i) \,,
 \label{eq:calXcov}
\end{equation}
is just a statement of Lorentz-covariance for $\mathcal X$.

To conclude our demonstration of Eq.~(\ref{eq:INcov}) we note that $\mathcal I_{\mathcal N} \equiv \mathcal I - \mathcal I_{\mathcal A}$ and that the two terms on the right-hand side each satisfy the functional form of $\mathcal X$, shown in Eq.~(\ref{eq:calXdef}). In the case of $\mathcal I$, the original integral defining $G$, one takes
\begin{equation}
\mathcal F(P_f, P_i, k) =  \Theta(k^0) (2 \pi) \delta(k^2 - m_2^2) \frac{1}
  {  (P_f-k)^2 - m_1^2 + i \epsilon  } 
\frac{1}
{  (P_i-k)^2 - m_1^2 + i \epsilon  } \,.
\end{equation}
This function is only invariant under orthochronous transformations, as is standard when one discards the anti-particle pole, but this is sufficient for the present argument. For both $\mathcal I$ and $\mathcal I_{\mathcal A}$ the spherical harmonics, as well as the cutoff function $H(\alpha, \textbf k)$ can be absorbed into the definition of $\mathcal G$. Again the key point is that these objects are frame-independent because they carry a frame label, $\textbf k^\star (k, P) = \textbf k^\star (k', P')$. We deduce that $\mathcal I_{\mathcal N}$ must satisfy Eq.~(\ref{eq:calXcov}). Multiplying both sides by $\Lambda_{-\pmb \beta}$, we conclude Eq.~(\ref{eq:INcov}).

To see the power of this identity we take $\pmb\beta = \pmb\beta_i$, implying
\begin{equation}
[\Lambda_{ \pmb{\beta}_i}]^\mu{}_\nu P_f^\nu \equiv P'_f{}^\nu = (E'_f,\mathbf{P}'_f) ,  \qquad  [\Lambda_{\pmb{\beta}_i}]^\mu{}_\nu P_i^\nu =(E_i^\star,\mathbf{0}),
\end{equation}
and thus that only one external direction enters the integral. In this case the integration coordinate is simply transformed to $\textbf k_i^\star$. We are then left with
\begin{multline}
\mathcal{I}^{ \mu_1 \cdots \mu_n}_{\mathcal N; \ell_f m_f;\ell_i m_i}(\alpha, P_f,P_i) =
 {[\Lambda_{-\pmb \beta_i}]^{\mu_1}}_{\nu_1}
 \cdots 
 {[\Lambda_{-\pmb \beta_i}]^{\mu_n}}_{\nu_n} \\ \times \sum_k  \int \frac{d^3 \textbf k_i^\star}{(2 \pi)^3}   \mathcal Q_k(\vert  \textbf k_i^\star \vert,  \textbf k_i^\star \cdot \textbf P_f' ) \mathcal Y_{\ell_f m_f}(\textbf k_f^\star)    k_i^{\star \nu_1} \cdots    k_i^{\star \nu_n}  \mathcal Y^*_{\ell_i m_i}(\textbf k_i^\star) \bigg \vert_{k^{\star 0}_i = \Omega_k} \,,
\end{multline}
where the sum over $k$ runs over all possible choices for the temporal component, as discussed above. 
The key point here is that, once the spherical harmonics and the factors of $k^{\star \mu}_i$ are factored out, the remaining integrand can only depend on $\textbf k_i^\star$ through its magnitude and a single angle. This is because no other direction is defined in the system once we have expressed all coordinates in the rest frame of the incoming state.

Defining $\cos \theta \equiv \hat { \textbf k}_i^\star \cdot \hat { \textbf P}_f'$ and picking an aribtrary orientation for the azimuthal angle $\phi$, we reach
\begin{multline}
\mathcal{I}^{ \mu_1 \cdots \mu_n}_{\mathcal N; \ell_f m_f;\ell_i m_i}(\alpha, P_f,P_i) =
 {[\Lambda_{-\pmb \beta_i}]^{\mu_1}}_{\nu_1}
 \cdots 
 {[\Lambda_{-\pmb \beta_i}]^{\mu_n}}_{\nu_n} \\ \times \sum_k     \int \frac{ d    k_i^\star  k_i^{\star 2}  d \cos \theta}{(2 \pi)^3}   \mathcal Q_k(     k_i^\star  ,  \cos \theta) \ \int_0^{2 \pi} d \phi \, \mathcal Y_{\ell_f m_f}(\textbf k_f^\star)    k_i^{\star \nu_1} \cdots    k_i^{\star \nu_n}  \mathcal Y^*_{\ell_i m_i}(\textbf k_i^\star) \bigg \vert_{k^{\star 0}_i = \Omega_k} \,.
\end{multline}
Here the $\phi$ integral can be evaluated analytically using the rotation properties of the spherical harmonics. Thus only the $k_i^\star$ and $\theta$ integrals need to be performed numerically.

\subsection{Evaluating $\mathcal I^{\chi}(P_f, P_i, m, \delta) $ \label{app:Jeval}}

In this subsection we show how the the generating functional, $\mathcal I^{\chi}(P_f, P_i, m, \delta)$, defined in Eq.~(\ref{eq:Ialphadef}) of the main text, reduces to the results given in Eqs.~(\ref{eq:ExpAlpha})-(\ref{eq:Mdef}). We begin by inserting Feynman-parameter integrals into the definition to reach
\begin{align}
\mathcal I^{\chi}(P_f, P_i, m, \delta)
&
=  2 \int_0^1dx\int_0^{1-x}dy
\int\frac{d^{4 - \delta} k_E}{(2\pi)^{4 - \delta}}
\frac{e^{i\chi_E\cdot k_E} }{[ k^2_E+m^2-2\,k_E\cdot(x P_{f,E}+y P_{i,E})+x P_{f,E}^2+y P_{i,E}^2]^3} \,, \\
&=
2 \int_0^1dx\int_0^{1-x}dy
\, e^{i\chi_E\cdot (x P_{f,E}+y P_{i,E})}
\int\frac{d^{4 - \delta} k_E}{(2\pi)^{4 - \delta}}
\frac{e^{i\chi_E\cdot k_E}}{[k_E^2+M(m,x,y)^2]^3} \,,
\label{eq:Ifeynpar}
\end{align}
where in the second line we have performed the shift $k_E \to k_E + (x P_{f,E}+y P_{i,E})$ and have introduced
\begin{equation}
\label{eq:MdefApp}
M(m,x,y)^2 \equiv m^2+x(1-x) P_{f,E}^2+y(1-y) P_{i,E}^2-2  x y P_{f,E}\cdot P_{i,E} \,.
\end{equation}

Next note that the denominator of the integrand of Eq.~(\ref{eq:Ifeynpar}) is invariant under $k_E \to - k_E$, implying that only even terms in $k_E$ contribute to the integral. Expanding the exponential and keeping only the even powers, we find
\begin{align}
\label{eq:IchiExpanded}
\mathcal I^{\chi}(P_f, P_i, m, \delta)
&
= 
2 \int_0^1dx\int_0^{1-x}dy
\, e^{i\chi_E\cdot (x P_{f,E}+y P_{i,E})}
\int\frac{d^{4 - \delta}k_E}{(2\pi)^{4 - \delta}}
\frac{1}{[k_E^2+M(m,x,y)^2]^3}\sum_{n=0}^\infty \frac{(i\chi_E\cdot k_E)^{2n}}{(2n)!} \,.
 \end{align}

 To further reduce the expression note that we can make the substitution $(i \chi_E \cdot k_E)^{2n} \longrightarrow A_n (-\chi_E^2)^n ( k_E^2)^{n}$, where $A_n$ is a normalization constant, to be determined. This holds because the integral over $k_{E, \mu_1} \cdots k_{E, \mu_{2n}}$ (multiplied by a function of $k_E^2$) must be proportional to $\delta_{\mu_1 \mu_2} \cdots \delta_{\mu_{2n-1} \mu_{2n}} + \cdots$ where the second ellipsis indicates a sum over all possible pairings. Contracting with $\chi_{E, \mu_1} \cdots \chi_{E, \mu_{2n}}$ then gives the claimed form.

To determine the normalization, first consider the case of $n=1$, corresponding to $k_{E,\mu} k_{E,\nu} \longrightarrow A_1 \delta_{\mu \nu} k_E^2$. Taking the trace on both sides then gives $A_1 = 1/(4 - \delta)$, where we are careful to consistently perform the calculation in $4 - \delta$ dimensions. Similarly, for $n=2$ one finds 
\begin{equation}
k_{E,\mu_1} k_{E,\mu_2} k_{E,\mu_3} k_{E,\mu_4} \longrightarrow A_2 (k_E^2)^2 \frac{ \delta_{\mu_1 \mu_2}  \delta_{\mu_3 \mu_4} + \delta_{\mu_1 \mu_3}  \delta_{\mu_2 \mu_4}  + \delta_{\mu_1 \mu_4}  \delta_{\mu_2 \mu_3} }{3}  \,.
\end{equation}
First summing over $\mu_1 = \mu_2$ and then over $\mu_3 = \mu_4$ gives $A_2^{-1} =  (1/3) [(4 - \delta)^2 + 2 (4 - \delta)] $. 

The result for general $n$ can be derived by first writing
\begin{align}
A_n  \int d \Omega_{3-\delta}
& \equiv  
\int d \Omega_{3-\delta} \frac{(i \chi_E \cdot k_E)^{2n}}{(k_E^2)^n (- \chi_E^2)^n}  
=
\partial_\alpha^{2n}
\int d \Omega_{3-\delta}
 \frac{ \exp[ \alpha (i \chi_E \cdot k_E)]}{(k_E^2)^n (- \chi_E^2)^n}  
 \bigg|_{\alpha=0} \,,
\end{align}
where in the second equality we have rewritten the integral with a dummy parameter $\alpha$, to be set to zero after differentiation.  Next we multiply both sides by $\exp[- k_E^2]$ and integrate with respect to $d k_E k_E^{4-\delta}$ to write
\begin{align}
A_n  \int d ^{4-\delta}k_E
\exp[  - k_E^2   ]
&=
A_n \,\pi^{2-\delta/2}
=
\partial_\alpha^{2n}
\int d ^{4-\delta}k_E
 \frac{ \exp[  - k_E^2  + i \alpha \chi_E \cdot k_E ]}{(k_E^2)^n (- \chi_E^2)^n}  
 \bigg|_{\alpha=0}.
\end{align}

Solving for $A_n$ and evaluating the remaining integral using Schwinger parameters, we deduce
\begin{align}
A_n 
&=  \frac{1  }{  \pi^{2 - \delta/2}  (- \chi_E^2)^n     } \partial_\alpha^{2n}   \int d^{4 - \delta} k_E  \int_0^\infty d \beta
 \frac{\beta^{n-1}}{\Gamma(n)} \exp[- \beta k_E^2]  \exp[  - k_E^2  + i \alpha \chi_E \cdot k_E ] \bigg|_{\alpha=0}\,,
\\
&=  \frac{1  }{  \pi^{2 - \delta/2}  (- \chi_E^2)^n     } \partial_\alpha^{2n}     \int_0^\infty d \beta
 \frac{\beta^{n-1}}{\Gamma(n)} 
 \int d^{4 - \delta} k_E  
\  \exp \! \bigg [ 
  - (1 +\beta) 
  \left(
  k_E
  +i \frac{\alpha \chi_E}{2(\beta +1)}
  \right)^2  
- 
\frac{\alpha^2 \chi_E^2}{4(\beta +1)  }
 \bigg ] \bigg|_{\alpha=0}\,,
\\
&=  \frac{(2n)!  }{   \Gamma(n) 4^{n} {n!}  }
\int_0^\infty d \beta
 {\beta^{n-1}}{(\beta+1)^{-n-2+\delta/2}} \,,
\\
&=  \frac{(2n)!  }{    4^{n} \, n!}
 \frac{\Gamma(2-\delta/2)}
 {\Gamma(2+n-\delta/2)}
\,.
\end{align}
In the second equality we have completed the square in $k_E$ and in the third we have integrated with respect to $k_E$ and also evaluated the $\alpha$ derivative and set $\alpha = 0$.

 Substituting into Eq.~(\ref{eq:IchiExpanded}) then gives
\begin{equation}
\mathcal I^{\chi}(P_f, P_i, m, \delta)
 = 
2 \int_0^1dx\int_0^{1-x}dy
\, e^{- i\chi\cdot (x P_{f}+y P_{i})}
  \sum_{n=0}^\infty 
A_n  \frac{ (\chi^2 )^{n}}{ (2n)! }   
  \mathcal J^n(P_f, P_i, m, \delta)  \,,  
  \label{eq:ExpAlphaApp}
 \end{equation}
  where we have returned to the Minkowski signature for $\chi$, $P_f$ and $P_i$ and have also defined
 \begin{align}
 \label{eq:Jdef}
  \mathcal J^n(P_f, P_i, m, \delta) & \equiv \int\frac{d^{4 - \delta}k_E}{(2\pi)^{4 - \delta}}
\frac{ k^{2n}_{E}}{[k_E^2+M(m,x,y)^2]^3}  \,.
 \end{align}

To conclude we simplify $\mathcal J^n(P_f, P_i, m, \delta) $ by evaluating the momentum integral 
\begin{align}
\mathcal J^n(P_f, P_i, m, \delta) 
&=
\int\frac{d\Omega_{3 - \delta}}{(2\pi)^{4 - \delta}}
\,
\int_0^\infty\,{dk_E\,k_E^{3-\delta}}
\frac{k^{2n}_{E}}{[k_E^2+M(m,x,y)^2]^3}  \,, \\
&=
 \frac{2\pi^{2 - \delta/2}}{(2\pi)^{4 - \delta}\,\Gamma(2 - \delta/2)}
\,
\int_0^1  \frac{d \zeta M(m,x,y)^2}{2 \zeta^2}  
 M(m,x,y)^{2n + 2 - \delta}  (1/\zeta - 1)^{n + 1 - \delta/2}   \frac{\zeta^3}{M(m,x,y)^6} \,.
\end{align}
In the second step we have integrated over $d\Omega_{3 - \delta}$ and then changed variables via $k^2_E \equiv M^2/\zeta - M^2$. The measure is modified as $2 k_E d k_E = - M^2 d \zeta/\zeta^2$ with the integral now running from $\zeta = 0$ to $\zeta = 1$. The final factor in the second line is just $\zeta^3/M^6 = [k_E^2 + M^2]^{-3}$.

Evalauting the $\zeta$ integral via
\begin{align}
\int_0^1 d \zeta
(1-\zeta)^{\alpha-1}
\,
\zeta^{\beta-1}=\frac{\Gamma(\alpha)\,\Gamma(\beta)}{\Gamma(\alpha+\beta)} \,,
\end{align}
we conclude
  \begin{equation}
  \mathcal J^n(P_f, P_i, m, \delta)   =  \frac{1}{2(4\pi)^{2 - \delta/2}} 
  \frac{
  \Gamma({n+2 - \delta/2})
\Gamma(1-n + \delta/2)
}{ \Gamma(2 - \delta/2)}  M(m,x,y)^{2n - 2 - \delta} \,.
\label{eq:JresultApp}
 \end{equation}
This directly gives Eq.~(\ref{eq:ExpAlpha}) in the main text.

\subsection{Determining $c_j$ and $\Lambda_j$ \label{sec:detcjLambdaj}}

As discussed in Secs.~\ref{sec:rho2to2} and \ref{sec:pipi2to2}, an immediate application of the proposed formalism is electromagnetic reactions coupling two-pion states: $(\pi^+ \pi^0)_{I_i} + j_\mu \to (\pi^+ \pi^0)_{I_f}$. For this case we require the function $G_{\sigma}$ for $\sigma = [\ell_f m_f;\ell_i m_i]$ (no current index) as well as $\sigma = {[\mu; \ell_f m_f;\ell_i m_i]}$ (one index) for $\ell_i, \ell_f \leq 1$. This requires evaluating $\mathcal I_{\mathcal A  }$ (with no indices) through $\mathcal I_{\mathcal A; \nu_1 \nu_2 \nu_3 }$. This set depends on only two scalar integrals, $\mathcal J^0$ and $\mathcal J^1$. %
The integral defining $\mathcal J^0$ is convergent so that we only need the $c_0=1$, $\Lambda_0=m$ term, i.e.~no subtraction is required. The integral defining $\mathcal J^1$, by contrast, has a logarithmic divergence (arising from $d^4 k_E k_E^2/k_E^6$). This is removed using the subtraction given in Eq.~(\ref{eq:n2}), corresponding to $c_1=-1$ with $\Lambda_1 = \Lambda$ equal to any value exceeding $2 m$.

 Though the integrals of direct interest are rendered convergent by (up to) one simple subtraction, we think it useful here to give the recipe for general $\mathcal I_{\mathcal A, \nu_1 \cdots \nu_N}$. In a nut shell the approach requires identifying the divergent part of $\mathcal J^n$ and, by substituting this into the relation to $\mathcal I_{\mathcal A, \nu_1 \cdots \nu_N}$, to identify an expression of the form
\begin{equation}
\mathcal I^{\text{div}}_{\mathcal A; \nu_1 \cdots \nu_{N}}(  P_f, P_i, \delta)   \equiv    \sum_{j=0}^{\left \lfloor {N/2} \right \rfloor } c_j \mathcal I^{\text{div}}_{\mathcal A; \nu_1 \cdots \nu_{N}}(  P_f, P_i,  \Lambda_j, \delta)  \,.
\end{equation}
The coefficients $c_j$ and the scales $\Lambda_j$ are then tuned such that this quantity vanishes. Note that $\left \lfloor {N/2} \right \rfloor $ terms must be tuned to vanish. This is because the integral $\mathcal I_{\mathcal A; \nu_1 \cdots \nu_{N}}$ depends on the scalar integrals up to $n = \left \lfloor {N/2} \right \rfloor$. The integral then scales as $d^4 k _E k^{2n}_E / k_E^6$ and generates a series of divergences of the form $\Lambda^{2 n - 2}, \Lambda^{2n - 4}, \cdots, \log(\Lambda)$, so a total of $n =  \left \lfloor {N/2} \right \rfloor $ terms. 

It turns out that one does not need to tune both the regularization scales $\Lambda_j$ and the coefficients $c_j$. We thus choose the recipe of setting $\Lambda_j = 2^{j-1} \Lambda$ [$\Lambda_1 = \Lambda\,, \ \Lambda_2 = 2 \Lambda \,, \ \Lambda_3 = 4 \Lambda \,, \cdots$] and tuning only the $n$ different $c_j$ terms. 

The latter  is achieved by studying the $1/\delta$ terms. In particular one can show that $\mathcal J^n \sim M(m,x,y)^{2n - 2}/\delta$. Thus, for $n=1$ the divergence is $M(m,x,y)$-independent and is removed by setting $c_1=-1$ as explained above. For $n=2$ the divergence scales as $M(m,x,y)^2$ leading to the linear combination $M(m,x,y)^2 + c_1 M(\Lambda,x,y)^2 - (1 + c_1) M(2 \Lambda,x,y)^2$ where we have already enforced $1 + c_1 + c_2 = 0$. This leads to a cancellation in all terms except for the explicit mass and $\Lambda$ dependence 
\begin{equation}
M(m,x,y)^2 + c_1 M(\Lambda,x,y)^2 - (1 + c_1) M(2 \Lambda,x,y)^2 = m^2 + c_1 \Lambda^2 - 4 (1 + c_1) \Lambda^2 \,,
\end{equation}
and requiring this to vanish gives $c_1 = [ m^2 - 4 \Lambda^2   ]/[3 \Lambda^2]$.  Here we restrict our attention to the degenerate case, $m_1 = m_2 = m$. While the subtraction is always mass-independent for integrals up to $\mathcal I_{\mathcal A; \nu_1 \nu_2 \nu_3 }$, beyond this the scheme in the non-degenerate case becomes more complicated. 

Note that the choices of coefficients that remove $\mathcal J^2$ divergences also automatically remove those in $\mathcal J^1$. This is because $\sum_j c_j M(\Lambda_j,x,y)^2$ contains both $\Lambda_j$-dependent and independent pieces. Thus the vanishing of $\sum_j c_j M(\Lambda_j,x,y)^2$ guarantees that the same holds for $\sum_j c_j $, and the latter is the condition for removing divergences in $\mathcal J^0$. 

As we now show, this pattern continues to all orders, so that it is always sufficient to determine $c_j$ by tuning away the highest $\mathcal J^n$ divergences. For the general-$n$, degenerate case, the system that we need to solve is
\begin{equation}
  \sum_{j=1}^n c_j    \big [ M(0,x,y)^2 + 4^{j-1} \Lambda^2 \big ]^n = -   \big [ M(0,x,y)^2 + m^2 \big ]^n \,,
\end{equation}
or equivalently
\begin{equation}
  \sum_{j=1}^n c_j  \sum_{k=0}^n C(n,k)   M(0,x,y)^{2k}   4^{(j-1)(n-k)} \Lambda^{2(n-k)}  = -  \sum_{k=0}^n C(n,k)     M(0,x,y)^{2k}   m^{2(n-k)}         \,,
\end{equation}
where $C(n,k)= \frac{n!}{(n-k)! k!}$ is the binomial coefficient. Defining $v_k \equiv C(n,k)   M(0,x,y)^{2k}$ and $A_{kj} \equiv   4^{(j-1)(n-k)} \Lambda^{2(n-k)} $, the above relation becomes $v_k A_{kj} c_j = v_k (- m^{2k})$ with repeated indices summed. Dropping the $v_k$ from both sides, we conclude that a solution is given by
\begin{equation}
\begin{pmatrix}
1 & 1 & 1 & \cdots & 1 \\[5pt]
\Lambda^2 & (2 \Lambda)^2 & (4 \Lambda)^2 & \cdots & (2^{n-1} \Lambda)^2 \\[5pt]
\Lambda^4 & (2 \Lambda)^4 & (4 \Lambda)^4 & \cdots & (2^{n-1} \Lambda)^4   \\[5pt]
\vdots & \vdots  & \vdots  & \ddots    &  \vdots \\[5pt]
\Lambda^{2n} & (2 \Lambda)^{2n} & (4 \Lambda)^{2n} & \cdots & (2^{n-1} \Lambda)^{2n} 
\end{pmatrix}
\begin{pmatrix}
c_1 \\[5pt]
c_2 \\[5pt]
c_3  \\[5pt]
\vdots \\[5pt]
c_n
\end{pmatrix}
=
\begin{pmatrix}
-1 \\[5pt]
-  m^2   \\[5pt]
-  m^4    \\[5pt]
\vdots \\[5pt]
- m^{2n}
\end{pmatrix} \,.
\end{equation}
It is straightforward to invert this matrix and read-off the values of $c_i$ to regulate an integral with any number of indices. In Table \ref{tab} we give the values up to $n=5$, assuming $\Lambda_j = 2^{j-1} (3m) $.

\begin{table}
\begin{center}
\begin{tabular}{ c | c | c | c | c | c}
$n$  & $c_1$  &  $c_2$ & $c_3$ & $c_4$ & $c_5$ \\ \hline 
$1$ & $  -1$  &   &  & & \\[5pt]
$2$ & $  -\frac{35}{27}  $   &   $  \frac{8}{27}  $   &  & & \\[5pt]
$3$ & $   - \frac{1001}{729}$ & $  \frac{286}{729}$ & $ - \frac{14}{729} $ & & \\[5pt]
$4$ & $- \frac{82225}{59049}$ & $\frac{16445}{39366}$ &  $- \frac{4025}{157464}$ & $\frac{143}{472392}$   &  \\[5pt]
$5$ &  $- \frac{37872835}{27103491}$ & $\frac{5410405}{12754584}$ & $- \frac{1853915}{68024448}$ & $\frac{329329}{816293376}$ & $- \frac{16445}{13876987392}$
\end{tabular}
\caption{Values of $c_j$ up to $n=5$, assuming $\Lambda_j = 2^{j-1} (3m) $. This information is sufficient to calculate $G_{[\mu_1 \cdots \mu_n; \ell_f m_f; \ell_i m_1]}$ for all indices satisfying $n+ \ell_f + \ell_i \leq 10$. \label{tab}}
\end{center}
\end{table}

Returning to the case of different masses, here one must instead solve
\begin{equation}
  \sum_{j=1}^n c_j  \sum_{k=0}^n C(n,k) \langle \langle      M(0,x,y)^{2k}   \rangle \rangle  4^{(j-1)(n-k)} \Lambda^{2(n-k)}  = - \langle \langle \big [  M(0,x,y)^2 +(1-x-y)m_2^2+ (x+y)m_1^2  \big ]^n \rangle \rangle \,,
\end{equation}
where
\begin{equation}
\langle \langle f(x,y) \rangle \rangle \equiv 2  \int_0^1dx\int_0^{1-x}dy f(x,y) e^{i\chi\cdot (x P_{f,E}+y P_{i,E})} \,.
\end{equation}
The key distinction is that $x$ and $y$ dependence now appears in both $M(0,x,y)^2$ and the mass terms on the right-hand side. Thus it is not possible to express the right-hand side in terms of a matrix product in which $ \langle \langle      M(0,x,y)^{2k}   \rangle \rangle $ is factored off. The upshot is that, in this case, the $c_j$ depend on the kinematics $s_i, s_f, q^2$, the masses $m_1$, $m_2$ and the generating parameter $\chi$. Again we stress that this is not of immediate concern as it is only relevant for form factors with many indices.

\subsection{Triangle singularities within $\mathcal I_{\mathcal A  }(P_f, P_i)$\label{app:triangle}}

In this appendix we give a more detailed discussion of the singularities that arise away from threshold in $\mathcal I_{\mathcal A  }(P_f, P_i)$, as summarized around Eqs.~(\ref{eq:Xdef}) and (\ref{eq:DiscVal}) of the main text and in Figs.~\ref{fig:eps_dep} and \ref{fig:diff_Eis}. 
The task is to study discontinuities of the integral
  \begin{equation}\label{eq:IA_doubleAPP}
\mathcal I_{\mathcal A}( P_f, P_i, m, 0)  %
=   \frac{1}{(4\pi)^{2}}     \int_0^1dx\int_0^{1-x}dy \,
  \frac{1}{M(m,x,y)^{ 2}  } \,.
 \end{equation}
These arise when the three particles in the triangle loop of Fig.~\ref{fig:iW_Gfunc}(b) can all go on shell. As Landau described in Ref.~\cite{Landau:1959fi}, the on-shell condition is realized at critical points $(x_c, y_c)$ of $M(m,x,y)^2$ defined by three conditions: ${\partial M(m,x_c,y_c)^2}/{\partial x}=0$, ${\partial M(m,x_c,y_c)^2}/{\partial y}=0$, and $M(m,x_c,y_c)^2=0$.  Since $M(m,x,y)^2$ is, at most, a second degree polynomial in $x$ and $y$, solutions to the three conditions can be found analytically.

To see the role of these critical points in practice, we integrate Eq.~(\ref{eq:IA_doubleAPP}) with respect to $y$, to reach
\begin{equation}
\mathcal I_{\mathcal A}( P_f, P_i, m, 0)  =
\int_0^1dx   F^{(1)}(x) \,,
\label{eq:IA_00APP}
 \end{equation}
where
 \begin{align}
F^{(1)}(x) & \equiv \frac{1}{(4\pi)^{2} s_i}   \frac{\mathcal L_{+ \epsilon}[y_+(x)] - \mathcal L_{- \epsilon}[y_-(x)]}{y_+(x)-y_-(x)} \,, \\[3pt]
\mathcal L_{\pm \epsilon}[f(x)] & \equiv \log \left \lvert \frac{1 - x -f(x)}{f(x)} \right \rvert + i \arctan \! \left( \frac{1 - x-\text{Re} f(x) }{ \text{Im} f(x)  \pm \epsilon} \right )  +  i \arctan \! \left( \frac{\text{Re} f(x) }{\text{Im}f(x)  \pm \epsilon} \right ) \,,
\end{align} 
with $y_{\pm}(x) \equiv (1/2)(A \pm \sqrt{A^2+B + i \epsilon})$. Here $A$ and $B$ are known functions of the kinematic variables and the Feynman parameter $x$, defined in Eqs.~(\ref{eq:Axdef}) and (\ref{eq:Bxdef}).

We next note that the three Landau conditions given above are satisfied whenever $A^2 + B = 0$ and, in addition $d [A^2 + B]/dx = 0$. Noting that $A^2 + B$ is a quadratic polynomial in $x$ we see that the conditions are equivalent to $A^2 + B \propto (x - x_c)^2$. Before considering this special case we take our general form and substitute $A^2 + B = 2 [(P_i \cdot P_f)^2/s_i^2 - s_f /s_i] (x-x_1)(x-x_2)$
\begin{equation}
\mathcal I_{\mathcal A}( P_f, P_i, m, 0)  = \frac{1}{32 \pi^2 \sqrt{\vert (P_i \cdot P_f)^2 - s_f  s_i } \vert}   \bigg [ \int_0^{x_1} + \int_{x_1}^{x_2}  +  \int_{x_2}^1   \bigg ]  \frac{\mathcal L_{+ \epsilon}[y_+(x)] - \mathcal L_{- \epsilon}[y_-(x)]}{\sqrt{   (x-x_1)(x-x_2) } } \,.
\end{equation}
 For concreteness, here we have assumed $(P_i \cdot P_f)^2/s_i^2 - s_f /s_i > 0$. In addition we have split the integral in $x$ into a sum over regions where $\text{Im} \sqrt{A^2 + B} =0$ ($x \in [0,x_1] \cup [x_2,1]$) as well as the region where it is non-zero ($x \in [x_1,x_2]$). This separation assumes $0 < x_1 < x_2 < 1$. If these are instead complex valued, or outside the range of integration, then one can directly evaluate the integral over the entire range. For general values of $x_1$ and $x_2$, the integral over each region is well defined even with $\epsilon = 0$ and can be directly evaluated.

The final case to consider is when the external kinematics are tuned to critical values $P_f = P_{f,c}$ and $P_i = P_{i,c}$ for which $x_1 = x_2 = x_c \in [0,1]$. This is equivalent to the Landau conditions mentioned above, and corresponds to the apex of the $(x - x_1)(x - x_2)$ parabola sitting on the $x$-axis. Perturbing the kinematics in one direction shifts the parabola down, opening a finite region of $[x_1, x_2 ]$ that must be integrated in isolation. Perturbing the kinematics in the other direction shifts the parabola upwards, causing the roots to become complex such that we can directly integrate $x$ from $0$ to $1$. We now demonstrate that, as one approaches $P_{f,c}$ and $P_{i,c}$ from the side of real $x_1, x_2$, the integral $\int_{x_1}^{x_2} dx$ has a nonzero limit due to a singularity in the integration range. As a consequence, the limit has a different value when approached from opposite sides. This manifests as a step singularity in the real part of $\mathcal I_{\mathcal A}$. In addition, the imaginary part diverges as $\log\lvert P_{f(i)}-P_{f(i),c}\rvert$.

The magnitude of the discontinuity is given by evaluating the integral between $x_1$ and $x_2$, for kinematics such that $0 < x_1 < x_2 < 1$, and then taking the limit $x_1, x_2 \to x_c$. This can easily be done by noting that, in this region, $y_+(x)$ and $y_-(x)$ are complex conjugates of each other so that the integrand simplifies to
\begin{align}
\text{Disc}(\mathcal I_{\mathcal A}) 
&= \int_{x_1}^{x_2}dx F^{(1)}(x) 
= 
\frac{1}{(4\pi)^{2}}
\frac{1}{s_{i}}
\int_{x_1}^{x_2}dx \frac{1}{i\text{Im} y_+}
i\left(
\arctan\left[\frac{1-x-\text{Re} y_+}{\text{Im} y_+}\right] +
\arctan\left[\frac{\text{Re} y_+}{\text{Im} y_+}\right]
\right)\,.
\end{align}
Next we note that, as $x_1$ approaches $x_2$, $\text{Im} y_{+}$ goes to zero. Thus it is natural to expand in this quantity
\begin{align}
\text{Disc}(\mathcal I_{\mathcal A})
& = 
\frac{1}{(4\pi)^{2}}
\frac{1}{s_{i}}
\int_{x_1}^{x_2}dx \frac{1}{\text{Im} y_{+}}
\left(\pi - \frac{\text{Im} y_{+}}{1-x-\text{Re} y_{+}} - \frac{\text{Im} y_{+}}{\text{Re} y_{+}} + \mathcal{O}\left((\text{Im}y_{+})^2\right)
\right)\,.
\end{align}
We see that only the first term will contribute in the limit $x_1 \to x_2$, and that it only contributes when $y_c = y_{+}(x_c)$ is in the $y$ integration region, i.e.~$0<y_c<1-x_c$. Evaluating the remaining integral, we conclude
\begin{align}
\text{Disc}(\mathcal I_{\mathcal A})
&= \frac{1}{(4\pi)^{2}}
\frac{1}{s_{i}}
\int_{x_1}^{x_2}dx \frac{\pi}{\text{Im} y_{+}} =
\frac{1}{(4\pi)^{2}}
 \frac{1}{\sqrt{(P_i\cdot P_f)^2 - \, s_f s_i}}
\int_{x_1}^{x_2}dx \frac{\pi}{\sqrt{(x-x_1)(x_2-x)}}\, ,  \\
&= \frac{1}{16\sqrt{(P_i\cdot P_f)^2 - \, s_f s_i}}\,.
\end{align}

\subsection{Evaluating $\mathcal I_{\mathcal A  }(P_f, P_i)$ through $\mathcal I_{\mathcal A; \nu_1 \nu_2 \nu_3 }(P_f, P_i)$ \label{sec:IAsigmas}}

Here we provide compact expressions here for the $\mathcal I_{\mathcal A; \sigma}(P_f, P_i)$ integrals with up to three Lorentz indices. Starting with Eqs.~(\ref{eq:ExpAlpha}) and (\ref{eq:DerWRTalphaMINK}) one can show
\begin{align}
\mathcal{I}_{\mathcal{A};\nu_1}(P_f,P_i ,m)  
&=
P_{f,\nu_1}\,\mathcal{I}^{(1,1)}(P_f,P_i ,m)
+ P_{i,\nu_1}\mathcal{I}^{(1,2)}(P_f,P_i ,m) \,,
\\[5pt]
\begin{split}
\mathcal{I}_{\mathcal{A};\nu_1\nu_2}(P_f,P_i ,m)  
&=
P_{f,\nu_1}P_{f,\nu_2}
\mathcal{I}^{(2,1)}(P_f,P_i ,m)  
+
P_{i,\nu_1}P_{i,\nu_2}
\mathcal{I}^{(2,2)}(P_f,P_i ,m)  
\\
& \hspace{0pt} +
P_{[i,\nu_1}P_{f,\nu_2]}
\mathcal{I}^{(2,3)}(P_f,P_i ,m)  
-
\frac{g_{\nu_1\nu_2} }{4}
\mathcal{I}^{(2,4)}(P_f,P_i ,m)   \,,
\end{split}\\[5pt]
\begin{split}
\mathcal{I}_{\mathcal{A};\nu_1\nu_2\nu_3}(P_f,P_i ,m)  
&=
P_{f,\nu_1}P_{f,\nu_2}P_{f,\nu_3}
\mathcal{I}^{(3,1)}(P_f,P_i ,m)  
+
P_{i,\nu_1}P_{i,\nu_2}P_{i,\nu_3}
\mathcal{I}^{(3,2)}(P_f,P_i ,m)  \\ & \hspace{0pt}
+
{\frac{1}{2}}P_{[f,\nu_1}P_{f,\nu_2}P_{i,\nu_3]}
\mathcal{I}^{(3,3)}(P_f,P_i ,m)  
+
{\frac{1}{2}}P_{[i,\nu_1}P_{i,\nu_2}P_{f,\nu_3]}
\mathcal{I}^{(3,4)}(P_f,P_i ,m)  \\ & \hspace{0pt}
-
\frac{1}{8}{g_{[\nu_1\nu_2}P_{f,\nu_3]} } 
\mathcal{I}^{(3,5)}(P_f,P_i ,m)  
-
\frac{1}{8}{g_{[\nu_1\nu_2}P_{i,\nu_3]} }
\mathcal{I}^{(3,6)}(P_f,P_i ,m)   \,,
 \end{split}
  \end{align}
where the brackets in the indices denote a sum over permutations, even when the indices are identical. {The definition is such that, for $n$ indices within a pair of square brackets, the sum runs over $n!$ terms (some of which may vanish).} Some examples include
\begin{align}
P_{[f,0}P_{i,0]}
&=2P_{f,0}P_{i,0},
\nn\\
P_{[f,0}P_{f,0}P_{i,0]}
&={6}P_{f,0}^2P_{i,0},
\nn\\
P_{[f,1}P_{i,2]}
&=P_{f,1}P_{i,2}+P_{f,2}P_{i,1},\nn\\
{g_{[00}P_{i,1]} }&=
2{g_{00}P_{i,1} }.
\end{align}

Here we have introduced the notation $\mathcal{I}^{(n,m)}$ where the indices just index the integrals needed and do not describe a property of the integrand (i.e.~$\mathcal{I}^{(n,m)}$ is just the $m$th integral needed to evaluate the $n$-index version of $\mathcal I_{\mathcal A}$). We now define the set of relevant quantities and also give useful expressions for evaluation. 

To evaluate $ \mathcal{I}_{\mathcal{A};\nu_1}(P_f,P_i ,m)$ we require
\begin{align}
\mathcal{I}^{(1,1)}(P_f,P_i ,m)
&={\Gamma(3)}\int_0^1dx\int_0^{1-x}dy
\,x  \,\mathcal J^0(P_f, P_i, m, 0)  =
 \int_0^1dx
\,{x\,F^{(1)}(x)} \,,
\\
\mathcal{I}^{(1,2)}(P_f,P_i ,m)
&={\Gamma(3)}\int_0^1dx\int_0^{1-x}dy
\,y  \,\mathcal J^0(P_f, P_i, m, 0) 
=
 \int_0^1dx
\,{F^{(2)}(x)}.
 \end{align}
 
 For $ \mathcal{I}_{\mathcal{A};\nu_1\nu_2}(P_f,P_i ,m)$ we need
 \begin{align}
\mathcal{I}^{(2,1)}(P_f,P_i ,m)
&={\Gamma(3)}\int_0^1dx\int_0^{1-x}dy
\,x^2  \,\mathcal J^0(P_f, P_i, m, 0)
=
 \int_0^1dx
\,{x^2 F^{(1)}(x)} \,,
\\
\mathcal{I}^{(2,2)}(P_f,P_i ,m)
&={\Gamma(3)}\int_0^1dx\int_0^{1-x}dy
\,y^2  \,\mathcal J^0(P_f, P_i, m, 0)
=
 \int_0^1dx {F^{(3)}(x)} \,,
\\
\mathcal{I}^{(2,3)}(P_f,P_i ,m)
&={\Gamma(3)}\int_0^1dx\int_0^{1-x}dy
\,y\,x  \,\mathcal J^0(P_f, P_i, m, 0)
=
 \int_0^1dx
\,{x\,F^{(2)}(x)} \,,
 \\
 \mathcal{I}^{(2,4)}(P_f,P_i ,m)
&={\Gamma(3)}\int_0^1dx\int_0^{1-x}dy
\,
\mathcal J^1(P_f, P_i, m, 0)
=\,
\int_0^1dx
\,F^{(5)}(x)
+\cdots \,,
\label{eq:I24}
\end{align}
 where the ellipses denote terms that will be canceled by the Pauli-Villars-like subtractions. These are terms that are independent of $m$. For example, in the last line above we are ignoring a term proportional to $\int\,dx\,dy\,y$.

  Finally, for $ \mathcal{I}_{\mathcal{A};\nu_1\nu_2\nu_3}(P_f,P_i ,m)$, six integrals appear
  \begin{align}
\mathcal{I}^{(3,1)}(P_f,P_i ,m)
&={\Gamma(3)}\int_0^1dx\,x^3 \int_0^{1-x}dy
 \,\mathcal J^0(P_f, P_i, m, 0)
 =
 \int_0^1dx
\,{x^3\,F^{(1)}(x)}
\,, \\
\mathcal{I}^{(3,2)}(P_f,P_i ,m)
&={\Gamma(3)}\int_0^1dx\int_0^{1-x}dy
\,y^3  \,\mathcal J^0(P_f, P_i, m, 0)
=  \int_0^1dx
\,{\,F^{(4)}(x)}
\,, \\
\mathcal{I}^{(3,3)}(P_f,P_i ,m)
&={\Gamma(3)}\int_0^1dx\,x^2\int_0^{1-x}dy
\,y  \,\mathcal J^0(P_f, P_i, m, 0)
=
 \int_0^1dx
\,{x^2\,F^{(2)}(x)}
\,, \\
\mathcal{I}^{(3,4)}(P_f,P_i ,m)
&={\Gamma(3)}\int_0^1dx\,x\int_0^{1-x}dy
\,y^2  \,\mathcal J^0(P_f, P_i, m, 0)
=
 \int_0^1dx
\,{x\,F^{(3)}(x)}
\,, \\
\mathcal{I}^{(3,5)}(P_f,P_i ,m)
&={\Gamma(3)}\int_0^1dx\,x\int_0^{1-x}dy
\,\mathcal J^1(P_f, P_i, m, 0)
=
 \int_0^1dx
\,{x\,F^{(5)}(x)}
+\cdots
\,, \\
\mathcal{I}^{(3,6)}(P_f,P_i ,m)
&={\Gamma(3)}\int_0^1dx\,\int_0^{1-x}dy
\,  y\,\mathcal J^1(P_f, P_i, m, 0)
=
 \int_0^1dx
\,{F^{(6)}(x)}
+\cdots.
 \end{align}
 Again, we have ignored terms that cancel after the Pauli-Villars-like subtraction. This includes terms that are proportional to the external momenta but independent of $m$, e.g.  $P_f^\nu\int\,dx\,dy\,x$.
 
In the final steps we have evaluated the $y$-integrals analytically and expressed the remaining $x$-integral in terms of $F^{(n)}(x)$. These, in turn, can be written
 \begin{align}
F^{(1)}(x) 
&=
 \frac{1}{(4\pi)^{2} s_i}   
 \frac{\mathcal L_{+ \epsilon}[y_+(x)] - \mathcal L_{- \epsilon}[y_-(x)]}{y_+(x)-y_-(x)} 
\,,
\label{eq:f1}
 \\
F^{(2)}(x) 
&=
 \frac{1}{(4\pi)^{2} s_i}   
 \frac{y_+(x)\mathcal L_{+ \epsilon}[y_+(x)] - y_-(x)\mathcal L_{- \epsilon}[y_-(x)]}{y_+(x)-y_-(x)} 
 \,,
 \label{eq:f2}
  \\
 F^{(3)}(x) 
&
 =
 \frac{1}{(4\pi)^{2} s_i}
 \left(   
 \frac{y_+(x)^2\mathcal L_{+ \epsilon}[y_+(x)] - y_-(x)^2\mathcal L_{- \epsilon}[y_-(x)]}{y_+(x)-y_-(x)}
 \right)  
 \,,
 \label{eq:f3}
 \\ 
F^{(4)}(x) 
& 
=
 \frac{1}{(4\pi)^{2} s_i}
 \left(   
 {(1-x)\left(y_+(x)+y_-(x)\right)}
+ \frac{y_+(x)^3\mathcal L_{+ \epsilon}[y_+(x)] - y_-(x)^3\mathcal L_{- \epsilon}[y_-(x)] }{y_+(x)-y_-(x)}
 \right)
 \,,
 \label{eq:f4}
 \\
 F^{(5)}(x)
&=
\frac{-1}{8\pi^{2}}
\bigg(
(1-x-y_-(x)) \log_{\epsilon}\left({1-x-y_-(x)}\right)
+(1-x-y_+(x)) \log_{-\epsilon}\left({1-x-y_+(x)}\right)
\nn\\
&+y_-(x) \log_{\epsilon}\left({-y_-(x)}\right)
+y_+(x) \log_{-\epsilon}\left({-y_+(x)}\right)
\bigg) \,,
\label{eq:f5}
\\
F^{(6)}(x)
&
=
\frac{-1}{(4\pi)^{2}}
\bigg(
-(1-x)(y_-(x)+y_+(x))
+
((1-x)^2-y^2_-(x))\, \log_{\epsilon}\left({1-x-y_-(x)}\right)
\nn\\
&
+
((1-x)^2-y^2_+(x))\, \log_{-\epsilon}\left({1-x-y_+(x)}\right)
+y^2_-(x) \log_{\epsilon}\left({-y_-(x)}\right)
+y^2_+(x) \log_{-\epsilon}\left({-y_+(x)}\right)
\bigg) \,,
\label{eq:f6}
 \end{align}
where $\mathcal L_{\pm \epsilon}[f(x)]$ is defined in Eq.~(\ref{eq:calLdef}). $\log_{\epsilon}(f(x))$ is defined to be the standard $\log$ with its branch-cut aligned on the negative real axis. Except if $f(x)$ is purely real and negative, in which case $\log_{\pm\epsilon}(f(x))=\pm i\,\pi\,+\log(|f(x)|)$. As discussed in the main text, these expressions allow one to work identically at $\epsilon=0$ in the numerical evaluation of the integrals with respect to $x$. The only memory of the non-zero value of $\epsilon$ that these functions carry are the sign.

\bibliography{bibi}
\end{document}